\documentclass[12pt]{iopart}
\usepackage[english]{babel}
\usepackage[utf8]{inputenc}
\usepackage{iopams}
\usepackage{float}
\usepackage{graphicx}

\usepackage{bm}
\usepackage{hyperref}
\hypersetup{
colorlinks,
citecolor=blue,
filecolor=blue,
linkcolor=blue,
urlcolor=blue}

\usepackage{subfigure}
\newcommand{\be}{\begin{equation}}
\newcommand{\ee}{\end{equation}} 
\newcommand{\bea}{\begin{eqnarray}}
\newcommand{\eea}{\end{eqnarray}}

\usepackage{makeidx}
\makeindex 
\begin{document}
\title{Anisotropy in Quasi-Static Magnetohydrodynamic Turbulence}

\author{Mahendra K. Verma}

\address{Department of Physics, Indian Institute of Technology Kanpur, Kanpur 208016, India}
%\ead{submissions@iop.org}
\vspace{10pt}
\begin{indented}
\item[]August 2016
\end{indented}
%\pagenumbering{roman}%

%\input{abstract.tex}

\begin{abstract}
In this review we summarise the current status of the quasi-static magnetohydrodynamic turbulence.  The energy spectrum is steeper than  Kolmogorov's $k^{-5/3}$ spectrum due to the decrease of the kinetic energy flux with wavenumber $k$ as a result of Joule dissipation. The  spectral index decreases  with the increase of interaction parameter. The flow is quasi two-dimensional with strong ${\bf U}_\perp$ at small $k$ and weak $U_\parallel$ at large $k$, where ${\bf U}_\perp$ and $U_\parallel$ are the perpendicular and parallel components of velocity relative to the external magnetic field.  For small $k$, the energy flux of ${\bf U}_\perp$ is negative, but for large $k$, the energy flux of  $U_\parallel$ is positive.  Pressure mediates the energy transfer from ${\bf U}_\perp$ to $U_\parallel$.   

\end{abstract}

\cleardoublepage

\section{Introduction}\label{sec:intro}

Magnetohydrodynamics (MHD) deals with the interactions between the flow of electrically conducting fluids and the associated  magnetic fields~\cite{Biskamp:book:MHD,Moreau:book:MHD}.  MHD flows involving plasma are observed in the Sun, stars, solar flares, Tokamak, etc., while those involving liquid metals are found in the core of the Earth; metallurgical applications like surface and stirring controls, instability suppression, liquid metal jets; in the heat exchanger of the proposed International Thermonuclear Experimental Reactor (ITER);  and laboratory experiments.  The liquid metal flows  in territorial experiments typically have low magnetic Reynolds number, and they are often described by quasi-static (QS) MHD.  In this review we present the current status of QS MHD turbulence.

In QS MHD, the imposed external magnetic field makes the flow anisotropic.  Also, the flow is strongly damped by the Joule dissipation that affects the system properties including the energy spectrum.  In addition, the conducting or insulating walls surrounding the fluid have a strong influence on the flow.  These effects have been discussed in excellent books~\cite{Davidson:book:MHD,Molokov:book_edited,Moreau:book:MHD,Muller:book} and review articles~\cite{Knaepen:ARFM2008,Zikanov:AMR2014}.  However, recent works in the field yield interesting insights into the  dynamics and anisotropy of QS MHD turbulence.  In this review article we cover these new developments.  To keep the review focussed, we limit our attention on the bulk flow, and ignore the complexities arising due to walls.

For small and moderate interaction parameters ($N$), the  QS MHD turbulence exhibits power-law energy spectrum ($E(k) \sim k^{-a}$), with the exponent $-a$ decreasing with $N$.   Earlier researchers~\cite{Hossain:PFB1991,Kit:MG1971,Kolesnikov:FD1974} had attributed the aforementioned steepening of the spectrum (compared to Kolmogorov's spectrum) to the two-dimensionalization of the flow, and related to it to the $k^{-3}$ spectrum of the two-dimensional hydrodynamic turbulence~\cite{Kraichnan:PF1967b}.  Several other models have been constructed to explain the steepening of the spectrum.  Recently, Verma and Reddy~\cite{Verma:PF2015b} argued that the above phenomena arises because of the decrease of the energy flux with $k$ due to the Joule dissipation; they also showed that the the energy spectrum is exponential ($E(k) \sim \exp(-bk)$) for very large $N$. 

Researchers have shown that the QS MHD flow is quasi two-dimensional with strong perpendicular component of velocity at large  length scales and relatively weaker parallel component of velocity at small length scales~\cite{Alemany:JdeM1979,Branover:book_chapter,Eckert:IJHFF2001,Favier:PF2010,Reddy:PF2014,Vorobev:PF2005}.   The anisotropy of the flow has been quantified using innovative measures such as energy spectrum of the perpendicular and  parallel components of the velocity field~\cite{Vorobev:TCFD2008}, ring spectrum~\cite{Burattini:PD2008,Favier:PF2010,Reddy:PF2014}, etc.  The   energy transfers such as energy flux and ring-to-ring energy transfer too provide valuable insights into the   quasi two-dimensional nature of QS MHD turbulence.   In this review we focus on the recent developments in the field, in particular the anisotropy of the QS MHD turbulence.   

The outline of the paper is as follows: In Section 2 we describe the governing equations of QS MHD in real and Fourier spaces.  Section 3 contains discussion on the past and current  models of QS MHD, while Sections 4 and 5 describe  the primary experimental and numerical results, respectively. In Section 6  we describe the measures of anisotropy in Fourier space using ring spectrum, while Section 7 contains descriptions of energy flux, shell-to-shell energy transfers, and ring-to-ring transfers.  In section 8, we present a model of QS MHD turbulence based on variable energy-flux, as well as  review  the older models in the light of new findings. Section 9 contains a brief discussion on QS MHD flows in channels and boxes.  We conclude in section 10.

\section{Governing equations}   \label{sec:equations}

\subsection{MHD equations}
The  equations of incompressible MHD are \cite{Biskamp:book:MHD,Roberts:book,Verma:PR2004} 
\begin{eqnarray}
\frac{\partial{\bf u}}{\partial t} + ({\bf u}\cdot\nabla){\bf u}
 &=& -\nabla({p}/{\rho}) + \frac{1}{ \rho}({\bf J}\times{\bf B}) +  \nu\nabla^2 {\bf u} + {\bf f},  \label{eq:MHD1} \\ 
\frac{\partial {\bf B}}{\partial t} + ({\bf u}\cdot\nabla){\bf B }
 &=& ({\bf B}\cdot\nabla){\bf u} + \eta\nabla^2 {\bf B}, \label{eq:MHD2}
 \end{eqnarray}
\begin{eqnarray}
\nabla \cdot {\bf u} &=& 0, \label{eq:MHD3}\\
\nabla \cdot {\bf B} &=& 0, \label{eq:MHD4}
\end{eqnarray}
where {\bf u} is the velocity field, {\bf B} is the  magnetic field,  {\bf J} is the current density, {\bf f} is the external forcing, $p$ is the  pressure of the fluid,   and $\nu$, $\eta$, and $\mu$ are respectively the kinematic viscosity, magnetic diffusivity, and magnetic permeability of the fluid.   We assume the density of the fluid, $\rho$,  to be a constant.   Note that $\eta = 1/(\mu \sigma)$, where $\sigma$ is the electrical conductivity of the fluid.  In this review, we employ the SI system of units.   

Under the MHD approximation, 
\begin{equation}
  {\bf J} = \frac{1}{\mu} \nabla \times {\bf B} .
 \label{eq:Jdef} 
\end{equation} 
Hence 
\begin{equation}
  {\bf J} \times {\bf B} = -\nabla \frac{B^2}{2\mu} + \frac{1}{\mu} ({\bf B} \cdot \nabla) {\bf B} ,
 \label{eq:Jdef1} 
\end{equation} 
substitution of which in Eq.~(\ref{eq:MHD1}) yields
\begin{equation}
\frac{\partial{\bf u}}{\partial t} + ({\bf u}\cdot\nabla){\bf u}
 = -\nabla({p}_\mathrm{tot}/{\rho}) + \frac{1}{\mu \rho}({\bf B}\cdot\nabla){\bf B} +  \nu\nabla^2 {\bf u} + {\bf f},  \label{eq:MHD1.1} 
\end{equation} 
 where $p_\mathrm{tot} = p + B^2/(2\mu)$ is the total pressure.  In the later discussion, we will drop the subscript $\mathrm{tot}$ from $p$ for brevity.  In addition, in the co-moving frame of a fluid element, ${\bf J} = \sigma {\bf E}^*$, where ${\bf E}^*$ is the electric field in the co-moving frame.  Using the Lorentz transformation under nonrelativistic limit, ${\bf E}^* = {\bf E} + {\bf u \times B}$, we obtain
 \begin{equation}
 {\bf J} =  \sigma ({\bf E} + {\bf u \times B}).
  \label{eq:Jdef1.1} 
\end{equation} 
Note that {\bf E} in the above discussion is the net (sum of internal and external) electric field. Equation~(\ref{eq:Jdef}) yields a constraint on {\bf J}:
  \begin{equation}
\nabla \cdot  {\bf J} =  0
\end{equation} 
 that helps us determine ${\bf E}$ given ${\bf u}$ and ${\bf B}$ using Eq.~(\ref{eq:Jdef1.1}).

 In the momentum equation [Eq.~(\ref{eq:MHD1.1})],  $({\bf u}\cdot\nabla){\bf u}$ is  the inertial term,  $ ({\bf B}\cdot\nabla){\bf B}/(\mu \rho)$ arises due to the Lorenz force, and $\nu\nabla^2 {\bf u}$ is the viscous dissipation term. In the induction equation [Eq.~(\ref{eq:MHD2})], $({\bf u}\cdot\nabla){\bf B }$ and $({\bf B}\cdot\nabla){\bf u}$ terms represent the advection and stretching of magnetic field respectively, and  $\eta\nabla^2 {\bf B}$ is the magnetic diffusion term.  The ratio of the nonlinear term,  $({\bf u}\cdot\nabla){\bf u}$, and the viscous term, $\nu\nabla^2 {\bf u}$, is the  Reynolds number 
\begin{equation}
\mathrm{Re} = \frac{U_0 L_0}{\nu},
\end{equation}
where $U_0$ and $L_0$ are the characteristic velocity and length scales respectively.  The ratio of the nonlinear term of the induction equation [either of $({\bf u}\cdot\nabla){\bf B }$ and $({\bf B}\cdot\nabla){\bf u }$] and the magnetic diffusion term, $\eta\nabla^2 {\bf B}$, is   the magnetic   Reynolds number  
\begin{equation}
\mathrm{Rm} = \frac{U_0 L_0}{\eta}.
\end{equation}

The magnetic Prandtl number $\mathrm{Pm}$, defined as 
\begin{equation}
\mathrm{Pm} = \frac{\nu}{\eta},
\end{equation}
is one of the most important parameters of MHD.  Note that $\mathrm{Rm} = \mathrm{Re} \mathrm{Pm}$. In Table~\ref{tab:Pm_system} we list these parameters for some of the important systems (for detailed discussion, refer to Braginskii~\cite{Braginskii:RPP1965}).   Ionised plasmas are hot, and their transport properties depend on the Coulomb interactions among the ions and electrons.  Hence the Prandtl number of plasmas depend critically on temperature, and  it could take wide range of values.   The kinematic viscosity of  liquid metals is quite close to that of water, i.e., $\nu \approx v \lambda \approx 10^{-6}~\mathrm{{m}^2/s}$, where $v$ is the speed of the molecules (sound speed $\sim 10^{3}$ m/s) and $\lambda$ is the mean free path length ($\sim 10^{-9}$ m).  The above formula, strictly valid for a dilute gas, provides a reasonable estimate for $\nu$ of water.  The electrical conductivity, $\sigma$, according to Drude's formula is $n e^2 \tau/m_e \approx 10^{8}~\mathrm{S/m}$, where $n$ is the number density of electrons in the fluid, $e, m_e$ are respectively the electric charge and mass of the election, and $\tau$ is the mean collision time.   Therefore, the magnetic diffusivity of liquid metals is 
\be
\eta = \frac{1}{\mu \sigma} \approx 10^{-2}~\mathrm{m^2/s}.
\ee
Hence the  Prandtl number of a liquid metal can be estimated as $\nu/\eta \approx 10^{-4}$.  We can also estimate the above Prandtl number using
\bea
\mathrm{Pm} & = & \frac{\nu}{\eta} \approx  v \lambda \mu \sigma \nonumber \\
	& \approx & \frac{\mu_0 \epsilon_0 e^2}{\epsilon_0 m_e \lambda }  n \lambda^3  \approx \frac{e^2}{c^2 \epsilon_0 m_e \lambda }  \nonumber \\
	& \approx &  \frac{e^2}{\hslash c \epsilon_0} \frac{\hslash}{m_e c} \frac{1}{\lambda } \approx \frac{4\pi}{137} \frac{L_\mathrm{Compton}}{\lambda} \approx 10^{-4}.
\eea
Here we use $v \tau = \lambda$, $\hslash = h/2\pi$ is the reduced Planck constant, $e^2/(4\pi \epsilon_0 \hslash c)$ is the fine structure constant, $L _\mathrm{Compton} \approx 10^{-12}$ m is the Compton wavelength, and $\lambda \sim 10^{-9}$ m is the mean free-path length.   In particular,  the respective Prandtl numbers of liquid Sodium,  Gallium, Galinstan,  Mercury, Molten Iron  are  approximately  $0.88\times10^{-5}$, $1.5\times10^{-6}$, $1.4\times10^{-6}$, $1.4\times10^{-7}$, $10^{-6}$.  

The flow behaviour depends quite crucially on the system parameters. We classify them in  four regimes:
\begin{enumerate}
\item $\mathrm{Re} \ll 1$ and $\mathrm{Rm} \ll 1$:  Dissipative MHD.
\item $\mathrm{Re} \gg 1$ and $\mathrm{Rm} \ll 1$:  Liquid-metal low-Rm MHD flows for which $\mathrm{Pm} =\mathrm{Rm} /\mathrm{Re}  \ll 1$. Typical laboratory systems come under this category.  The quasi-static  MHD is a limiting case of such flows when $\mathrm{Rm} = 0$.  In this review, we will focus on this regime.
\item $\mathrm{Re} \ll 1$ and $\mathrm{Rm} \gg 1$: Laminar plasma flows for which $\mathrm{Pm} \gg 1$. Such flows are observed in laminar dynamos~\cite{Moffatt:book}.
\item $\mathrm{Re} \gg 1$ and $\mathrm{Rm} \gg 1$:   Turbulent MHD, examples of which are the Earth's outer core, solar wind, solar convection zone, sunspots, and interstellar medium~\cite{Verma:PR2004,Moffatt:book}   (refer to Table~\ref{tab:Pm_system} for the parameters).  Note that such systems could exhibit self-induced magnetic field (dynamo) since the magnetic Reynolds number is greater than unity for them \cite{Moffatt:book,Monchaux:PRL2007}.
\end{enumerate}

\begin{table}[htbp]
\begin{center}
\caption{For some important systems, the Prandtl number $\mathrm{Pm}$, the Reynolds number $\mathrm{Re}$, and magnetic Reynolds number $\mathrm{Rm}$. }
\vspace{0.5cm}
\begin{tabular}{|l| c| c| c| }
\hline 
System & $\mathrm{Pm}$ & $\mathrm{Re}$ &  $\mathrm{Rm}$       \\
\hline \hline
 liquid metal experiments (terrestrial)        & $10^{-7}$--$10^{-6}$    & $10^{3}$--$10^4$  & $10^{-4}$--$10^{-2}$    \\
 Earth's outer core & $10^{-6}$ & $10^9$ & $10^3$ \\
  Sunspots       & $10^{-3}$      & $10^{12}$   & $10^{9}$  \\
  Interstellar media       & $10^{12}$      & $10^{3}$   & $10^{15}$  \\
  \hline
%\botrule
\end{tabular}
\label{tab:Pm_system}
\end{center}
\end{table}
%%%% figure
\begin{figure}[htbp]
\begin{center}
\includegraphics{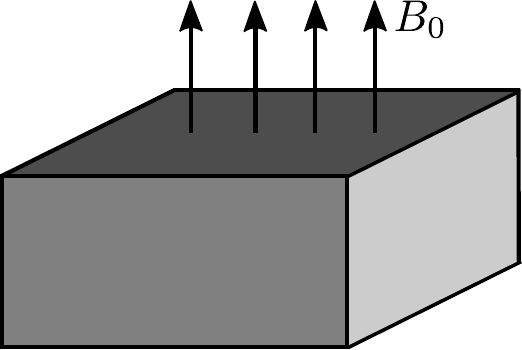}
\caption{ A schematic diagram exhibiting a magnetofluid under the influence of a constant external magnetic field $B_0 \hat{z}$.  The velocity field at the walls could satisfy periodic or no-slip boundary condition.  }
\label{fig:schmatic_real_space}
\end{center}
\end{figure}

On many occasions, plasmas or liquid metals are subjected to a constant external  magnetic field (denoted by ${\mathbf B}_0 $).  In the schematic diagram shown in Fig.~\ref{fig:schmatic_real_space}, ${\mathbf B}_0 $ points along $\hat{z}$.  For such systems, the  magnetic field of Eqs.~(\ref{eq:MHD1.1}, \ref{eq:MHD2}) can be  decomposed into its mean, $\mathbf B_0$, and  fluctuation, $\mathbf b$, i.e., ${\mathbf B} = {\mathbf B}_0 + \mathbf b$.   We rewrite Equations~(\ref{eq:MHD2}--\ref{eq:MHD4},\ref{eq:MHD1.1}) in terms of ${\mathbf B}_0$ and  $\mathbf b$ as
\begin{eqnarray}
\frac{\partial{\bf u}}{\partial t} + ({\bf u}\cdot\nabla){\bf u}
 &=& -\nabla({p}/{\rho}) + \frac{1}{\mu \rho}({\bf B}_0\cdot\nabla){\bf b} + 
\frac{1}{\mu \rho}({\bf b} \cdot\nabla){\bf b} \nonumber \\
&&  + \nu\nabla^2 {\bf u} + {\bf f},  \label{eq:momentum}  \\
\frac{\partial{\bf b}}{\partial t} + ({\bf u}\cdot\nabla){\bf b}
 &=& ({\bf b}\cdot\nabla){\bf u} + 
({\bf B}_0 \cdot\nabla){\bf u} + \eta\nabla^2 {\bf b},\label{eq:induction}
\end{eqnarray}
\begin{eqnarray}
\nabla \cdot {\bf u} &=& 0,  \label{eq:div_u_0}\\
\nabla \cdot {\bf b} &=& 0. \label{eq:div_b_0}
\end{eqnarray}
Shebalin~\cite{Shebalin:JPP1983} and Teaca {\em et al}~\cite{Teaca:PRE2009} analyzed the induced anisotropy  by the mean magnetic field ${\mathbf B}_0$.  One of the features of such flows is that the velocity along ${\mathbf B}_0$ is suppressed. This is a common feature of  anisotropic MHD turbulence and anisotropic QS MHD turbulence.

The aforementioned equations get  simplified further in the presence of a strong external magnetic field, and when $\mathrm{Rm} \rightarrow 0$.  This system is the {\em quasi-static MHD}, a topic of this review. We will quantify $B_0$-induced anisotropy in such flows.  

In the next subsection we will describe the governing equations of QS MHD.

\subsection{QS MHD} 
The magnetic Reynolds number $\mathrm{Rm}  = U_0 L_0/\eta$.  Hence, $\mathrm{Rm} \rightarrow 0$  when  $\eta \rightarrow \infty$ (or  $\mathrm{Pm} \rightarrow 0$), and  $U_0$ and $L_0$ take moderate values (in contrast, large $U_0$ and $L_0$ in planetary or astrophysical systems yield large $\mathrm{Rm}$).   However, the Reynolds number $\mathrm{Re} = \mathrm{Rm} /\mathrm{Pm} $ is nonzero.  Liquid metal flows with large $\mathrm{Re} $ are turbulent.

The magnetic Reynolds number is the ratio of the nonlinear term of the induction equation and the magnetic diffusion.  Hence in the limit $\mathrm{Rm}   \rightarrow 0$, the nonlinear terms of the induction equation [Eq.~(\ref{eq:induction})] can be ignored compared to the diffusion term, thus Eq.~(\ref{eq:induction}) reduces to
\begin{equation}
\frac{\partial{\bf b}}{\partial t} = ({\bf B}_0  \cdot\nabla){\bf u} + \eta\nabla^2 {\bf b}.\label{eq:QSMHD1}
\end{equation}
The Fourier representation  of Eq.~(\ref{eq:QSMHD1}) is
 \begin{equation}
\frac{\partial \hat{\bf b} ({\bf k}) }{\partial t} + \eta k^2 \hat{\bf b}({\bf k}) = \widehat{ [({\bf B}_0  \cdot\nabla){\bf u} ]} ({\bf k}) =  \hat{\bf f}({\bf k}, t)\label{eq:qsl}
\end{equation}
where $\hat{.}$ represents the Fourier transform.  The solution of the above equation is
 \begin{equation}
  \hat{\bf b} ({\bf k},t) = \left(  \hat{\bf b} ({\bf k},0) -  \frac{ \hat{\bf f}({\bf k}, t)}{\eta k^2} \right) \exp\left(- \eta k^2 t \right)  + \frac{ \hat{\bf f}({\bf k}, t)}{\eta k^2},   
  \label{eq:b_qs_with_transients}
\end{equation}
where $\hat{\bf b} ({\bf k},0)$ is the initial magnetic field.  For large $\eta$,  $\exp\left(- \eta k^2 t \right) \rightarrow 0$, and hence
\begin{equation}
  \hat{\bf b} ({\bf k},t) =  \frac{ \hat{\bf f}({\bf k}, t)}{\eta k^2} = \frac{\widehat{ [({\bf B}_0  \cdot\nabla){\bf u} ]} ({\bf k}) }{\eta k^2},  
\end{equation}
which is the solution of Eq.~(\ref{eq:qsl}) with $\partial \hat{\bf b} ({\bf k})/\partial t =0$. This is  the  {\em quasi-static approximation}~\cite{Knaepen:ARFM2008,Moreau:book:MHD}.  Physically, the large magnetic diffusivity quickly suppresses the transients  [the first term of Eq.~(\ref{eq:b_qs_with_transients})], and hence the induced magnetic field is proportional to  ${\bf f}({\bf k})$.  In real space, the  resulting induction equation can be written as
\begin{equation}
\eta \nabla^2{\bf b} = -({\bf B}_0 \cdot\nabla){\bf u}, \label{eq:QS}
\end{equation}
 which is Poisson's equation that yields a unique solution for ${\bf b}$ given the source term, $ -({\bf B}_0 \cdot\nabla){\bf u}$, and the boundary condition.  We write the solution (${\bf b }$) symbolically as
\begin{equation}
{\bf b } = -\Delta^{-1} \left[ \frac{1}{\eta }({\bf B}_0\cdot\nabla){\bf u} \right],
\label{eq:b_poisson}
\end{equation}
where $\Delta^{-1} $ is the inverse of the Laplacian operator. Here $\eta$ is considered to be a constant in space.  

Thus, under QS approximation, for the lowest wavenumber ($k \sim 1/L$) or large length scales,
\begin{equation}
\frac{b}{B_0}  \approx  \frac{U L }{\eta}  = \mathrm{Rm} \ll 1.
\end{equation}
Since $b \ll B_0$, in Eq.~(\ref{eq:momentum}), we  ignore the $({\bf b}\cdot\nabla){\bf b}/(\mu \rho) $ term compared to the $({\bf B}_0 \cdot\nabla){\bf b}/(\mu \rho) $. Hence Eq.~(\ref{eq:momentum}) becomes
\begin{equation}
\frac{\partial{\bf u}}{\partial t} + ({\bf u}\cdot\nabla){\bf u} = -\nabla{(p/\rho)} + \frac{ 1 } {\mu \rho}  ({\mathbf B}_0 \cdot \nabla) {\bf b}  + \nu\nabla^2 {\bf u} + {\bf f}. \label{eq:sec2_NS0}
\end{equation} 
Substitution of Eq.~(\ref{eq:b_poisson}) in Eq.~(\ref{eq:sec2_NS0}) yields the QS MHD equations:
\begin{eqnarray}
\frac{\partial{\bf u}}{\partial t} + ({\bf u}\cdot\nabla){\bf u} = -\nabla{(p/\rho)} - \frac{ \sigma } {\rho} \Delta^{-1} \left[ ({\mathbf B}_0 \cdot \nabla)^2 {\bf u}  \right] + \nu\nabla^2 {\bf u} + {\bf f},  \label{eq:NS} \\
\nabla \cdot {\bf u}=  0. \label{eq:continuity} 
\end{eqnarray}
If  $\mathbf B_0$ is  along the $z$-direction, then the Lorentz force term of Eq.~(\ref{eq:NS})  can be written as~\cite{Knaepen:ARFM2008,Moreau:book:MHD}
\begin{equation}
 -\frac{ \sigma } {\rho} \Delta^{-1} \left[ ({\mathbf B}_0 \cdot \nabla)^2 {\bf u}  \right]  =  -\frac{ \sigma B_0^2} {\rho} \Delta^{-1} \left[    \frac{\partial^2 {\bf u}}{\partial z^2}  \right].
\end{equation}  
We solve Eq.~(\ref{eq:NS},\ref{eq:continuity}) given boundary condition and initial condition.

 Under the quasi-static limit, 
\be
\nabla \times {\bf E} = -\frac{\partial {\bf b}}{\partial t} \approx 0,
\ee
hence we can write ${\bf E} = -\nabla \phi$, where $\phi$ is the electric potential. Substitution of ${\bf E} = -\nabla \phi$ in Eq.~(\ref{eq:Jdef1.1}) yields the current density
 \begin{equation}
 {\bf J} =    \sigma (-\nabla \phi+ {\bf u \times B}_0).
  \label{eq:Jdef1.2} 
\end{equation} 
Using the constraint $\nabla \cdot  {\bf J} =  0$ we obtain
\begin{equation}
\nabla^2   \phi = \nabla \cdot ({\bf u \times B}_0),
\label{eq:phi}
\end{equation} 
which is  Poisson's equation.  We solve the above equation for a given boundary condition that yields $\phi$, substitution of which in Eq.~(\ref{eq:Jdef1.2}) yields ${\bf J}$.  Once $ {\bf J} $ has been determined, we can solve for the velocity field using the following equation:
\be
\frac{\partial{\bf u}}{\partial t} + ({\bf u}\cdot\nabla){\bf u}
= -\nabla({p}/{\rho}) + \frac{1}{ \rho}({\bf J}\times{\bf B}_0) +  \nu\nabla^2 {\bf u} + {\bf f}
\label{eq:QSMHS_eqn_realspace}
 \ee
This general strategy is followed for solving bounded QS MHD flows.  

{\color{blue} Note that Eq.~(\ref{eq:b_poisson}) yields ${\bf b}$ in terms of ${\bf u}$; this  ${\bf b}$ is substituted in Eq.~(\ref{eq:sec2_NS0}), whose solution yields ${\bf u}(t)$.  The second approach, which is based on scalar potential $\phi$, is slightly different.  Here, ${\bf J}$, computed using Eq.~(\ref{eq:Jdef1.2}), is substituted for the Lorentz force term ${\bf J \times B}_0$ of Eq.~(\ref{eq:QSMHS_eqn_realspace}).  The boundary conditions for wall-bounded flows are handled somewhat differently in these two approaches.  The formulation based on ${\bf b}$ allows some freedom in the choice of boundary condition for wall-bounded flows.   This issue and the uniqueness of the induced currents are discussed in a recent paper by Bandaru {\em et al}~\cite{Bandaru:JCP2016}.}

For QS MHD, we define another important  nondimensional number called the {\em interaction parameter}, $N$, which is the ratio of the Lorentz force $(\sigma/\rho) \Delta^{-1} \left[ ({\mathbf B}_0 \cdot \nabla)^2 {\bf u}  \right] $ and the nonlinear term $({\mathbf u} \cdot \nabla) {\mathbf u}$, i.e.,
\begin{equation}
N =  \frac{\sigma {B^2_0} L_0}{\rho u_\mathrm{rms}}.
\end{equation}
This parameter plays an important role in determining the flow properties.  The diffusion time of the kinetic energy due to the Lorentz force is $t_J = \rho/(\sigma B_0^2)$, and the eddy turnover time is $t_\mathrm{eddy} = L_0/u_\mathrm{rms}$. Hence the interaction parameter can also be written as
\begin{equation}
N =  \frac{t_\mathrm{eddy}}{t_J}.
\label{eq:N_timeratio}
\end{equation}

We nondimensionalize the above equations using the characteristic velocity $U_0$ as the velocity scale, the size of the box $L_0$ as the length scale, and $L_0/U_0$ as the time scale, which yields
\begin{eqnarray}
\frac{\partial{\bf U}}{\partial t'} + ({\bf U}\cdot\nabla'){\bf U} = -\nabla'{P} -   {B^{\prime }_0}^2  \Delta^{-1} \left[  \frac{\partial^2{\bf U}}{\partial Z^2} \right] + \nu^\prime \nabla^{\prime 2} {\bf U} + {\bf F}, \label{eq:NS2}\\
\nabla^\prime \cdot {\bf U} = 0, \label{eq:continuity2} 
\end{eqnarray}
\noindent
where the nondimensionalized variables are $\mathbf U = \mathbf u/U_0$, $ \nabla' = L_0 \nabla $, $t' = t(U_0/L_0)$,  $B_0^{\prime 2}  = \sigma B_0^2 L_0 /(\rho U_0)$, ${\bf F = f}L/U_0^2$, $P=p/(\rho U_0^2) $, and $\nu^\prime=\nu/(U_0 L_0)$.   In terms of the nondimensional variables, the interaction parameter is
\begin{equation}
N = \frac{ {B'^2_0} L'}{U'}  
\end{equation}
where $U'$ is the rms value of ${\bf U'}$, and $L'$ is the correlation or integral length of the flow in the nondimensional box; both $U'$ and $L'$ will be defined subsequently.  It is important to note that $B'^2_0$ is not same as $N$, but they are of the same order since $U'$ and $L'$ are of the order unity.  We remark that in the subsequent discussion we drop the prime from $t'$.

In this review, we treat $B'_0$ as an input parameter, while $N$ as the response parameter.  The difference becomes significant particularly for decaying turbulence where $N$ can change significantly with time.  Some authors characterise the interaction parameter using the initial values of rms velocity and integral length scale (see, e.g., ~\cite{Favier:PF2010,Zikanov:JFM1998}).  Note however that for decaying turbulence, the instantaneous $N$ will differ from the initial $N$.  In this review we report the values of $N$ at the steady-state, rather than $N$ at $t=0$~\cite{Reddy:PF2014}.   We denote initial interaction parameter using a separate parameter $N_0$. We also remark that the Reynolds number in terms of nondimensional variable is  
\begin{equation}
{Re} = \frac{U'L'}{\nu^\prime}.
\end{equation}  

\subsection{QS MHD equations in the Fourier space }\label{sec:Fourier_QSMHD}

\begin{figure}[htbp]
\begin{center}
\includegraphics{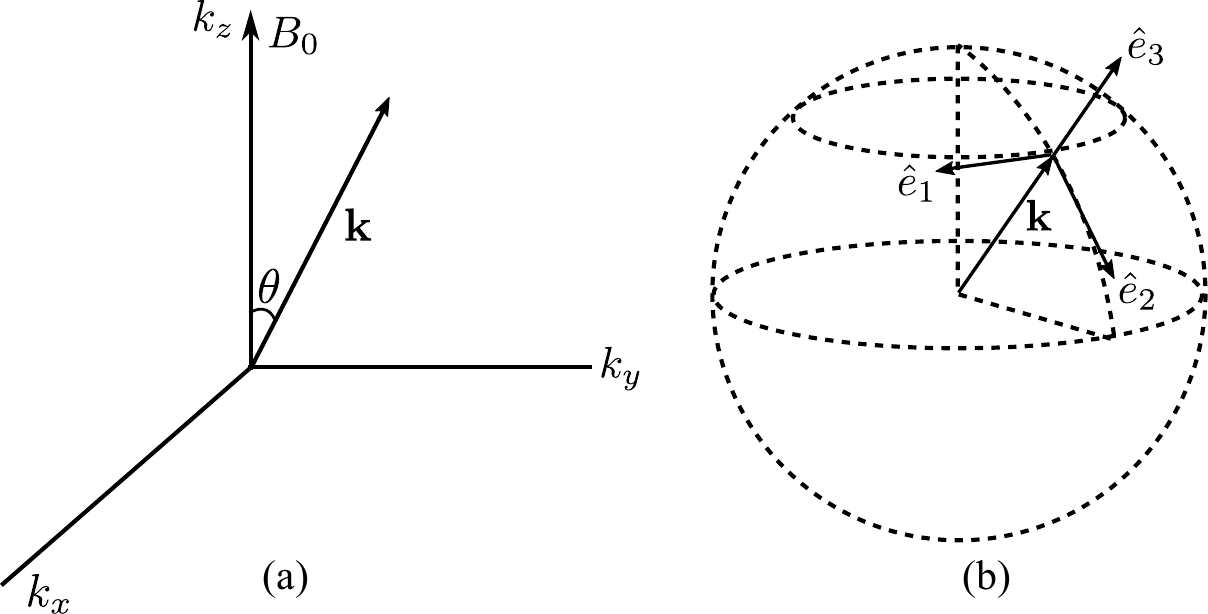}
\caption{ (a) A wavenumber {\bf k} in Fourier space.  The velocity field $\hat{\bf U}({\bf k})$ is perpendicular to ${\bf k}$. (b) Toroidal and poloidal  decomposition of a Fourier mode.  The external magnetic field is along the $z$ axis.}
\label{fig:schmatic_k_space}
\end{center}
\end{figure}

Fourier space representation is often employed in turbulence research since it captures the scale-by-scale interactions of the flow. It is also useful to quantify the energy contents at various scales. Transformation of Eqs.~(\ref{eq:NS2}) and (\ref{eq:continuity2})  to  Fourier space  yields~\cite{Knaepen:JFM2004,Knaepen:ARFM2008,Moreau:book:MHD,Schumann:JFM1976,Zikanov:JFM1998}:
\begin{eqnarray}
\frac{\partial{\hat{U}_i(\bf{k})}}{\partial t'} & = & - ik_j \sum_{\bf q} \hat{U}_j({\bf q}) \hat{U}_i({\bf{k}-\bf{q}})   - ik_i \hat{P}({\bf k}) - {B'_0}^2{\mathrm{cos^2}}(\theta)\hat{U}_i({\bf k}) \nonumber\\ & &- \, \nu' k^2 \hat{U}_i({\bf k})+\hat{F}_i({\bf k}),\label{eq:k_NS} \\
k_i  \hat{U}_i(\mathbf{k}) & = & 0,\label{eq:k_NS2}
\end{eqnarray}
where $\hat{U}_i(\mathbf{k}), \hat{F}_i(\mathbf{k})$ are  the Fourier transforms of the $i^\mathrm{th}$ components of the velocity and force fields respectively,  $ \hat{P}(\mathbf{k})$ is the Fourier transform of the pressure field, and  $\theta$ is the angle between the wavenumber vector ${\bf k}$ and the external magnetic field $\mathbf B_0$.  Refer to Fig.~\ref{fig:schmatic_k_space}(a) for an illustration.   The convolution term, $- i k_j \sum \hat{U}_j({\bf q}) \hat{U}_i({\bf{k}-\bf{q}})$, arises due to the nonlinear interactions, and it is responsible for the energy transfers from one scale to another.   Here we assume Einstein convention for the indices according to which the repeated indices are summed.  {\color{blue} Also, for brevity we drop the prime of $t'$ in subsequent discussion.}

It is convenient and insightful to decompose the velocity field using the basis function $(\hat{e}_1, \hat{e}_2,\hat{e}_3)$ shown in Fig.~\ref{fig:schmatic_k_space}(b):
\be
 \hat{e}_3 = \hat{k};~~\hat{e}_1=\hat{k}\times \hat{z};~~ \hat{e}_2=\hat{e}_3\times \hat{e}_1;
\ee
where $\hat{k}$ is the unit vector along ${\bf k}$, and $\hat{z}$ is the unit vector along ${\bf B}_0$. Due the incompressibility condition, ${\bf k \cdot \hat{U}(k)} =0$,  the velocity component along $\hat{e}_3$ vanishes, and
\be
{\bf \hat{U}(k)} = \hat{U}^{(1)}({\bf k}) \hat{e}_1 + \hat{U}^{(2)}({\bf k}) \hat{e}_2.
\ee
The components $\hat{U}^{(1)}$ and $\hat{U}^{(2)}$ are called {\em toroidal} and {\em poloidal} modes of the field. 

%%%% figure

The energy of a Fourier mode ${\bf k}$, also called {\em modal energy},  is 
\begin{equation}
E({\bf k}) = \frac{1}{2} |\hat{\mathbf U}({\mathbf k})|^2, 
\label{eq:Ek}
\end{equation}
and its evolution in Fourier space is
\begin{equation}
\frac{\partial{E({\bf k})}}{\partial t} = T({\bf k}) - 2 {{B'_0}^2}  {\mathrm{cos^2}}(\theta) E({\bf k}) - 2\nu' k^2 E({\bf k}) + \mathcal{F}({\bf k}) ,
\label{eq:dEk_dt}
\end{equation}
where $T({\bf k})$ is the rate of the nonlinear energy transfer to the mode ${\bf k}$, and $\mathcal{F}({\bf k})$ is the energy supply rate by the external forcing ${\bf F}$:
\begin{eqnarray}
T({\bf k})   = \Re[\{- ik_j \sum \hat{U}_j({\bf q}) \hat{U}_i({\bf{k}-\bf{q}})  \}   \hat{U}^{*}_i({\bf k}) ] \\
\mathcal{F}({\bf k})  = \Re \left[ \hat{F}_i({\bf k})  \hat{U}^{*}_i({\bf k}) \right],
\end{eqnarray}
where $\Re$ stands for the real part of the argument. 
Note that the pressure does not appear in the energy equation due to the incompressibility condition ${\bf k \cdot \hat{U}(k)} = 0$~\cite{Verma:PR2004}.   

The other two terms of Eq.~(\ref{eq:dEk_dt}) are the dissipative terms---the Joule dissipation rate
\begin{equation}
\epsilon_J({\bf k})  = 2 {{B^\prime_0}^{2}}\cos^{2}(\theta)E({\bf k}),\label{eq:JD}
\end{equation}
and the viscous dissipation rate
\begin{equation}
\epsilon_\nu({\bf k})  = 2 \nu' k^2 E({\bf k}).  \label{eq:VD}
\end{equation}
Note that $\epsilon_J$ is the energy transferred from the velocity field to the magnetic field, which is instantaneously dissipated due to large magnetic diffusivity $\eta$.  Also,  the Joule dissipation is active at all scales, unlike the viscous dissipation rate that  dominates at small scales.  

We also define one-dimensional energy spectrum $E(k)$ using~\cite{Lesieur:book:Turbulence}
\begin{equation}
E =  \int_0^{\infty}E(k)dk = \int E({\bf k}) d{\bf k} = \frac{3}{2}U^2.
\end{equation}
The above equation also implies that
\be
E(k) = \sum_{k-1 < k' \le k} E({\bf k}).
\label{eq:Ek_def}
\ee
The  integral length scale of the system, which is a measure of the velocity correlation length, is defined as~\cite{Burattini:PD2008,Vorobev:PF2005}
\begin{equation}
L = \frac{\pi}{(2{U}^2)} \int_0^{\infty}(E(k)/k)dk.
\end{equation}
The eddy-turnover time is defined as $\tau=L/U$. 

In a three-dimensional hydrodynamic turbulence, the term $T(k)$ facilitates energy transfer from small wavenumber modes to large wavenumber modes.  A collective effect of this transfer is a net energy flux from a wavenumber sphere, which is defined as~\cite{Kraichnan:JFM1959,Lesieur:book:Turbulence}
\begin{equation}
\Pi(k) = -\int_0^k T(k') dk'
\end{equation}
or
\begin{equation}
T(k) = -\frac{d \Pi(k)}{dk} .  
\label{eq:Tk_dpidk}
\end{equation}
We assume that the flow is in a steady state ($dE({\bf k})/dt = 0$).  Substitution of Eq.~(\ref{eq:Tk_dpidk}) in Eq.~(\ref{eq:dEk_dt})  and a summation over the modes in a shell of radius $k$ yield
\begin{equation}
 \frac{d \Pi(k)}{dk}   = - \epsilon_\nu(k)  -\epsilon_J(k) +F(k).
\label{eq:dPi_dk}
\end{equation}
Note that the external force ${\bf F}(k)$ is expected to be active only at small wavenumbers or large scales.  Thus, for $k > k_f$, where $k_f$ is the forcing wavenumber, $F(k) = 0$.  In this regime, the flux $\Pi(k)$ will decrease with $k$ since $\epsilon_J(k)$ is active at all scales~\cite{Reddy:PP2014,Reddy:PF2014,Verma:EPL2012,Verma:PF2015b}. This result is  contrary to the constant energy flux observed in fluid turbulence in which $\epsilon_\nu$ is effective only at large $k$'s~(also see Sec.~\ref{subsec:hydro_turbulence}).  The aforementioned decrease of $\Pi(k)$ has major impact on the energy spectrum of QS MHD, as well as on the anisotropy of the flow (to be discussed in Sections~\ref{sec:numerical} and \ref{sec:model} respectively).     This kind of steepening of the energy flux and spectrum are also observed in hydrodynamic turbulence with Ekman friction.  Verma~\cite{Verma:EPL2012} showed that in the presence of Ekman friction, the enstrophy flux of two-dimensional hydrodynamic turbulence decreases with $k$, while the energy spectrum $E(k)$ is steeper than $k^{-3}$ corresponding to that of 2D hydrodynamic turbulence in the constant enstrophy-flux regime.

In this review we focus on the description of anisotropy in QS MHD arising due to the external magnetic field.  We will quantify anisotropy using the energy spectrum and energy transfer diagnostics.  Our work will be focussed on a Fourier space description since it captures scale-by-scale anisotropy; this quantity is inaccessible in a real space representation.  In the Fourier space, we study the angular-dependent ring spectrum and ring-to-ring to energy transfers., in addition to standard diagnostics like energy spectrum and flux.

It is important to state the energy equation in dimensional form since many analytical works use this equation:
\begin{equation}
\frac{\partial{E({\bf k})}}{\partial t} = T({\bf k}) -  \frac{2 \sigma B_0^2}{\rho}   {\mathrm{cos^2}}(\theta) E({\bf k}) - 2\nu k^2 E({\bf k}) + F({\bf k}).
\label{eq:Ek_dimensional}
\end{equation}
Also, in Fourier space Eq.~(\ref{eq:b_poisson}) translates to 
\be
\hat{\bf b}({\bf k}) = \frac{i {\bf (B_0 \cdot k)}}{\eta k^2} \hat{\bf u}({\bf k}).
\ee

After a detailed discussion on the formalism of QS MHD, we provide a qualitative description of its dynamics. 

\subsection{Dynamics in QS MHD: a qualitative picture}
The Lorentz force on a fluid element in QS MHD is
\be
{\bf f}_L = {\bf J \times B} \approx \sigma ({\bf E + u \times B_0}) \times {\bf B}_0.
\ee
Clearly ${\bf f}_L$ is perpendicular to ${\bf B}_0$.  Equation~(\ref{eq:k_NS}) however appears to indicate that ${\bf f}_L$ is in the direction of $-{\bf u}$, but it is not the case due do the $-\nabla B^2/(2\mu)$ term [see Eq.~(\ref{eq:Jdef1})].  In Fourier space
\bea
{\bf f}_L({\bf k}) & = & {\bf \hat{J}(k) \times B}_0 \nonumber \\
	& = & i {\bf (k \times \hat{b}(k)) \times B}_0 \nonumber \\
	& = & - \frac{{\bf (B_0 \cdot k)}}{\eta k^2} {\bf (k \times \hat{u}(k)) \times B}_0 \nonumber \\
	& = & \frac{\bf (B_0 \cdot k)}{\eta k^2} \left[ {\bf k (\hat{u}(k) \cdot B_0)} -  {\bf \hat{u}(k) ( k \cdot B_0)} \right].
	\label{eq:f_L}
\eea
Clearly ${\bf f}_L({\bf k}) \cdot {\bf B}_0 = 0$, hence ${\bf f}_L({\bf r}) \cdot {\bf B}_0 = [\sum_{\bf k} {\bf f}_L({\bf k}) \exp(i {\bf k \cdot r})] \cdot {\bf B}_0  = 0$.  Thus ${\bf f}_L({\bf r})$ (in real space) is perpendicular to ${\bf B}_0$.  For large $B_0$ or $N$,  ${\bf f}_L$ dominates the nonlinear term ${\bf u \cdot \nabla u}$ and the pressure gradient.  The Lorentz force being in the $xy$ plane is one of the primary reasons for the quasi two-dimensionalization of QS MHD turbulence.   

Equation~(\ref{eq:f_L}) shows that   ${\bf f}_L({\bf k}) \propto{\bf B}_0  \cdot {\bf k}$, hence  ${\bf f}_L(k_x, k_y, 0) = 0$, i.e., ${\bf f}_L({\bf k}) $ in the $k_z=0$ plane vanishes.  Thus, in the $k_z =0$ plane, the nonlinear term $\widehat{\bf u \cdot \nabla u}({\bf k})$  dominates the other terms, and  the flow behaviour has similarities with  those in two-dimensional (2D) hydrodynamic turbulence.  We caution however that $u_z \ne 0$ in QS MHD turbulence, thus making the flow quasi two-dimensional.  Hence, QS MHD turbulence is more complex than 2D hydrodynamic turbulence.  Also note that the Fourier modes ${\bf u(k)}$ with $k_z \ne 0$ are suppressed by the Joule dissipation term that increases with $N$.  A combination of the  aforementioned effects leads to quasi two-dimensionalization of the QS MHD flow for large $N$.

The QS MHD turbulence differs significantly from Alfv\'{e}nic turbulence.  For example, Alfv\'{e}nic turbulence has large $\mathrm{Rm}$ or very small $\eta$, contrary to QS MHD turbulence for which $\mathrm{Rm} \rightarrow 0$ or $\eta \rightarrow \infty$.  The linearized QS MHD equation is
\be
\frac{\partial{\hat{U}_i(\bf{k})}}{\partial t} = - {B'_0}^2{\mathrm{cos^2}}(\theta)\hat{U}_i({\bf k}) ,
\ee
whose solution yields the following  linear mode of QS MHD:
\be
\hat{U}_i({\bf k}) = \exp( - {B'_0}^2 t{\mathrm{cos^2}} \theta).
\ee
  This dissipative mode is very different from an Alfv\'{e}n wave, which is a solution  of the linear Alfv\'{e}nic MHD with $\nu = \eta = 0$~\cite{Biskamp:book:MHD,Moreau:book:MHD}.   In  QS MHD, the kinetic energy is directly transferred to the Joule dissipation, and it does not support any MHD wave.

In Alfv\'{e}nic turbulence, $\eta \rightarrow 0$ or $\sigma \rightarrow \infty$.  Hence, 
\be
{\bf E} + {\bf u \times B}_0 = \frac{{\bf J}}{\sigma} \approx  0.
\ee
However for QS MHD, 
\be
{\bf E} + {\bf u \times B}_0 = \frac{{\bf J}}{\sigma}  \ne 0
\ee
since $\sigma$ is finite.

 In this review we will discuss energy spectrum and flux of the bulk flow of QS MHD turbulence.  The phenomena to be discussed are strongly motivated from the hydrodynamic  turbulence. Hence, in the next subsection we introduce the phenomenology of hydrodynamics turbulence briefly.

\subsection{A brief introduction to hydrodynamic turbulence phenomenology}
\label{subsec:hydro_turbulence}
Most of the flows in laboratory experiments and terrestrial atmosphere can be considered to be incompressible since the density variation in such flows is only a small fraction of the mean density.  Therefore, such hydrodynamic flows are described by Navier-Stokes equation:
\bea
\frac{\partial{\bf u}}{\partial t} + ({\bf u}\cdot\nabla){\bf u}
= -\nabla({p}/{\rho}) +  \nu\nabla^2 {\bf u} + {\bf f}, \\
\nabla \cdot {\bf u}  = 0.
 \eea
 The flow becomes turbulent when the nonlinear term is much larger than the viscous term, or when $\mathrm{Re} = UL/\nu \gg 1$.  Without loss of generality, we take $\rho=1$.
 
Modelling turbulent flow has been a key problems of physics.  One of the most acclaimed theory of turbulence is by Kolmogorov~\cite{Kolmogorov:DANS1941a}. Here external force ${\bf f}$ is assumed to be active at large length scales (of the order of system size), i.e. for  $k = k_f  \sim 1/L$.  The energy supplied by ${\bf f}$  cascades to smaller scales, and finally it is dissipated at the dissipation scales.  When we employ Eq.~(\ref{eq:dPi_dk}) to hydrodynamic turbulence for $k > k_f$, we have $\epsilon_J =0$, and the energy supply rate by the external force $F(k) = 0$.  Therefore
\be 
 \frac{d \Pi(k)}{dk}   = \epsilon_\nu =- 2\nu k^2 E(k).
\label{eq:dPi_dk_hydro}
\ee
For turbulent flows, the viscous dissipation dominates in the dissipation range, i.e. for $k>k_d$, where $k_d$ is the dissipation wavenumber. Therefore, in the wavenumber band $k_f < k < k_d$, called the {\em inertial range}, $F(k), D(k) \rightarrow 0$, and hence Eq.~(\ref{eq:dPi_dk}) yields $ d \Pi/dk  \approx 0$.  Therefore the energy flux remains an approximate constant in the inertial range, i.e.,
\be
\Pi(k) = \Pi = \mathrm{const},
\label{eq:Kolm_Pik}
\ee
and it equals the total dissipation rate. 

Now, using dimensional analysis, one can derive  the one-dimensional energy spectrum as
\be
E(k) = K_{K_o} \Pi^{2/3} k^{-5/3}
\label{eq:Kolm_Ek}
\ee
where $ K_{K_o}$ is the Kolmogorov's constant.  Numerical simulations, experiments, and analytical tools report that $ K_{K_o} \approx 1.6$. 

The space dimension does not appear explicitly in the above set of arguments, hence we may expect Eqs.~(\ref{eq:Kolm_Pik},\ref{eq:Kolm_Ek}) to describe both two-dimensional and three-dimensional (3D) flows.  Three-dimensional hydrodynamic turbulence exhibits $\Pi(k)$ and $E(k)$ given by Eqs.~(\ref{eq:Kolm_Pik},\ref{eq:Kolm_Ek}) respectvely, but these formulae are not valid in 2D hydrodynamic turbulence.  In inviscid 2D hydrodynamics (with $\nu=0$), the total energy, $u^2/2$, and the total enstrophy, $\omega^2/2$, are conserved, contrary to 3D hydrodynamics in which only the total energy is conserved.  The aforementioned conservation laws for 2D hydrodynamics leads to very different turbulence phenomenology in 2D~\cite{Kraichnan:PF1967b}. Here the fluid is forced at $k_f \gg 1/L$. Kraichnan~\cite{Kraichnan:PF1967b} showed that in 2D hydrodynamic turbulence, for $k < k_f$,
\be
E(k) =  K_{2D} \Pi^{2/3} k^{-5/3},~~~\Pi = \mathrm{const} < 0,
\ee
where $K_{2D}$ is a constant, and $\Pi$ is the energy flux.  However for $k > k_f$,
\be
E(k) =  K'_{2D} \Pi_\omega^{2/3} k^{-3},~~~\Pi_\omega = \mathrm{const} > 0,
\ee
where $\Pi_\omega$ is the enstrophy flux, and $K'_{2D}$ is another constant.  Note that in 2D hydrodynamic turbulence, the kinetic energy exhibits an inverse cascade, while the enstrophy flux shows a forward cascade.
 
In the paper, we show that  for small interaction parameter $N$ ($N \lessapprox 1$),  QS MHD turbulence has similarities with 3D hydrodynamic turbulence, with the spectral exponent  close to $-5/3$.  However for large $N$, the energy spectrum is steeper than $k^{-5/3}$, and the spectral exponent decreases with $N$, reaching as low as $\approx (-5)$ for intermediate $N$ (e.g. $N=27$).  For very large $N$,  the energy spectrum of QS MHD turbulence becomes exponential, i.e. $E(k) \sim \exp(-bk)$ where $b$ is a constant.

In engineering applications and in planetary interiors,  the QS magnetofluid is often confined between walls~\cite{Moreau:book:MHD,Muller:book} that have significant effects on the flow, which will be discussed briefly in Sec.~\ref{sec:wall}. The present section does not contain any discussion on the walls.    In this review we focus on the bulk flow in QS MHD where the aforementioned equations provide adequate  description.

In the next section we will describe some of the analytical models of QS MHD turbulence.

\section{Analytical Models of QS MHD turbulence}   \label{sec:analytic}

The equations of QS MHD are nonlinear, hence their general analytical solution is  not available.  In the past, researchers have constructed models for QS MHD turbulence some of which will be described below.  Keeping in mind the theme of the review, we will focus on the models for the bulk flow.

  Moffatt~\cite{Moffatt:JFM1967} and Schumann~\cite{Schumann:JFM1976} were one of the first to model the energy distribution in QS MHD.  They imagined an isotropic magnetofluid in which an external magnetic field is suddenly turned on.  They studied how the fluid energy in such flows decays with time.  In the early stages, the velocity correlation function is described by an isotropic second-rank tensor: 
\begin{equation}
\langle  \hat{u}_i({\mathbf k},t)  \hat{u}_j^*({\mathbf k},t)  \rangle =  \phi_{ij}({\mathbf k},t) = \left( \delta_{ij} - \frac{k_i k_j}{k^2} \right) E({\bf k}),
\label{eq:u_tensor}
\end{equation}
where $E({\bf k})$ is as defined in Equation~(\ref{eq:Ek}), and $\delta_{ij}$ is the Kronecker delta function.   Moffatt~\cite{Moffatt:JFM1967} assumed that for sufficiently large $N$ and small $\nu$ (large Reynolds number), the nonlinear energy transfer $T(k)$ is weak compared to the Lorentz force.  Hence he modelled the energy equation for the decaying QS MHD turbulence as [see Eq.~(\ref{eq:Ek_dimensional})]
\begin{equation}
\frac{\partial{E({\bf k})}}{\partial t} =  - \frac{{2\sigma B_0}^2}{\rho} E({\bf k}) {\mathrm{cos^2}}\theta.
\label{eq:Ek_decaying}
\end{equation}
Note that $\theta$ is a function of ${\bf k}$.  According to the above, the energy is dissipated more strongly near the polar region ($\theta \approx 0$) than the equatorial region.  The solution of the above equation is
\begin{equation}
E({\bf k}, t) =  E({\bf k}, 0) \exp\left(- (2 {\mathrm{cos^2}}\theta) \frac{t}{t_J} \right), 
\label{eq:Ek_decaying_soln}
\end{equation}
where $t_J =  \rho/(\sigma B_0^2)$ is the kinetic-energy diffusion time-scale due to the Lorentz force.   

At  time $t$, the spectrum is effectively damped for $\theta < \theta_c$ where 
\begin{equation}
\cos \theta_c = \sqrt{t_J/t}.
\label{eq:cos_theta_c}
\end{equation}
Moffatt~\cite{Moffatt:JFM1967}  derived the evolution of the total energy  as
\begin{eqnarray}
E(t) & = &  \int E({\bf k}, 0) \exp\left(- (2 {\mathrm{cos^2}}\theta) \frac{t}{t_J} \right) d\mathbf k \nonumber \\
 & = & \int k^2 dk d(\cos\theta) d\phi E({\bf k}, 0) \exp\left(- (2 {\mathrm{cos^2}}\theta) \frac{t}{t_J} \right) \nonumber \\
  & = &  \sqrt{\frac{t_J}{t}} \int k^2 dk d\left(\cos\theta  \sqrt{\frac{t}{t_J}} \right) d\phi E({\bf k},0) \exp\left(- (2 {\mathrm{cos^2}}\theta) \frac{t}{t_J} \right) \nonumber \\
 & = & K \sqrt{\frac{t_J}{t}},
\label{eq:E_decaying}
\end{eqnarray}
where $K$ is value of the integral of the third line, whose dimension is $u^2$. In the last step of the above equation, we make a change of variable from $\cos \theta$ to $\cos \theta\sqrt{t/t_J}$.  Thus, Moffatt~\cite{Moffatt:JFM1967} argued that the total energy of QS MHD decays as $t^{-1/2}$.  Since Moffatt~\cite{Moffatt:JFM1967} and Schumann~\cite{Schumann:JFM1976} ignored the nonlinear term in the above derivation, the above decay law is said to be applicable in the {\em linear regime}. The above assumption is clearly  invalid at $\theta \approx \pi/2$, where the nonlinear term ${\bf u \cdot \nabla u}$ is the most dominant term.  We will discuss these issues in Sec.~\ref{sec:model}.  

Using Eq.~(\ref{eq:u_tensor}), Moffatt~\cite{Moffatt:JFM1967}  concluded that 
%\begin{align}
\begin{eqnarray}
E_\parallel  & = & \frac{1}{2}u_z^2 = \frac{1}{2} \left(1-\frac{k_z^2}{k^2}\right)  E(k) =  \frac{1}{2} (1-\cos^2 \theta)  E(k), \label{eq:Moffatt_E1} \\
E_\perp &  = &  \frac{1}{2} (u_x^2+u_y^2) = \frac{1}{2} \left[2-\frac{k_x^2+k_y^2}{k^2}\right]  E(k) =  \frac{1}{2} (2-\sin^2\theta) E(k).  \label{eq:Moffatt_E2} 
\end{eqnarray}
Due to two-dimensionalization of QS MHD flows, $\theta \approx \pi/2$.  Hence 
\be
E_\parallel = E_\perp= E/2. \label{eq:Moffatt_Erelation}
\ee
  However, the numerical simulations of QS MHD turbulence exhibit very different behaviour, as will be shown in Sec.~\ref{sec:energy}.  Moreover, the velocity-velocity correlation tensor of QS MHD is anisotropic, hence it cannot be described by Eq.~(\ref{eq:u_tensor}). See Sections~\ref{subsubsec:tensorial} and \ref{sec:model}
 for further discussion.

Using Eqs.~(\ref{eq:Moffatt_E1}, \ref{eq:Moffatt_E2}), Sommeria and Moreau~\cite{Sommeria:JFM1982}  (also see Knaepen and Moreau~\cite{Knaepen:ARFM2008}) computed the ratio of the length scales parallel and perpendicular to the mean magnetic field as
\begin{equation}
\left(\frac{l_\parallel}{l_\perp}\right)^2 = \frac{\int d{\bf k} k_\perp^2 E({\bf k})}{2 \int d{\bf k} k_\parallel^2 E({\bf k})} = \frac{\int d{\bf k} (1-\cos^2\theta) E({\bf k})}{2 \int d{\bf k} (\cos^2\theta)  E({\bf k})} .
\end{equation}
Using Eq.~(\ref{eq:cos_theta_c}) and setting $\theta \approx \pi/2$, they argued that
\begin{equation}
\left(\frac{l_\parallel}{l_\perp}\right)^2 \approx \frac{1}{\langle \cos^2\theta \rangle} \approx \frac{t}{t_J}
\end{equation}
or
\begin{equation}
 \frac{l_\parallel}{l_\perp}  \sim \sqrt{\frac{t}{t_J}}.
\end{equation}
Thus, $l_\parallel$ elongates with time as $t^{1/2}$, and $l_\parallel/l_\perp$ saturates at approximately one eddy turnover time.  Hence using Eq.~(\ref{eq:N_timeratio}), Sommeria and Moreau~\cite{Sommeria:JFM1982}  obtained
\begin{equation}
\frac{l_\parallel}{l_\perp} \sim \sqrt{N}.
\end{equation}

Using dimensional analysis Sreenivasan and Alboussi\`{e}re~\cite{Sreenivasan:EJMB2000,Sreenivasan:JFM2002} derived the time evolution of QS MHD turbulence in the {\em nonlinear regime} under the assumption of conservation of  total angular momentum.  They obtained the following set of equations:
\bea
E^{1/2} l^2_\perp l^{1/2}_\parallel  =  \mathrm{const},  \\
\frac{\mathrm{d}E}{\mathrm{d}t}   \sim    -\frac{E}{t_J}\left(\frac{l_\perp}{l_\parallel}\right)^2, \\
N_t  =   \frac{l^2_\perp l^{1/2}_\parallel}{t_J E^{1/2}} \left(\frac{l_\perp}{l_\parallel}\right)^2 \sim 1,
\eea
where $N_t$ was called the {\em true interaction parameter}~\cite{Sreenivasan:EJMB2000,Sreenivasan:JFM2002}.  The solutions of the above equations are
\bea
\frac{E}{E_0}  \sim \left[1+\frac{1}{N_0}\frac{t}{t_J}\right]^{-1}, \\
\frac{l_\parallel}{l_0} \sim  N_0^{2/5} \left[1+\frac{1}{N_0}\frac{t}{t_J}\right]^{3/5}, \\
\frac{l_\perp}{l_0}  \sim  N_0^{-1/10} \left[1+\frac{1}{N_0}\frac{t}{t_J}\right]^{1/10},
\eea
which yield $E \sim t^{-1}$ for large $t$, contrary to the Moffatt's decay law according to which $E \sim t^{-1/2}$ [see Eq.~(\ref{eq:E_decaying})].   Sreenivasan and Alboussi\`{e}re~\cite{Sreenivasan:JFM2002} argued that their decay law is similar to that of Alemany {\it et al}~\cite{Alemany:JdeM1979}.    

Experiments and numerical simulations of QS MHD show that the energy spectrum of the flow is steeper than 3D hydrodynamic turbulence, which is described by Kolmogorov's $k^{-5/3}$ spectrum.   Kit and Tsinober~\cite{Kit:MG1971}, Kolesnikov and Tsinober~\cite{Kolesnikov:FD1974}, and Hossain~\cite{Hossain:PFB1991} invoked 2D hydrodynamic turbulence phenomenology and argued that $E(k)$ is near $k^{-3}$ due to two dimensionalization of the QS MHD turbulence.  Verma and Reddy~\cite{Verma:PF2015b} however argued that the steepening of the energy spectrum is due to the loss of energy flux $\Pi(k)$ due to the Joule dissipation (to be detailed in Sec.~\ref{sec:model}).   In a related work, Ishida and Kaneda~\cite{Ishida:PF2007} derived an expression for the velocity spectrum of QS MHD using perturbation  method  and showed that $E(k) \sim k^{-7/3}$. 

The QS MHD flow is quasi two-dimensional with strong ${\bf U}_\perp$ and small $U_\parallel$. Researchers have attempted to compute the energy exchange among ${\bf U}_\perp$ and  $U_\parallel$ using various mechanisms.  Thess and Zikanov~\cite{Thess:JFM2007} performed linear stability analysis of QS MHD in a triaxial ellipsoid and unbound QS MHD flows to model the transition from two-dimensional flows to three-dimensional flows.   They observed that the two-dimensional flows  become three-dimensional abruptly with a sudden burst.     Klein and Poth{\'e}rat~\cite{Klein:PRL2010} and Poth{\'e}rat~\cite{Potherat:EPL2012} proposed that  {\em barrel effect} is  responsible for the transformation of a quasi-2D flow to a 3D flow in wall-bounded geometries; here two-dimensional rotational currents play an important role.  Favier {\em et al}~\cite{Favier:PF2010,Favier:JFM2011}  and Reddy {\em et al}~\cite{Reddy:PP2014} argued that ${\bf U}_\perp$ feeds energy to $U_\parallel$, thus making the flow quasi two-dimensional.  Favier {\em et al}~\cite{Favier:JFM2011} performed EDQNM (Eddy-Damped Quasi-Normal Markovian) closure to QS MHD turbulence and found results similar to their numerical work~\cite{Favier:PF2010}.   It will be interesting to find detailed connections between the  anisotropy mechanisms proposed by Thess and Zikanov~\cite{Thess:JFM2007}, Klein and Poth{\'e}rat~\cite{Klein:PRL2010}, Favier {\em et al}~\cite{Favier:PF2010}, and Reddy {\em et al}~\cite{Reddy:PP2014}.

The theoretical arguments described in the present section are  inspired by several experiments.  In addition, experiments have been preformed to test some of the aforementioned turbulence models.  We will describe some key experiments of QS MHD turbulence in the next section.

\section{Experiments of QS MHD turbulence} \label{sec:experiment}

In this section we will describe some of the leading experimental results on QS MHD turbulence. In all these experiments, a turbulent flow is subjected to an external magnetic field.  Turbulence is typically provided by the interaction of the flow with a grid, as in laboratory experiments involving hydrodynamic turbulence.  The velocity fluctuations are measured by potential probes.  The frequency spectrum is computed from the velocity time series, and the wavenumber spectrum $E(k)$ is interpolated from the frequency spectrum using Taylor's hypothesis~\cite{Pope:book}. 

Alemany {\it et al}~\cite{Alemany:JdeM1979} performed an experiment in which turbulence is generated by a moving grid in a mercury column, and computed the energy spectrum for various interaction parameters. For small interaction parameters ($N<3$), $ E(k) \sim k^{-5/3}$, but for large interaction parameters,  $E(k) \sim k^{-3}$.   Alemany {\it et al} argued that the $k^{-3}$ energy spectrum for large $N$ is  due to the quasi-equilibrium between the Joule dissipation and the angular energy transfers.  They also showed that the decay rate of the kinetic energy follows $E(t) \sim t^{-1.7}$. 
  
Branover \textit{et al}'s~\cite{Branover:book_chapter} performed experiment on a mercury channel in the presence of a constant magnetic field.  They generated turbulence in the flow using a honey-comb grid.  For Branover \textit{et al}, the range of $N$ is 0.15--27  and that of the Hartman number (defined in Sec.~\ref{sec:wall}) is 60--1200.  They reported that for small interaction parameters ($N\sim 1$), the spectral index is approximately $-5/3$, but for moderate and large interaction parameters, the  spectral exponents range from $-7/3$ to $-11/3$.   

\begin{figure}
\begin{center}
\includegraphics[scale=1.2]{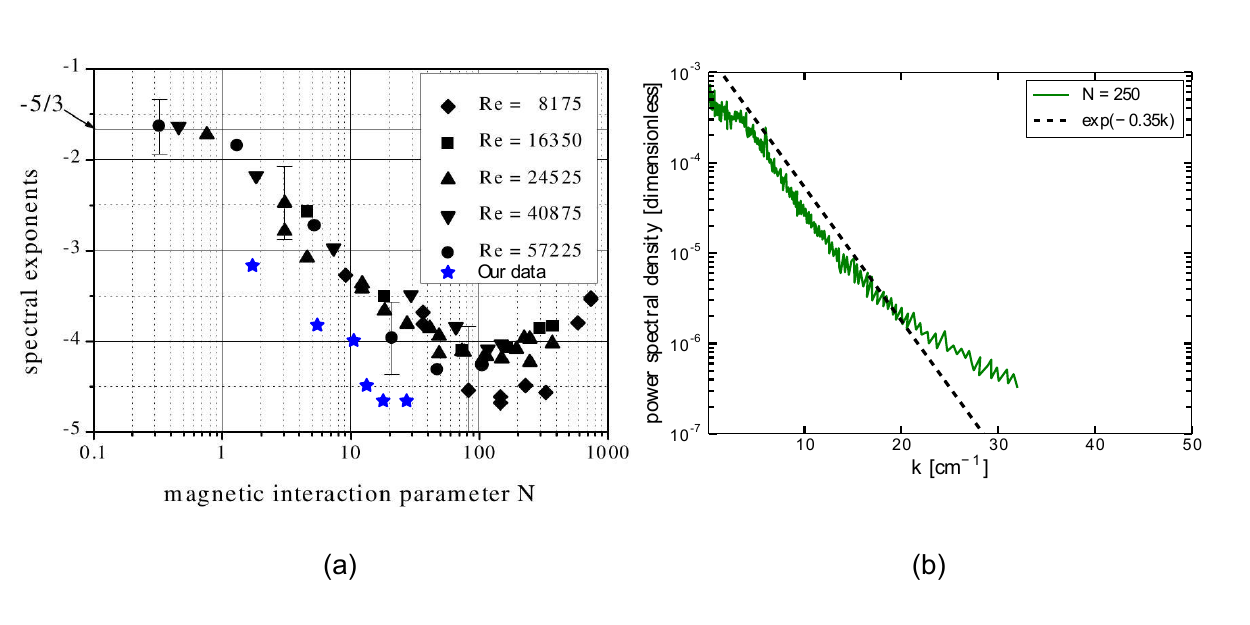}
 \end{center}
\caption{  (a) A plot of the spectral exponents vs. $N$ as reported by Eckert \textit{et al}~\cite{Eckert:IJHFF2001} in liquid-sodium experiment. (b) A plot of $E(k)$ vs.~$k$ for $N=250$.   Figure (a) is adopted from Eckert \textit{et al}~\cite{Eckert:IJHFF2001}.  From Reddy {\em et al}~\cite{Reddy:PF2014}. Reprinted with permission from AIP Publishing. }
 \label{fig:Eckart}
 \end{figure}

Eckert \textit{et al}~\cite{Eckert:IJHFF2001} studied the energy spectrum in a  liquid sodium  channel with turbulence enhancers  to reduce the effects of M-profiles~\cite{Muller:book} in the flow.   Between the Hartmann layers they observed quasi 2D vortices  aligned along the external magnetic field.    Eckert \textit{et al}  showed that the spectral exponent  decreases  from $-5/3$ to $-5$ as interaction parameter $N \in [0.3, 1000]$ is increased, as illustrated in Fig.~\ref{fig:Eckart}(a).   For very large $N$, the spectral exponent of approximately $-5$ is too steep, and it is better described by exponential spectrum, i.e. $E(k) \sim \exp(-a k)$, as indicated by Reddy \textit{et al}~\cite{Reddy:PF2014}.  In Fig.~\ref{fig:Eckart}(b) we illustrate how the experimental $E(k)$ of Eckert \textit{et al} for $N=250$ is better described by an exponential spectrum than the power law spectrum. This issue will be revisited in Sections~\ref{sec:numerical} and \ref{sec:model}.

Kolesnikov and Tsinober~\cite{Kolesnikov:FD1974},  and Kit and Tsinober~\cite{Kit:MG1971} performed  experiment of QS MHD turbulence and observed steepening of the energy spectrum with the increase of $N$.  They attribute the above feature to two-dimensionalization of the flow.  Sreenivasan and Alboussi\`{e}re~\cite{Sreenivasan:JFM2002} performed an experiment on mercury in a channel subjected to a uniform external magnetic field.   They reported that the duration of the initial {\em linear} decaying phase  of the MHD flow increases with an increase of the interaction parameter.  Klein and Poth{\'e}rat~\cite{Klein:PRL2010} and Poth{\'e}rat and Klein ~\cite{Potherat:JFM2014} performed QS MHD experiment to explore how the flow becomes three-dimensional.  They showed that the inertia and two-dimensional rotational currents make the flow three-dimensional.  

The liquid metal flows in engineering and planetary interiors typically involve walls that affect the flow due to the induced wall currents.  However in this review we focus on flows where the effects of walls are negligible.  This is to study the bulk properties of QS MHD turbulence away from the walls.  In Sec.~\ref{sec:wall} we  briefly describe the behaviour of QS MHD with walls, as well as  pattern formation  in a box containing liquid metal~\cite{Sommeria:JFM1986,Herault:EPL2015}.

In numerical simulations we observe similar features as above.  We will describe them in the  next section.

\section{Simulation of QS MHD turbulence}  \label{sec:numerical}

 In this section, we will describe how QS MHD turbulence is simulated using computers.  The equations of QS MHD are solved in a given volume for a given boundary condition and initial condition.  The equations are solved in real space using the finite difference, finite volume, and finite element methods, or in Fourier space using the pseudo-spectral method. 

Liquid metal flows in industrial applications and laboratory experiments involve complex geometries where the walls play important role.  Such flows are best solved in real space using finite difference, finite volume, and finite element methods.   Equations~(\ref{eq:Jdef1.2}, \ref{eq:phi}, \ref{eq:QSMHS_eqn_realspace}) are solved in this scheme.  For a given boundary condition on $\phi$, the Poisson's equation, Eq.~(\ref{eq:phi}), yields $\phi$, which is used to compute ${\bf J}$.  After this, Eq.~(\ref{eq:QSMHS_eqn_realspace}) is used to solve for the velocity field.  In this review we focus on bulk flows in QS MHD turbulence for which spectral method is more appropriate.  Hence we do not discuss the finite difference, finite volume, and finite element methods in detail.   Refer to Vantieghem {\em et al}~\cite{Vantieghem:TCFD2009} and references therein for details.

Often we try to understand the properties of the bulk flow by ignoring the boundary effects.  For such studies, periodic boundary condition is employed, and the equations are   solved conveniently using  pseudo-spectral method.  {\color{blue} This scheme also allows us to explore structures, energy, and anisotropy at different scales;  this exploration is the main objective of the review.}   Researchers  have simulated QS MHD turbulence using pseudospectral method for various values of $\mathrm{Re}$ and $N$, and studied energy spectrum and flux. 

In a channel flow, to be discussed in \ref{sec:wall} in somewhat more detail,  typically no-slip boundary condition (${\bf u} =0$) is employed at the walls.  Such systems are solved using finite difference, finite volume,  finite element methods, or pseudospectral method with Chebyshev polynomials or other special basis functions satisfying no-slip boundary condition~\cite{Dymkou:TCFD2009}.  One example of  Chebyshev implementation is by Boeck {\em et al}~\cite{Boeck:PRL2008}  who simulated low-Rm MHD flow in a channel with no-slip boundary conditions for Reynolds number of 8000.  Note that  the flows with very thin boundary layers (called Hartmann layer for QS MHD turbulence) are very expensive to compute using conventional spectral methods due to extreme resolution required to simulate the sharp velocity gradients in the Hartmann layer. The cost of such computation increases with the increase of Ha. Dymkou and Poth\'{e}rat~\cite{Dymkou:TCFD2009}  overcame this difficulty by formulating a new basis function  based on the least dissipative modes.  Using this method Kornet and Poth\'{e}rat~\cite{Kornet:JCP2015} performed direct numerical simulations of MHD flows in a channel.

Here we present a short summary of the numerical results of QS MHD turbulence.  In Secs.~\ref{sec:energy} and \ref{sec:ET} we will report the anisotropic spectra and anisotropic energy transfers deduced using the numerical data obtained from spectral studies. 

\subsection{Quasi 2D nature of QS MHD  flow}
 
Numerical simulations (e.g.~\cite{Hossain:PFB1991,Schumann:JFM1976,Zikanov:JFM1998}) show that the QS MHD flow is nearly isotropic for small interactions $N$, but it is quasi 2D for large  $N$.  The degree of two-dimensionality increases with the increase of  $N$. Schumann~\cite{Schumann:JFM1976} simulated decaying QS MHD turbulence in a periodic box for $N$ ranging from 0 to 50. For large $N$,  he  reported that  the velocity fluctuations along $B_0$ are strongly suppressed, and the flow becomes quasi 2D.   Zikanov and Thess~\cite{Zikanov:JFM1998} performed forced simulations and observed that the flow remains three-dimensional and turbulent for a low interaction parameter ($N \approx 0.1$), but becomes quasi-two-dimensional with sporadic bursts for a moderate interaction parameter ($N \approx 0.4$), and purely quasi-2D for a high interaction parameter ($N \approx 10$).   Burattini \textit{et al}~\cite{Burattini:PF2008} and Vorobev et al~\cite{Vorobev:PF2005}   studied anisotropy of QS MHD turbulence.   Vorobev et al~\cite{Vorobev:PF2005} quantified the flow anisotropy  and showed that for $N = 5$,  $E_\perp(k)/E_\parallel(k) > 1$ at low wavenumbers, and $E_\perp(k)/E_\parallel(k) < 1$ at higher wavenumbers.  We will details these results in Sec.~\ref{sec:energy}. 

\begin{figure}
\begin{center}
\includegraphics{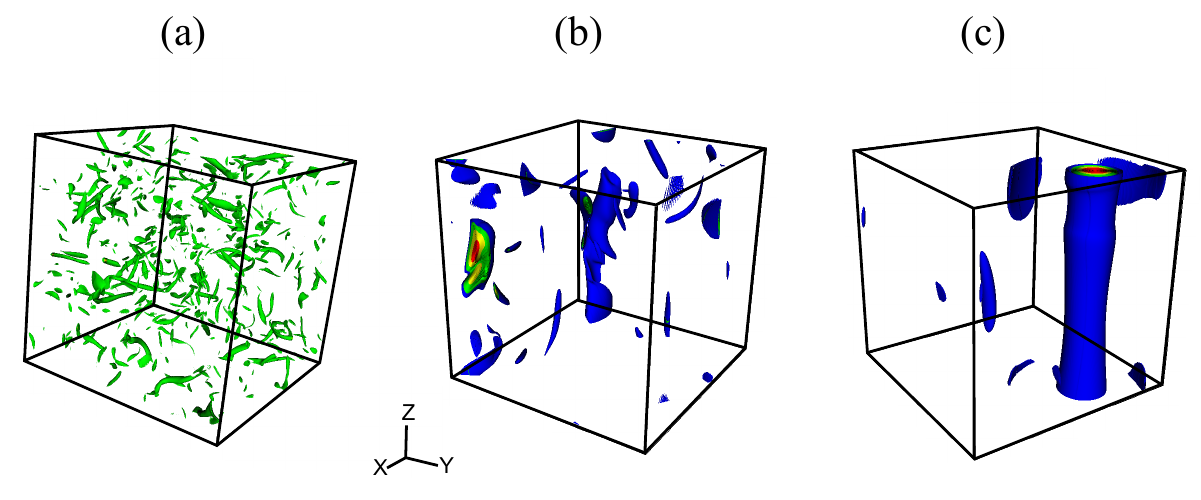}
\end{center}
\caption{Isosurfaces of the absolute value of vorticity, $|\nabla \times {\bf u}|$, for (a) $N = 0$ (isotropic), (b)  $N = 5.5$, and (c)  $N = 18$. The flow field is anisotropic for $N\ne 0$.  From Reddy {\em et al}~\cite{Reddy:PF2014}. Reprinted with permission from AIP Publishing.  }    
 \label{fig:isosurface_vorticity}
\end{figure}

\begin{figure}
\begin{center}
\includegraphics{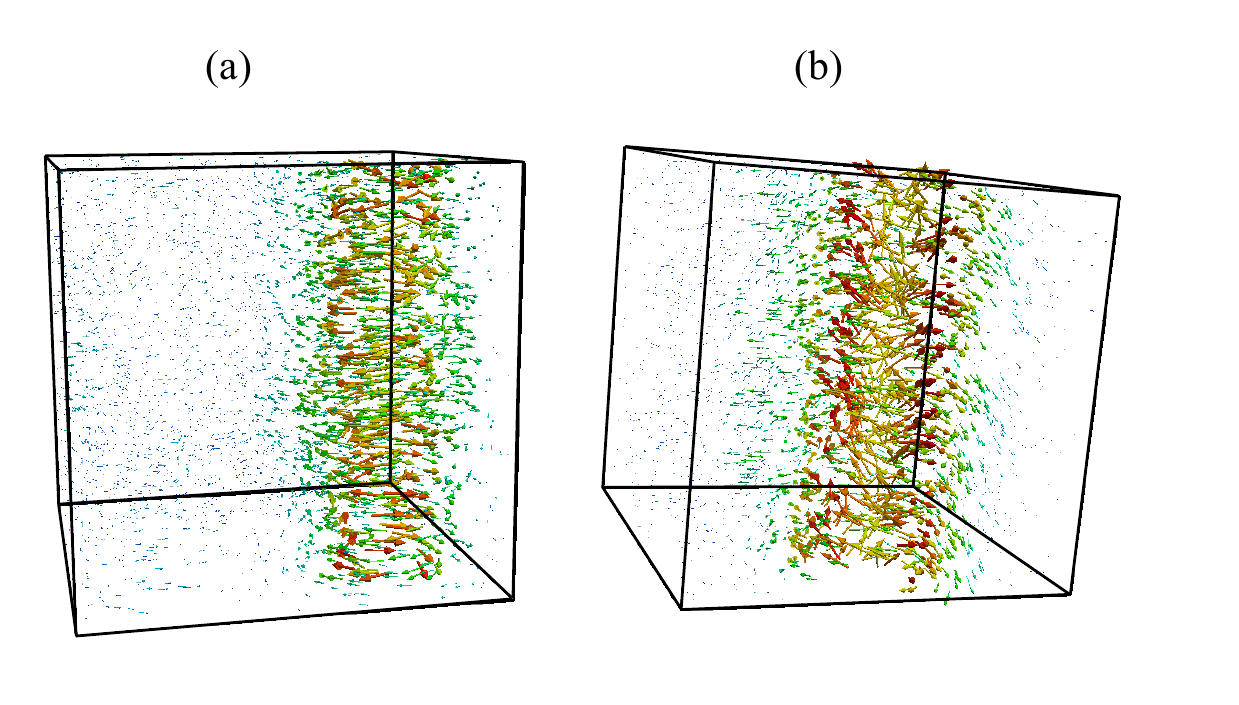}
 \end{center}
\caption{Plot of the velocity field for (a) $N=18$  and (b) $N = 130$ exhibiting quasi 2D flow.  $U_z$ For $N=130$ is stronger than that for $N=18$. From Reddy {\em et al}~\cite{Reddy:PF2014}. Reprinted with permission from AIP Publishing.}  
 \label{fig:N132_vectors}
 \end{figure}

Favier \textit{et al}~\cite{Favier:PF2010} performed  simulation of decaying QS MHD for $N$ ranging from 1 to 5 and studied anisotropy.   They  showed that the flow is  two-dimensional with three-components (2D-3C).  Later,  Reddy and Verma~\cite{Reddy:PF2014}, and Reddy {\em et al}~\cite{Reddy:PP2014} performed  spectral simulations of forced QS MHD turbulence for $N$ from 0 to 220 and observed similar behaviour.  We illustrate these features using several flow profiles. In Fig.~\ref{fig:isosurface_vorticity} we illustrate the isocontours of the vorticity field for $N=0$, $5.5$, and $18$, with ${\bf B}_0$ along $\hat{z}$, and in Fig.~\ref{fig:N132_vectors} we show the velocity vector for $N=18$ and 130.  The flow is isotropic for $N=0$, but it starts to become anisotropic as $N$ takes larger values. For $N=18$ and 130, the flows have strong ${\bf U}_\perp =  U_x \hat{x} + U_y \hat{y} $ and weak $U_\parallel = U_z$. Clearly the flow is not two-dimensional, but quasi two-dimensional, or two-dimensional with three-components (2D-3C)~\cite{Favier:PF2010,Reddy:PF2014}.   For $N=132$, the strength of $U_z$ is larger than that for $N=18$, as shown in Fig.~\ref{fig:N132_vectors}. In later sections we will investigate how the anisotropy changes with $N$, and explore  the reasons for the quasi-2D nature of  QS MHD turbulence.

%\begin{figure}[htbp]
 
 Another important aspect of spectral simulation is the quantification of energy spectrum, which is described in the next subsection.
  
\subsection{One-dimensional energy spectrum} \label{subsec:Ek}
Most spectral works report one-dimensional energy spectrum $E(k)$ which is defined using Eq.~(\ref{eq:Ek_def}).   The aforementioned one-dimensional energy spectrum describes the average energy in a shell.  This is useful since it describes the energy contents at different scales.  It also helps us contrast QS MHD turbulence with isotropic hydrodynamic turbulence.

As described in Sec.~\ref{subsec:hydro_turbulence}, for 3D hydrodynamic turbulence ($N=0$),  $E(k) \sim \Pi^{2/3} k^{-5/3}$ [see Eq.~(\ref{eq:Kolm_Ek})].  However for 2D hydrodynamic turbulence, $E(k) \sim \Pi^{2/3} k^{-5/3}$ for $k < k_f$ and $E(k) \sim  \Pi_\omega^{2/3} k^{-3}$ for $k>k_f$, where $k_f$ is the forcing wavenumber band, and $\Pi_\omega$ is the enstrophy flux.  The numerical results  described in the present section and and the experiments results of Sec.~\ref{sec:experiment}  reveal that $E(k)$ is steeper than  Kolmogorov's $k^{-5/3}$ power law, but it also differs from  2D hydrodynamic turbulence. Understanding $E(k)$ of  QS MHD turbulence is one of the major topics of this review.

\begin{figure}
\begin{center}
\includegraphics[scale=1]{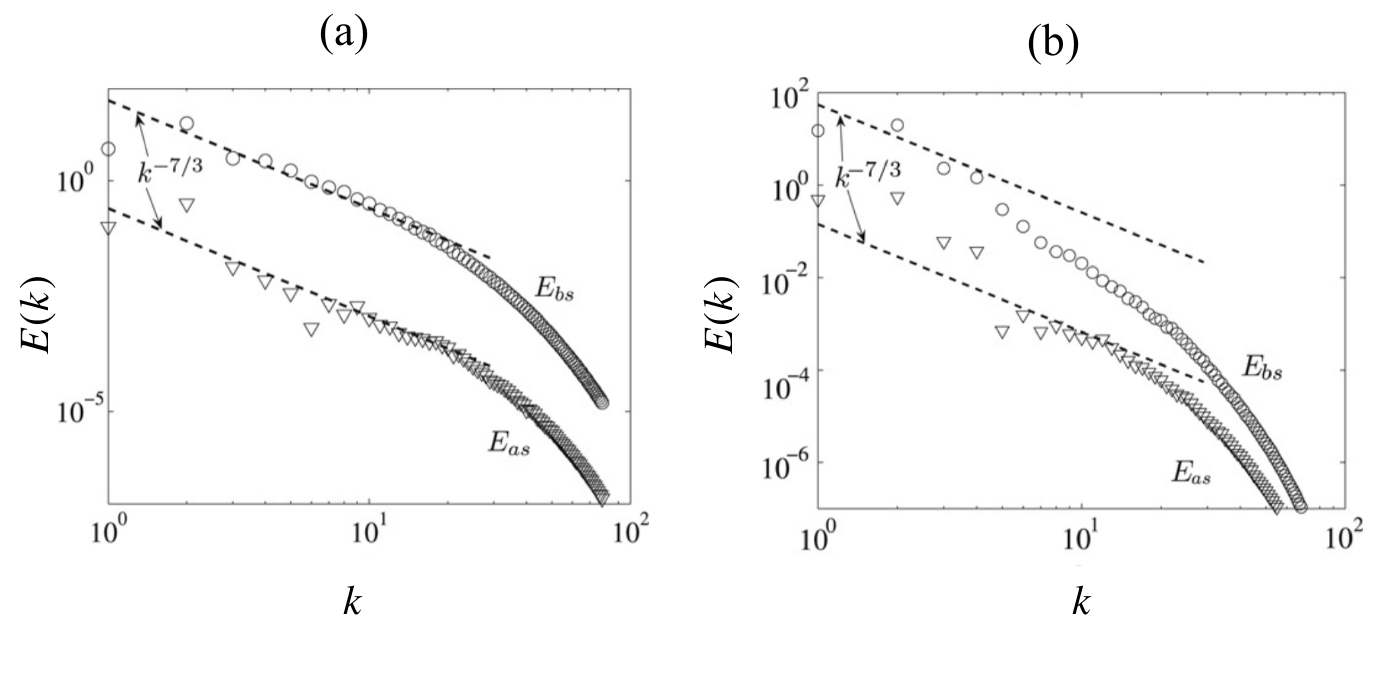}
\end{center}
\caption{{\color{blue} Anisotropic energy spectrum $E_{as}(k)$ and $E_{bs}(k)$ reported by   Burattini \textit{et al}~\cite{Burattini:PD2008} for (a) $N=1$ and (b) $N=5$.  $E(k) \sim k^{-7/3}$ for $N=1$, but it is steeper than $ k^{-7/3}$ for $N=5$.  The definitions of  $E_{as}(k)$ and $E_{bs}(k)$ are given by Eqs.~(\ref{eq:Ea},\ref{eq:Eb}) respectively}. From Burattini \textit{et al}~\cite{Burattini:PD2008}.  Reprinted with permission from Elsevier.}  
 \label{fig:Burattini}
\end{figure}

\begin{figure}
\begin{center}
\includegraphics[scale=0.3]{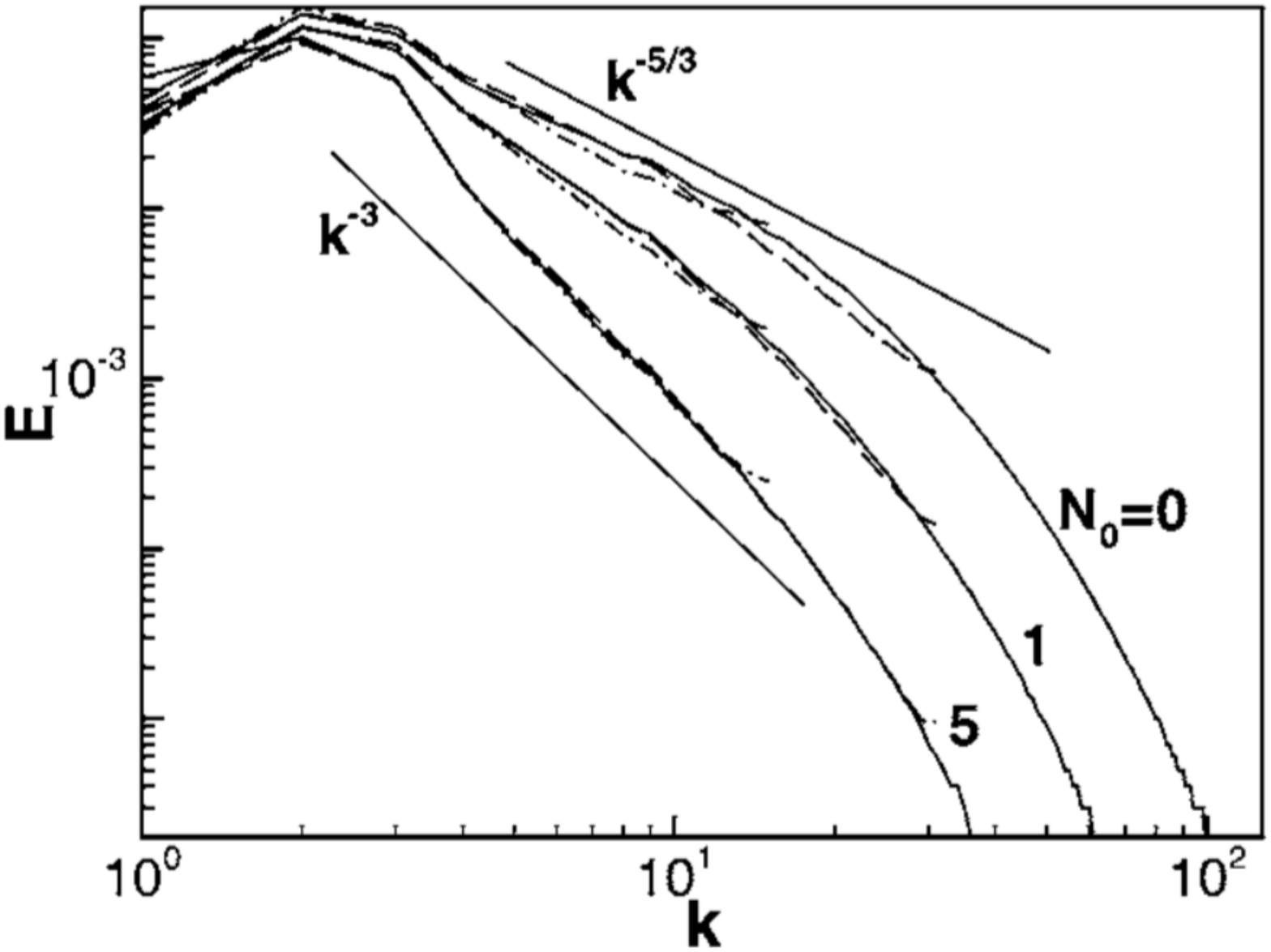}
\end{center}
\caption{In the forced QS MHD turbulence simulation of Vorobev {\em et al}~\cite{Vorobev:PF2005}, steepening of the energy spectrum $E(k)$ as the interaction parameter $N$ is increased from 0 to 5.  From Vorobev {\em et al}~\cite{Vorobev:PF2005}. Reprinted with permission from AIP Publishing.   }    
 \label{fig:vorobev}
\end{figure}

Using numerical simulations, Hossain~\cite{Hossain:PFB1991}  showed that for low interaction parameters $N$ ($\sim 0.1$), the flow is three-dimensional and it exhibits a forward cascade of energy (from small wavenumbers to higher wavenumbers). However, for large ($N = 10$), he reported that $E(k) \sim  k^{-3}$, and related it to 2D hydrodynamic turbulence~\cite{Kraichnan:PF1967b}.   {\color{blue} Ishihara {\em et al}~\cite{Ishihara:PRL2002} employed tensorial and dimensional analysis to compute anisotropic corrections in a turbulent shear flow, and showed that the velocity correlation function can be approximated as
\be
Q_{ij}({\bf k}) = \frac{K_\mathrm{Ko}}{4\pi k^2} \Pi^{2/3} k^{-5/3}  P_{ij}({\bf k}) + Q_{ij}^{(1)}({\bf k}),
\ee
where $\Pi$ is the energy flux,  $P_{ij}({\bf k}) = \delta_{ij} - k_i k_j/k^2$, and  $ Q_{ij}^{(1)}({\bf k})$ is the anisotropic tensor.  Ishihara {\em et al}~\cite{Ishihara:PRL2002} modelled $Q_{ij}^{(1)}({\bf k})$ as
\be 
Q_{ij}^{(1)}({\bf k}) = C_{ij\alpha \beta} S_{\alpha \beta},
\ee
where $S_{\alpha \beta}$ is the traceless tensor representing the shear stress, and 
\be 
C_{ij\alpha \beta} = \frac{E_{as}(k)}{4\pi k^2} (P_{i\alpha}({\bf k}) P_{j \beta}({\bf k}) 
+ P_{i\beta}({\bf k}) P_{j\alpha}({\bf k}) ) +   \frac{E_{bs}(k)}{4\pi k^2} \frac{k_\alpha k_\beta}{k^2}
\ee
with 
\bea 
E_{as}(k) & = & A \Pi^{1/3}k^{-7/3}, \label{eq:Ea}\\
E_{bs}(k) & = & B \Pi^{1/3}k^{-7/3}, \label{eq:Eb}
\eea
where $A,B$ are constants.  Ishida and Kaneda~\cite{Ishida:PF2007} extended Ishihara {\em et al}'s arguments to QS MHD turbulence and computed the modification in the inertial-range energy spectrum for low interaction parameters. They showed that for low interaction parameters, $E(k) \sim k^{-7/3}$, in similar lines as the results of Ishihara {\em et al}~\cite{Ishihara:PRL2002};   Ishida and Kaneda~\cite{Ishida:PF2007} confirmed this result with direct numerical simulations.}  Burattini \textit{et al}~\cite{Burattini:PD2008} studied the nonlinear energy transfers and the energy flux using numerical simulations for $N=0,1$ and 5.  They observed that the anisotropic energy spectra  are proportional to $k^{-7/3}$ for $N=0$ and 1  (see Fig.~\ref{fig:Burattini}(a)), consistent with the predictions of Ishida and Kaneda~\cite{Ishida:PF2007}.  They also observed that for large $N$, $E(k)$ is steeper than $ k^{-7/3}$ (see Fig.~\ref{fig:Burattini}(b)).  Vorobev {\em et al}~\cite{Vorobev:PF2005} performed direct numerical simulations and large-eddy simulations of QS MHD turbulence and showed that the spectral exponent varies for $-5/3$ to $-3$ as $N$ is increased from 0 to 5. In Fig.~\ref{fig:vorobev}, we illustrate $E(k)$ of  Vorobev {\em et al}'s DNS.

\begin{figure}
\begin{center}
\includegraphics{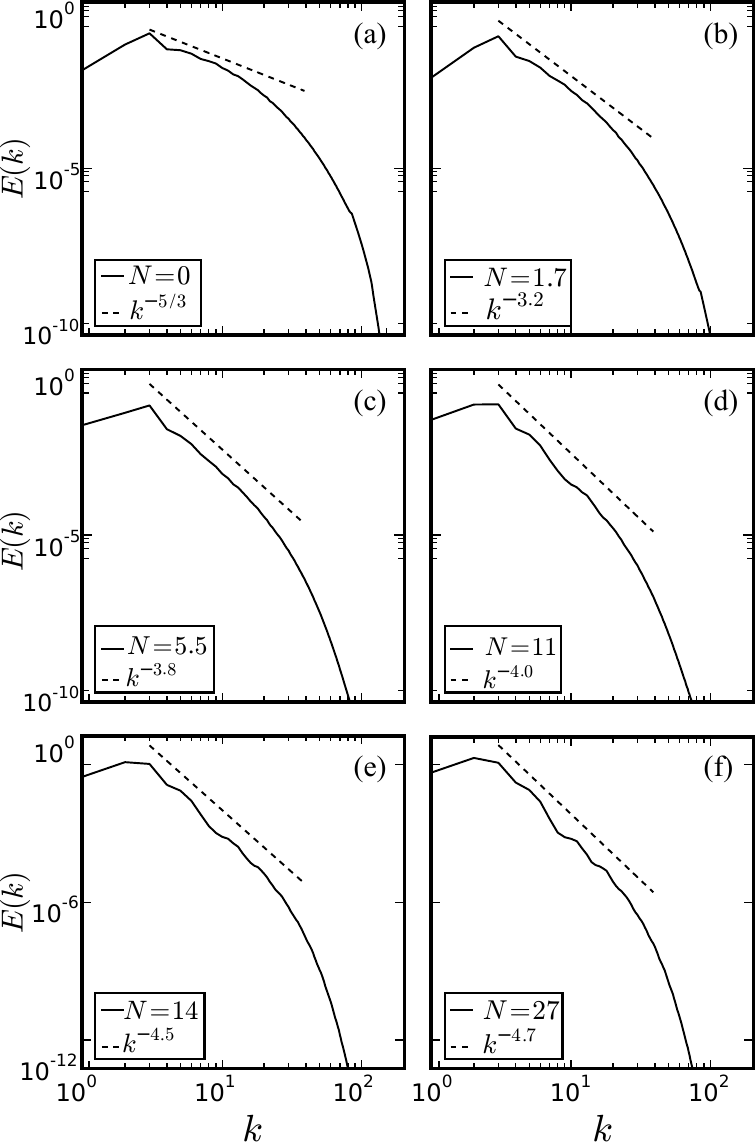}
\end{center}
\caption{Kinetic energy spectra for (a) $N = 0$, (b) $N=1.7$, (c) $N=5.5$, (d) $N=11$, (e) $N = 14$, and (f) $N=27$.  In the inertial range $E(k) \sim k^{-\alpha}$ with the spectral indices $\alpha = 5/3, 3.2, 3.8, 4.0, 4.5, 4.7$ respectively. Reprinted with permission from Reddy~\cite{Reddy:thesis}.}  
\label{fig:KE_spectrum_powerlaw}
\end{figure}

\begin{figure}
\begin{center}
\includegraphics[scale=0.8]{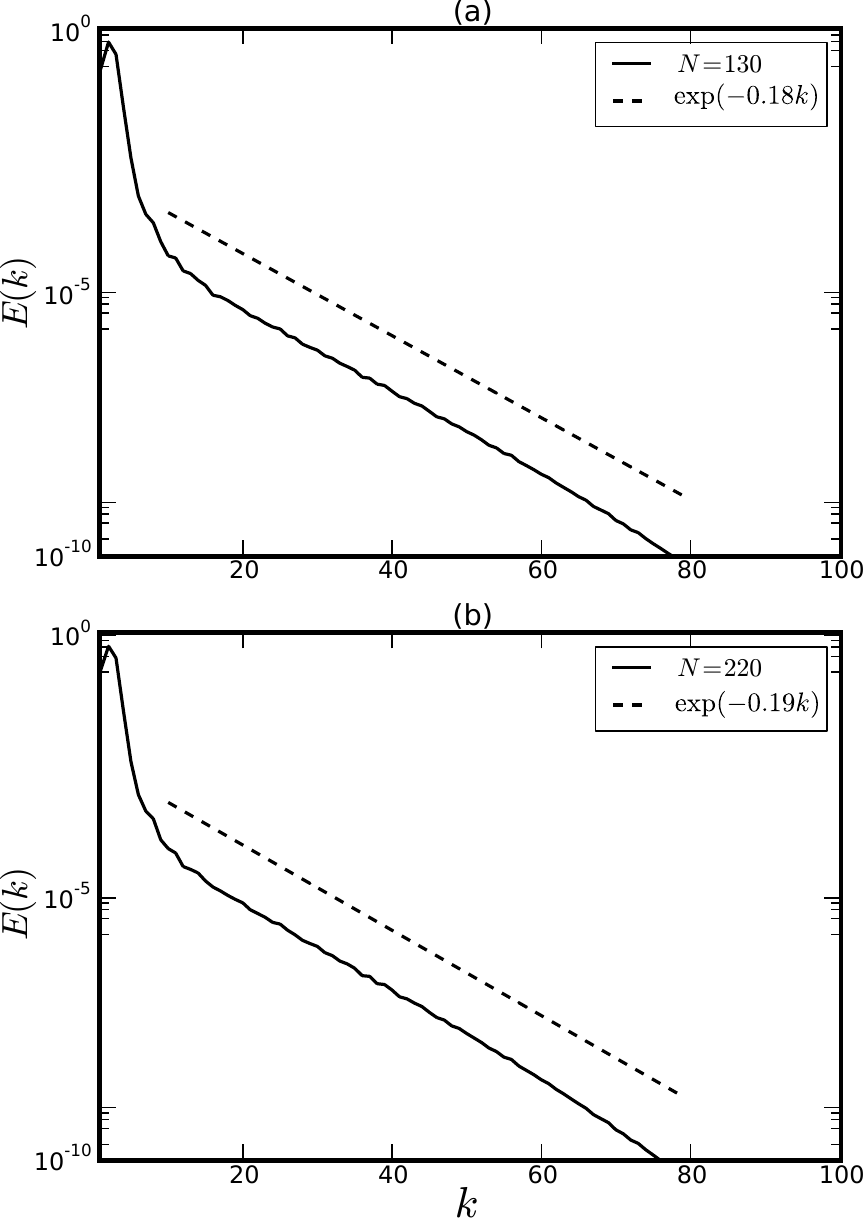}
\end{center}
\caption{Kinetic energy spectra for (a) $N=130$ and (b) $N=220$ (very large  values of  $N$) that exhibit $E(k) \sim \mathrm{exp}(-0.18k)$ and $\mathrm{exp}(-0.19k)$ respectively.  Reprinted with permission from Reddy~\cite{Reddy:thesis}.}  
\label{fig:KE_spectrum_semilog}
\end{figure}

 Many researchers~\cite{Kolesnikov:FD1974,Kit:MG1971,Hossain:PFB1991}  attribute the aforementioned steepening  of the energy spectrum  to two-dimensionalization.  Note that two-dimensional hydrodynamic turbulence has $E(k) \sim k^{-3}$ for large $k$~\cite{Kraichnan:PF1967b}.  Verma and Reddy~\cite{Verma:PF2015b} however argue that the steepening of $E(k)$ in QS MHD turbulence occurs due to the Joule dissipation.  According to Eq.~(\ref{eq:dPi_dk}),   in QS MHD turbulence, the Joule term  dissipates kinetic energy at all scales, hence the flux $\Pi(k)$  decreases with wavenumber $k$, unlike constant $\Pi(k)$ in the inertial range of hydrodynamic turbulence.  As a result, $E(k)$ of QS MHD turbulence is steeper than the hydrodynamic $k^{-5/3}$ spectrum.  In Figs.~\ref{fig:KE_spectrum_powerlaw}(a-f) and Fig.~\ref{fig:KE_spectrum_semilog}, we exhibit the energy spectra reported by Reddy and Verma~\cite{Reddy:PF2014} for $N = 0, 1.7, 5.5, 11, 14, 27, 130$ and 220.  These spectra are for the statistical steady-state data of  forced run (forcing applied at $k_f = 1$ to 3).   For $N = 0, 1.7, 5.5, 11, 14, 27$,  $E(k) \sim k^{-\alpha}$ with the spectral indices $\alpha = 5/3, 3.2, 3.8, 4.0, 4.5, 4.7$ respectively.  But for $N=130$ and 220, the spectrum follows exponential behaviour---$\mathrm{exp}(-0.18k)$ and $\mathrm{exp}(-0.19k)$ respectively.  The errors in the coefficients are of the order of 10\%.  These results are summarised in Table~\ref{tab:energy}.  The decrease of the energy flux $\Pi(k)$ with $k$ leads to the aforementioned steepening of the energy spectrum.

%Before closing this section we remark that Kornet and Poth\'{e}rat~\cite{Kornet:JCP2015} solved fluid flows in a channel using a spectral method with special basis functions suitable for resolving the Hartmann layer.  The present review focusses on the bulk flow, hence we will not discuss the algorithm of Kornet and Poth\'{e}rat~\cite{Kornet:JCP2015} here.  

 %\begin{sidewaystable}[htbp]
\begin{table}[htbp]
\begin{center}
\caption{\label{tab:param_3}Parameters of the simulation: the grid size, the interaction parameter $N$ computed at the steady state, the interaction parameter $N_0$ computed at the instant when external magnetic field is applied, the integral length scale $L$, the anisotropy ratio $A=E_\perp/2E_\parallel$,  the ratio of the Joule dissipation and the viscous dissipation $\epsilon_J/\epsilon_{\nu}$, and the spectral laws.} 
%\begin{tabular}{ p{1.0cm}  p{1.1cm}  p{1.0cm}  p{0.9cm} p{1.0cm}  p{1.2cm}  p{2.4cm}  p{1.9cm}  p{0.7cm}}
\begin{tabular}{|c|c|c|c |c |c |c |c |c |c|c| } 
 \hline 
Grid & $N$ &~\ $N_0$ & $L$ &  $A =E_{\perp}/2E_{\parallel}$  & ~\ $\epsilon_J/\epsilon_{\nu}$	& spectral law \\
 \hline \hline 
$256^3$ & 0		& 0		& 0.095	& 1.0   &  	0	& $k^{-5/3}$ 	\\
$512^3$	& 0		& 0		& 0.096 	& 0.99 &  	0 	&  $k^{-5/3}$\\
$512^3$	& 0.1		& 0.1		& 0.095 	& 1.01 &  	0.28 	&  $k^{-1.8}$\\
$512^3$	& 0.64	& 0.5		& 0.105 	& 1.01 & 	2.07 &  $k^{-2.0}$\\
$512^3$	& 1.6		& 1.0		& 0.11  	& 1.05 & 	4.4 	&  $k^{-2.8}$\\
$256^3$ 	& 1.7		& 1.0		& 0.12 	& 1.1  &	4.2 	& $k^{-3.2}$	\\
$256^3$ 	& 5.5		& 2.5		& 0.14	& 1.5  &	9.7	& $k^{-3.8}$	\\
$256^3$ 	& 11		& 5.0		& 0.15	& 4.5  &	11	& $k^{-4.0}$	\\
$256^3$ 	& 14		& 7.5		& 0.15	& 8.0  &	11	& $k^{-4.5}$	\\
$256^3$ & 18		& 10.0	& 0.15	& 16	  &	9.8	& $k^{-4.7}$	\\
$256^3$ & 27		& 20.0	& 0.15	& 1.6  &	6.9	& $k^{-4.7}$	\\
$256^3$ & 130		& $-$	& 0.17	& 3.0  & 	4.1	& $\mathrm{exp}(-0.18k)$\\
$256^3$ & 220		&$-$		& 0.17	& 1.7 &	3.9	& $\mathrm{exp}(-0.19k)$\\ 
\hline
\end{tabular}
\label{tab:energy}
\end{center}
\end{table}

\subsection{Miscellaneous numerical results of QS MHD turbulence}
  Thess and Zikanov~\cite{Thess:JFM2007} performed linear stability analysis to study the flow transition from 2D to 3D. They performed their study for inviscid flows in a triaxial ellipsoid and observed that the two-dimensional flows abruptly become three-dimensional with a sudden burst.   In another development, for low-Rm flows, Poth\'{e}rat et al~\cite{Potherat:JFM2000} observed three-dimensionalization of quasi-two-dimensional quasi-static MHD flows due to the {\em barrel effect}~\cite{Potherat:EPL2012}.    In a channel flow, Boeck {\em et al}~\cite{Boeck:PRL2008}  also observed recurring transitions between two-dimensional  and three-dimensional  states.   In decaying QS MHD flows, Burattini \textit{et al}~\cite{Burattini:JFM2010} reported $t^{-1/2}$ decay law for the kinetic energy.  

There have been simulations of MHD turbulence at low magnetic Reynolds number.  Knaepen \textit{et al}~\cite{Knaepen:JFM2004} showed that the behaviour of low-Rm MHD is similar to QS MHD turbulence ($\mathrm{Rm} \rightarrow 0$). We also remark that the quasi 2D nature and steepening of $E(k)$ of QS MHD turbulence observed in numerical simulations are consistent with similar findings in experiments discussed in Sec.~\ref{sec:experiment}.

 In the next section we will describe angular distribution of energy in Fourier space.

\section{Anisotropic energy distribution in QS MHD turbulence} \label{sec:energy}

Quantification of anisotropy in turbulent flow is a challenge.  For  two-dimensional magnetohydrodynamic turbulence, Shebalin {\em et al}~\cite{Shebalin:JPP1983} proposed a measure of anisotropy for any quantity $Q$ as
\begin{equation}
\theta_Q = \tan^{-1} \frac{k_z^2 Q({\bf k})}{k_x^2 Q({\bf k})}
\label{eq:thetaQ}
\end{equation}
that can be easily generalised to three dimensions.    Researchers~\cite{Burattini:PD2008,Favier:PF2010,Sagaut:book,Teaca:PRE2009} decomposed the Fourier space into rings and quantified the energy contents in rings as {\em ring spectrum}.  For a more detailed measure, some researchers have studied the energy contents in toroidal and poloidal components of a vector field~\cite{Favier:PF2010,Sagaut:book}. We also remark that spherical harmonics have been used to quantify anisotropy~\cite{Biferale:PR2005}. 

In QS MHD, a strong external magnetic field $B_0$ induces Joule dissipation $N E(k) \cos^2 \theta$ that is dependent on the polar angle $\theta$ (see Fig.~1(b)).  Consequently, the energy distribution in QS MHD is anisotropic, in  contrast to the hydrodynamic turbulence for which the energy distribution is isotropic in the inertial range.     Zikanov and Thess~\cite{Zikanov:JFM1998} reported the energy contents of the perpendicular and parallel components of the velocity field.   Vorobev \textit{et al}~\cite{Vorobev:PF2005}  showed that for $N=5$, $E_{\perp}(k)/E_{\parallel}(k) > 1$ at low wavenumbers, and $E_{\perp}(k)/E_{\parallel}(k) < 1$ at higher wavenumbers. 

 Burattini \textit{et al}~\cite{Burattini:PF2008,Burattini:PD2008} computed the ring spectrum, as well as the energy spectra of the perpendicular and components of velocity.  Favier {\em et al}~\cite{Favier:PF2010,Favier:JFM2011} quantified anisotropy in QS MHD turbulence using the toroidal and poloidal components of the velocity field.  Recently  Reddy and Verma~\cite{Reddy:PF2014} studied the ratio $E_\perp/(2 E_\parallel)$, ring spectrum, Joule dissipation spectrum etc.  Their analysis is for a wide range of $N$---from 0 to 220.  In the following discussion, we will summarise the results on the anisotropy in QS MHD turbulence.

\subsection{Anisotropy of total energy and dissipation rates of QS MHD turbulence} \label{subsec:anis_global}
Reddy and Verma~\cite{Reddy:PF2014} quantified the anisotropy of the total energy using an anisotropic parameter
\begin{equation}
A = \frac{E_\perp}{2 E_\parallel}
\end{equation}
where $E_\perp = (U_x^2+U_y^2)/2$ and $E_\parallel=U_z^2/2$.   In Table~\ref{tab:energy} we list the values of $A$ as well as $\epsilon_J/\epsilon_\nu$ under steady state of the forced turbulence runs.   By definition, for isotropic flows ($N=0$), $A=1$.  The flow continues to be nearly isotropic till $N=1.7$ for which $A \approx 1.1$.  Beyond $N = 1.7$, $A$ increases with $N$ till $N=18$ after which it decreases with the increase of $N$.  These observations show that the flow is quasi 2D for large $N$, consistent with the illustrations of Figs.~\ref{fig:isosurface_vorticity} and \ref{fig:N132_vectors}.    In the following discussion, we demonstrate  that in QS MHD turbulence, the energy dissipates more strongly in the polar region than in the equatorial region due to the $\cos^2 \theta$ term of Eq.~(\ref{eq:dEk_dt}).  The two-dimensionalisation of QS MHD further strengthens $U_\perp$ due to the inverse cascade of kinetic energy.  

The maximum value of $E_\perp/(2 E_\parallel)$ occurs at $N\approx 18$.  This is due to the fact that the strength of  $U_\perp$ increases with $N$ until $N\approx 18$.  However for $N > 18$, $U_\perp$ transfers energy to $U_\parallel$ via pressure thus making $E_\parallel$ significant.  This feature is evident in Fig. \ref{fig:N132_vectors} where $U_\parallel$ is stronger for $N=130$ than for $N=18$.   Interestingly, the ratio of the Joule dissipation rate $\epsilon_J$ and the viscous dissipation rate $\epsilon_\nu$  also peaks near $N \approx 14$.   For $N > 0.64$,  $\epsilon_J$ dominates $\epsilon_\nu$ because the large-scale velocity is dissipated  more effectively by $\epsilon_J$ than $\epsilon_\nu$.  In addition, for very large $N$, the increased strength of $E_\parallel$  makes $\epsilon_\nu$ significant.     We  revisit these connections in Sec.~\ref{subsec:ring_Ek}.
%
%\vspace{4cm} 
%\begin{figure}[H]
%\begin{center}
%%\includegraphics[scale=0.8]{C1/shell_sector}
%\end{center}
%\caption{Plot $\epsilon_J/\epsilon_\nu$ and $E_\perp/E_\parallel$ vs.~$N$. }  
%\label{fig:eps_vsN}
%\end{figure}
%%
%%

\subsection{Anisotropy in energy spectrum in QS MHD turbulence}  \label{subsec:E_perp(k)/E_pll(k)}
 
\begin{figure}
\begin{center}
\includegraphics{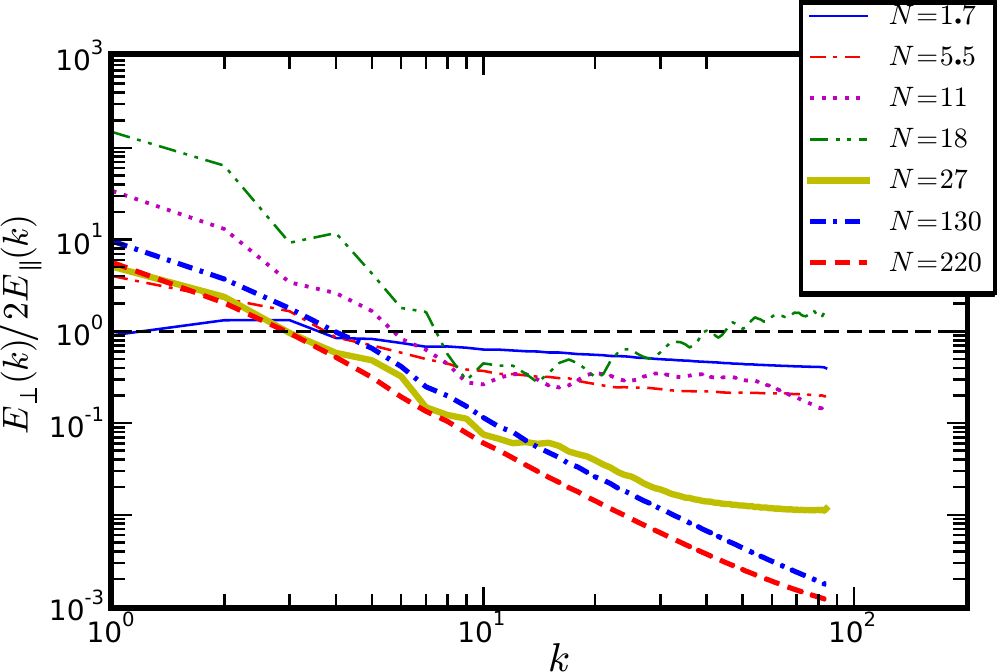}
\end{center}
\caption{Variation of $A=E_{\perp}(k)/2E_{\parallel}(k)$ with $k$ for various values of the interaction parameter $N$. From Reddy and Verma~\cite{Reddy:PF2014}. Reprinted with permission from AIP Publishing.}  
\label{fig:Eperp_Epar}
\end{figure}

To explore the nature of the anisotropy at different length scales, a wavenumber-dependent  anisotropy measure, $E_{\perp}(k)/2E_{\parallel}(k)$, has been proposed~\cite{Burattini:PF2008,Reddy:PF2014,Vorobev:PF2005}.  The ratio plotted in  Fig.~\ref{fig:Eperp_Epar}  shows that $E_{\perp}(k) > E_{\parallel}(k)$  at low wavenumbers, which is due to the inverse cascade of $U_\perp$ at small $k$.  The ratio $E_{\perp}(k)/2E_{\parallel}(k)$ increases with $N$ upto $N=18$, after which it decreases.  The  occurrence of peak at $N\approx 18$ is due to energy transfer from $U_\perp$ to $U_\parallel$ for $N>18$, as discussed in previous subsection. 

 For large wavenumbers, the ratio $E_{\perp}(k)/2E_{\parallel}(k)$  is near unity till $N =18$, after which it decreases monotonically with $N$~\cite{Favier:PF2010,Reddy:PF2014,Vorobev:PF2005}. For large $N$, the significant increase of $E_{\parallel}(k)$ at large $k$ is due to the energy transfer from $U_\perp$ to $U_\parallel$, and subsequent forward cascade of $U_\parallel$.  These  observations  lend strong credence to  the observed 2D-3C nature of QS MHD turbulence, first proposed by Favier \textit{et al}~\cite{Favier:PF2010}.  We will demonstrate these energy exchanges in Sec.~\ref{sec:ET}.
 
%\begin{figure}[htbp]

Further insights into the anisotropy and angular dependence of the energy distribution in the spectral space are obtained by  the ring spectrum, which will be discussed below.

\subsection{Angular dependence of energy spectrum in QS MHD turbulence} 
\label{subsec:angular-dependence-Ek}

Researchers have devised measures to quantify the angular dependence of $E({\bf k})$ in QS MHD turbulence.  In the following discussion, we list four measures: the angular variation of the energy of the toroidal and poloidal components of the velocity field (see Sec.~\ref{sec:Fourier_QSMHD} and Fig.~\ref{fig:schmatic_k_space} for definition); ring spectrum;  decomposition of the ring spectrum using Legendre polynomials; and tensorial representation.

\subsubsection{Toroidal and Poloidal decomposition} \label{subsubsec:toriodal_poloidal}
\begin{figure}
\begin{center}
\includegraphics{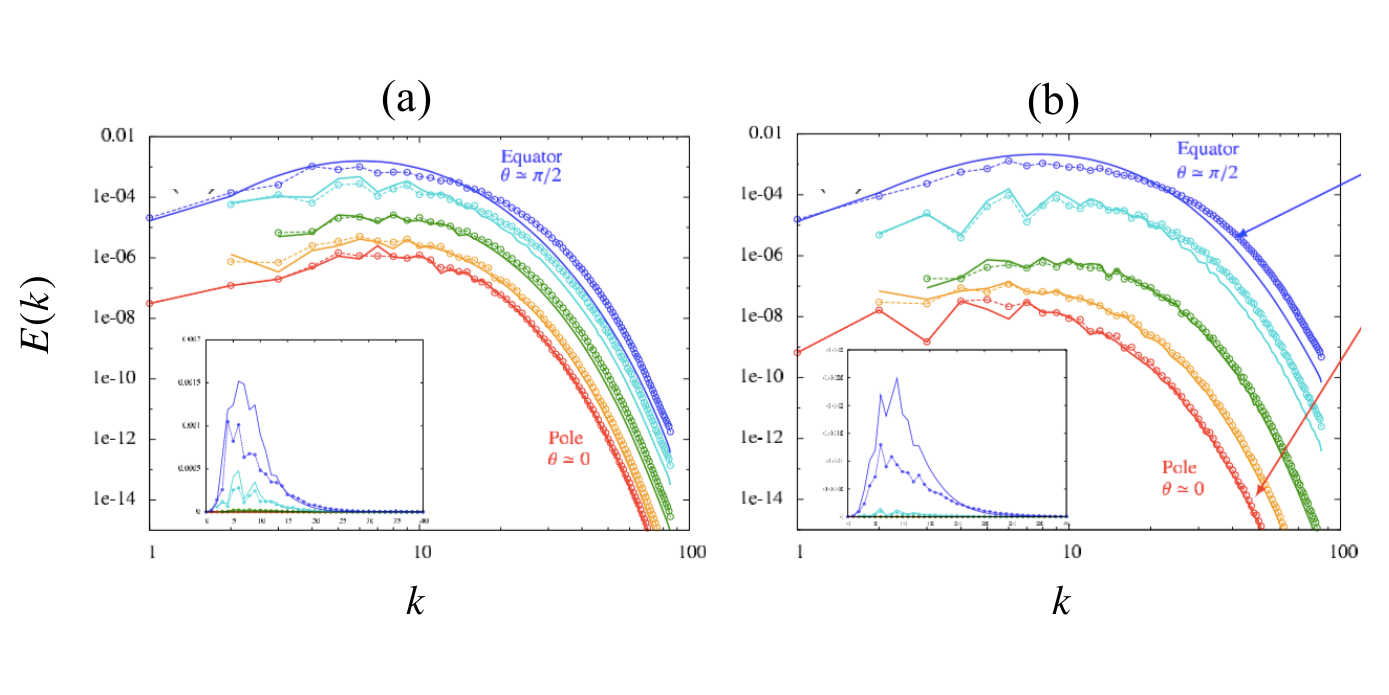}
\end{center}
\caption{ Angular energy spectra $E^\mathrm{tor}(k)$ (solid line) and $E^\mathrm{pol}(k)$ (dotted lines)  of the five angular rings from the pole (red curve) to the equator (blue curve): (a) for $N=1$ at $t=5 t_J$, and (b) for $N=5$ at $=20 t_J$, where $t_J = \eta/B_0^2$ is the diffusive dime due to the Lorentz force. The embedded figure represent the same plots in linear scale.  From Favier {\em et al}~\cite{Favier:PF2010}. Reprinted with permission from AIP Publishing. }  
\label{fig:Favier}
\end{figure}

Favier {\em et al}~\cite{Favier:PF2010} divided the Fourier space into 5 angular regions (rings) of equal angular widths, and computed the cumulative energy contents of the toroidal and Poloidal components in these rings (see Sec.~\ref{sec:Fourier_QSMHD} and Fig.~\ref{fig:schmatic_k_space}). They performed two decaying QS MHD simulations for $N=1$ and 5 and computed the toroidal and poloidal angular energy spectra, which are denoted by $E^\mathrm{tor}(k)$ and $E^\mathrm{pol}(k)$.   In Fig.~\ref{fig:Favier}, $E^\mathrm{tor}(k)$ and $E^\mathrm{pol}(k)$ are represented using solid and dotted lines respectively. The plots show that the energy of QS MHD turbulence is concentrated near the equator.  The energy contents in the polar region is much  smaller than that in the equatorial region.

%\begin{figure}[htbp]

\subsubsection{Ring  spectrum for QS MHD turbulence} \label{subsec:ring_Ek}

\begin{figure}[H]
\begin{center}
\includegraphics[scale=0.3]{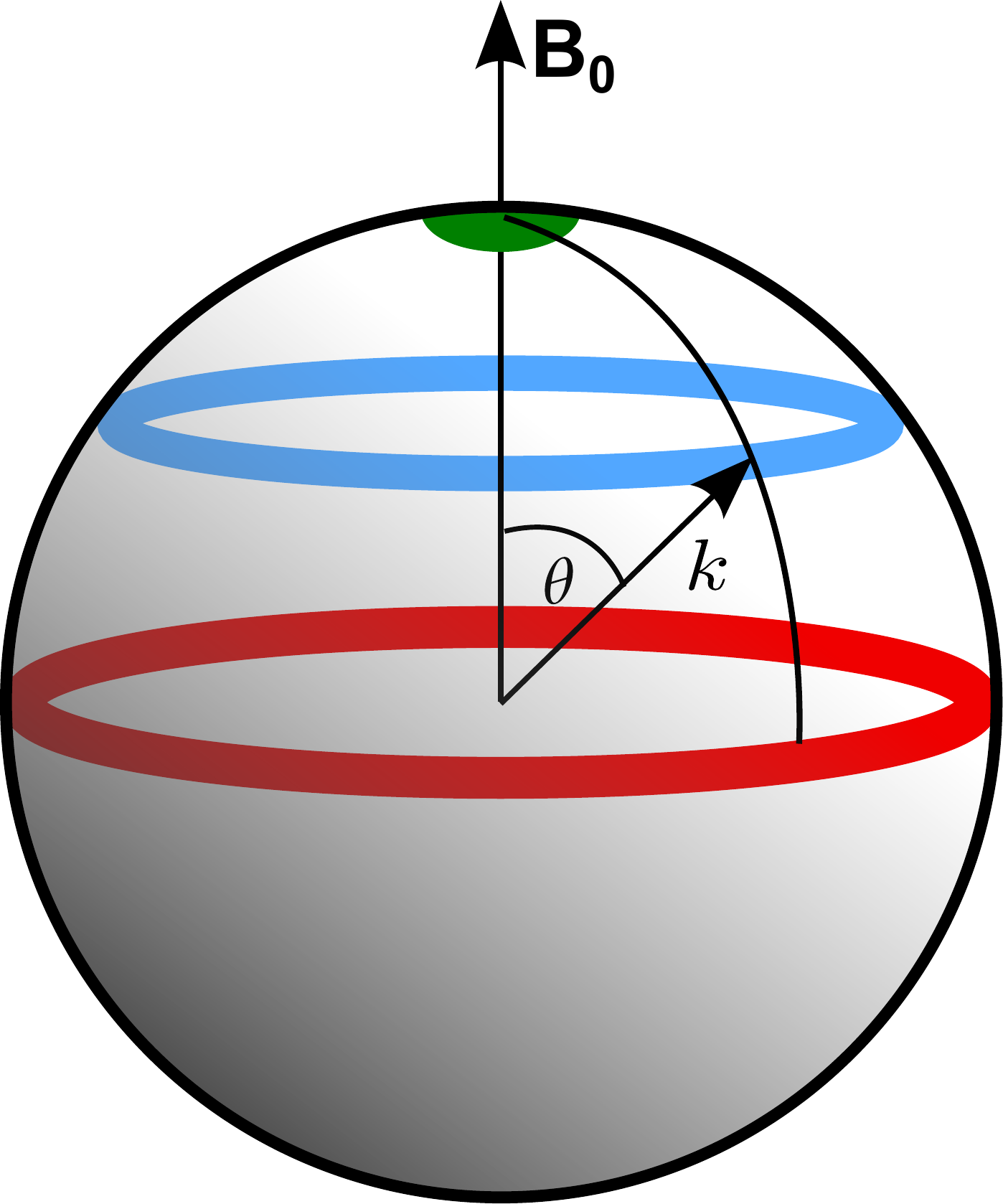}
\includegraphics[scale=0.8]{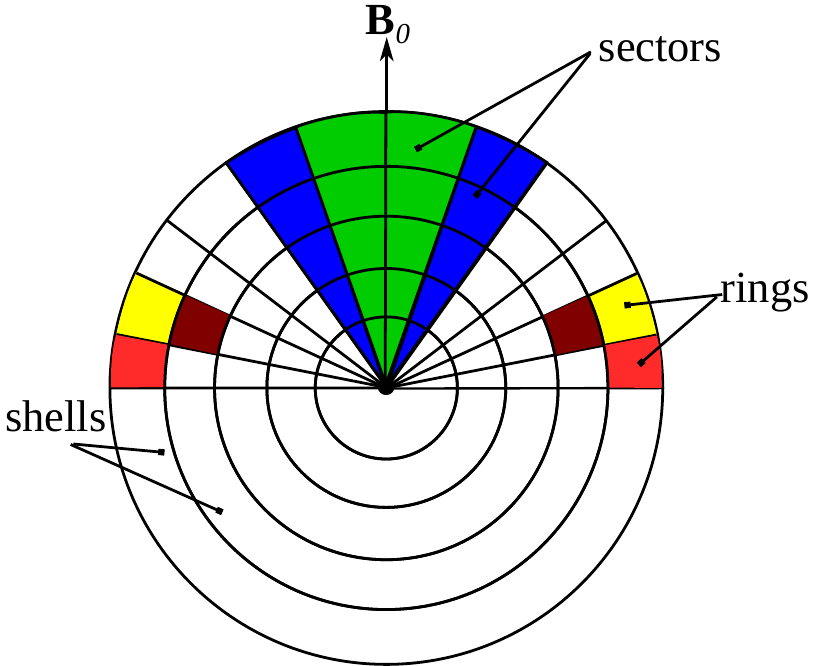}
\end{center}
\caption{(a) Illustration of the  ring decomposition in the spectral space. (b) A cross-sectional view of the wavenumber shells, sectors, and rings. }  
\label{fig:ring_decomposition}
\end{figure}

A given wavenumber shell is divided into rings which are indexed using shell index $m$ and sector index $\alpha$ (see Figures~\ref{fig:ring_decomposition} for an illustration)~\cite{Burattini:PF2008,Burattini:PD2008,Reddy:PF2014}.  A ring is an intersection of a shell and a sector~\cite{Teaca:PRE2009}.  This scheme is similar to that of Favier {\em et al}~\cite{Favier:PF2010}; they chose 5 sectors in their decomposition. Note that the mean magnetic field is aligned along $\theta=0$.  Also, the average $E({\bf k})$ of QS MHD turbulence is independent of the azimuthal angle $\phi$, hence $E({\bf k})$ is function only of $k$ and $\theta$.

  The ring spectrum $E(k,\theta)$ is defined as
\begin{equation}
E(k,\theta) = \frac{1}{C_{\alpha}} \sum_{k \le |{\bf k'}| < k+1;\mathrm{\angle}({\bf k^\prime})\in [\theta_\alpha,\theta_{\alpha+1})} \frac{1}{2} |{\mathbf U}({\mathbf k'})|^2,
\end{equation}
where $\mathrm{\angle}({\bf k^\prime})$ is the angle between ${\bf k^\prime}$ and ${\bf B_0}$,    and $\alpha$ is the index of the sector whose  angular range is from $\theta_{\alpha}$ to $\theta_{\alpha+1}$.  Reddy and Verma~\cite{Reddy:PF2014} divided the sum with a normalization factor
\begin{equation}
C_{\alpha} = |\cos(\theta_{\alpha}) - \cos(\theta_{\alpha+1})|
\label{eq:ringEk_normalization}
\end{equation}
to compensate  for the effects of a larger number of modes in the rings with larger $\theta$.  This factor is related to the $d\cos\theta$ factor in the volume integral in spherical geometry.  After normalization, $E(k,\theta) $ is a measure of the average energy per mode  in a given ring.  A corollary, for a given $k$ in an isotropic flow,  the ring spectra of all the rings are equal in a statistical sense.
 
%\begin{figure}[htbp]
\begin{figure}[]
\begin{center}
\includegraphics{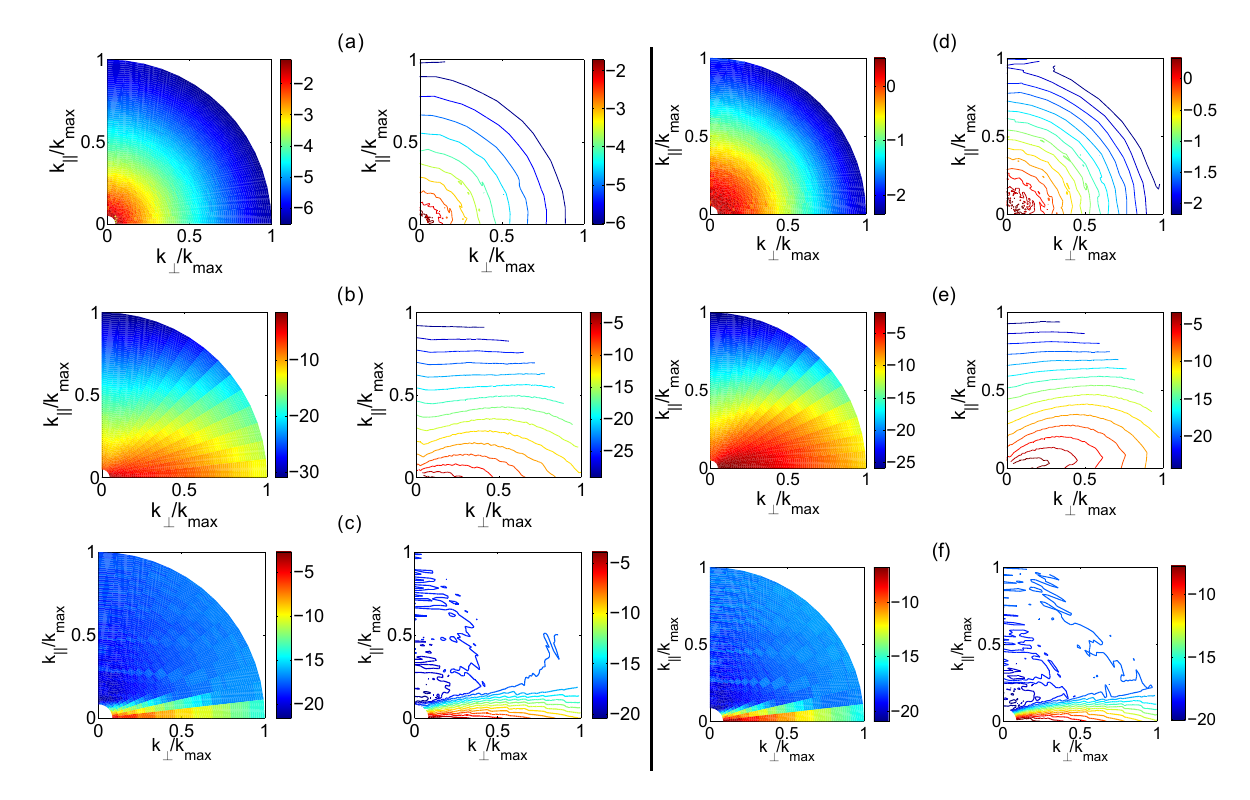}
\end{center}
\caption{Angular distribution of the logarithm of the kinetic energy ($\log(E(k,\theta)$) for (a) $N = 0$, (b) $N = 18$ and (c) $N = 130$ exhibited as density and contour plots.  The corresponding  angular distribution of the Joule dissipation rate ($\mathrm{log}(\epsilon_J(k,\theta))$) for (d) $N = 0$, (e) $N = 18$, and (f) $N=130$.  Reprinted with permission from Reddy~\cite{Reddy:thesis}.}  
\label{fig:ring_spectrum}
\end{figure}

 Burattini {\em et al}~\cite{Burattini:PF2008} plotted three-dimensional $E(k_x, k_y, k_z)$ of QS MHD turbulence and showed that the energy is suppressed along the $z$ direction.  Note that the ring spectrum averages the energy in the $k_x k_y$ plane for a given $k_z$ by exploiting the azimuthal symmetry.

Reddy and Verma~\cite{Reddy:PF2014}  divided the spectral space in the northern hemisphere into thin shells of unit widths.  The shells in turn were further divided into 15 thin rings from $\theta=0$ to $\theta=\pi/2$ with sector widths of $\pi/30$.  Note that the southern hemisphere has the same ring spectrum as the northern hemisphere   due to  $\theta \rightarrow \pi-\theta$ symmetry, hence we compute the energy spectrum only for the northern hemisphere.   Figure~\ref{fig:ring_spectrum}(a,b,c)  exhibits the density and contour plots of the energy spectrum $E(k,\theta)$ for $N=0$, $18$, and $130$ respectively~\cite{Reddy:PF2014}. The energy spectrum for  $N=0$ is isotropic, but those for $N=18$ and $130$ are anisotropic, with the degree of anisotropy increasing with $N$.  Since the viscous dissipation rate $\epsilon_\nu(k,\theta) \propto E(k,\theta)$, $\epsilon_\nu(k,\theta)$ has the same distribution as $E(k,\theta)$.   However, the  Joule dissipation rate $\epsilon_J (k,\theta) = 2{B'_0}^2 E(k,\theta)\cos^2\theta$  has an additional $\cos^2 \theta$ dependence.  As a result, $\epsilon_J (k,\theta)$ does not peak at $\theta = \pi/2$, but before $\theta = \pi/2$. In Fig.~\ref{fig:ring_spectrum}(d,e,f) we exhibit the $\epsilon_J (k,\theta)$  that exhibits the above properties for $N=0,18,130$ respectively~\cite{Reddy:thesis} .

\begin{figure}[]
\begin{center}
\includegraphics[width=8cm]{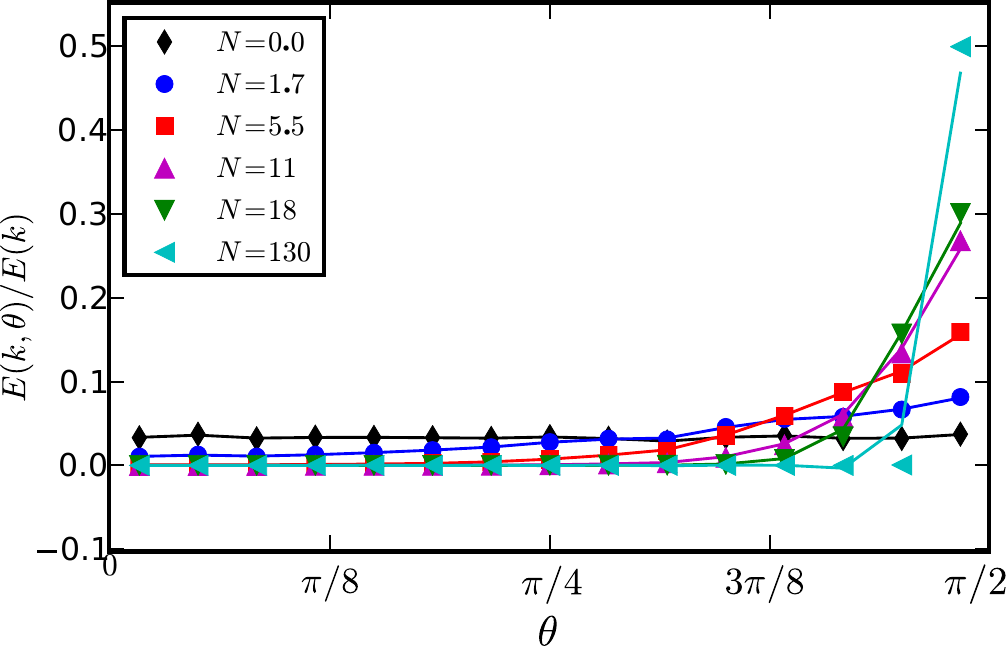}
\includegraphics[width=8cm]{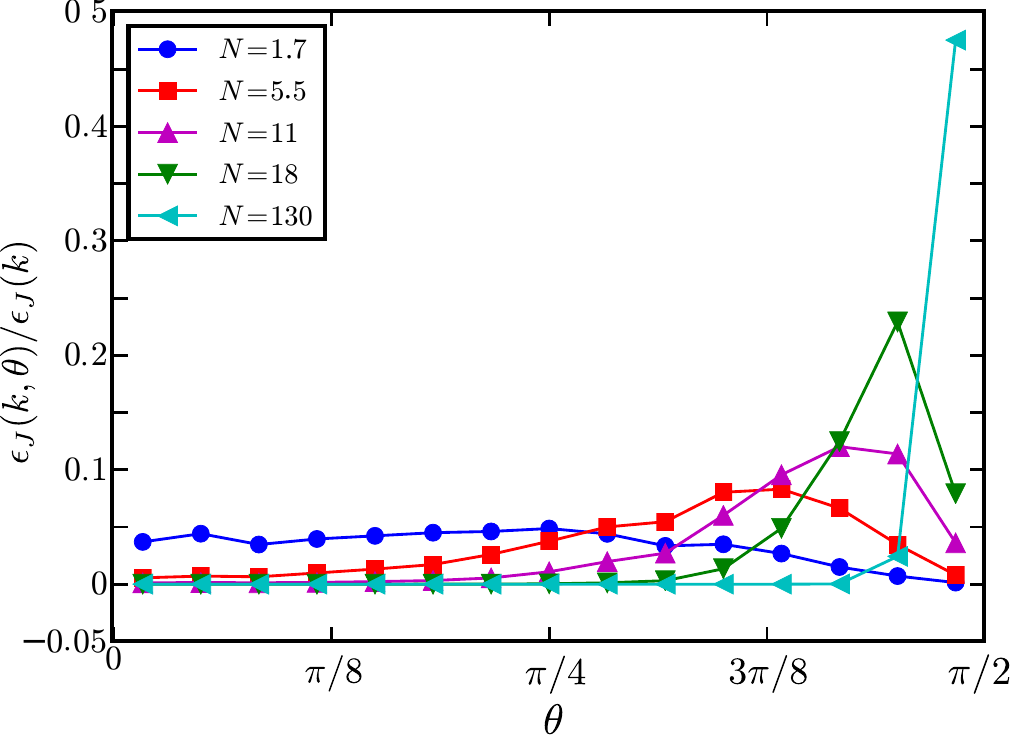}
\end{center}
\caption{For various $N$: (a) Plot of $E(k=20,\theta)/E(k=20)$ vs. $\theta$. (b) Plot of the normalized Joule dissipation rate $\epsilon_J(k=20,\theta)/\epsilon_J(k=20)$. From Reddy and Verma~\cite{Reddy:PF2014}. Reprinted with permission from AIP Publishing.} 
\label{fig:EkDk_20}
\end{figure}

The ring spectra demonstrates that the flow is strongly anisotropic for large $N$ with strong concentration of energy near $k_\parallel \approx 0$ plane.  Using the information of Figs.~\ref{fig:Eperp_Epar} and \ref{fig:ring_spectrum}(a,b,c) we  conclude that low wavenumber modes have significant ${\bf U}_\perp$, but the intermediate and large wavenumber modes have dominant $U_z$~\cite{Reddy:PF2014}. This feature is consistent with the real-space profile shown in Figs.~\ref{fig:isosurface_vorticity} and \ref{fig:N132_vectors}.   In Fig.~\ref{fig:EkDk_20} we exhibit the normalised ring spectrum $E(k=20,\theta)/E(k=20)$ and $\epsilon_J(k=20,\theta)/\epsilon_J(k=20)$ that confirms the above behaviour. 

The Lorentz force vanishes at $\theta = \pi/2$ due to the $\cos^2\theta$ factor, hence the nonlinear term ${\bf U \cdot \nabla U}$ dominate in the $k_z=0$ plane. Therefore QS MHD turbulence has behaviour similar to 2D hydrodynamics in the $k_z=0$ plane.  These observations yield dynamical perspectives to the findings of earlier  researchers \cite{Alemany:JdeM1979,Burattini:PD2008,Eckert:IJHFF2001,Kit:MG1971,Kolesnikov:FD1974,Moreau:book:MHD} who reported that QS MHD turbulence exhibits behaviour similar to 2D hydrodynamics. 

%\begin{figure}[htbp]
%\begin{figure}[]
%\begin{center}
%\includegraphics[width=10cm]{figures/Dissipation_N0_N10_N75_n}
%\end{center}
%\caption{Angular distribution of the logarithm of the Joule dissipation rate ($\mathrm{log}(\epsilon_J(k,\theta))$) for (a) $N = 0$, (b) $N = 18$, and (c) $N=130$.   Left panel: density plot, right panel: contour plot.  Adoption of a figure of Reddy~\cite{Reddy:thesis}. }  
%\label{fig:Dissipation_spectrum}
%\end{figure}

\begin{figure}[ht]
\begin{center}
\includegraphics[width=8.0cm,angle=0]{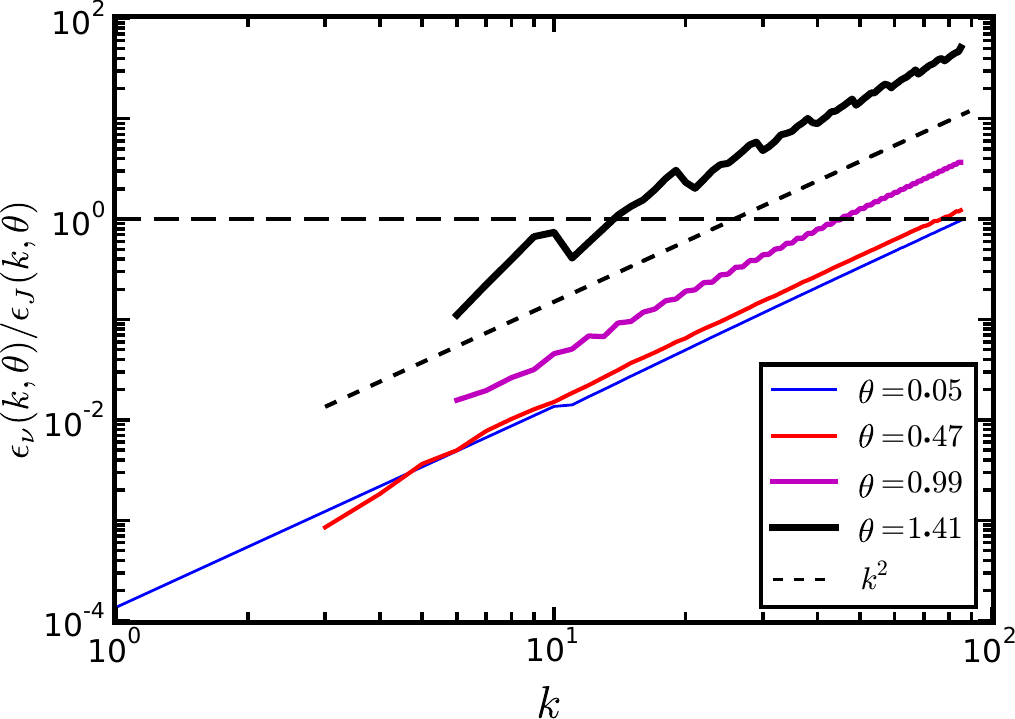}
\end{center}
\caption{For $N=27$,  $\epsilon_\nu(k,\theta)/\epsilon_J(k,\theta)$ vs.~$k$ for various sectors. $\epsilon_\nu(k,\theta)/\epsilon_J(k,\theta) \sim k^2$.   From Reddy {\em et al}~\cite{Reddy:PP2014}. Reprinted with permission from AIP Publishing. }
\label{fig:diss_spec_theta}
\end{figure}

\begin{figure}[ht]
\begin{center}
\includegraphics[width=8.0cm,angle=0]{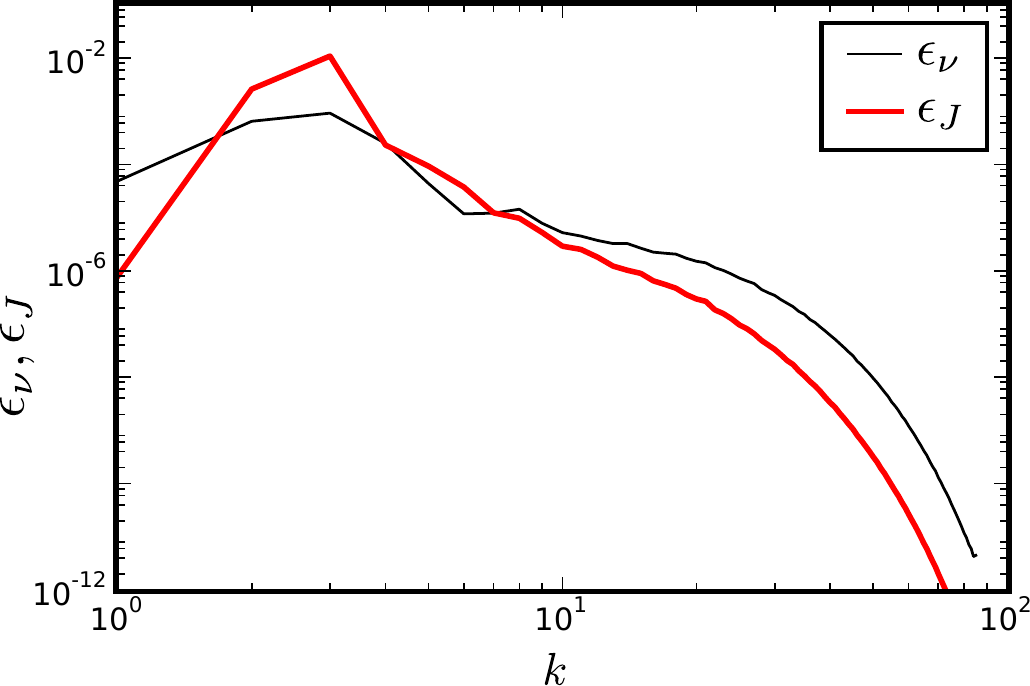}
\end{center}
\caption{The Joule dissipation spectrum $\epsilon_J$ and viscous dissipation spectrum $\epsilon_\nu$ for $N=27$.  Reprinted with permission from Reddy~\cite{Reddy:thesis}. }
\label{fig:diss_spec}
\end{figure}

The ratio of the angular distribution of the dissipation rates is given by~\cite{Reddy:PP2014}
\begin{equation}
\frac{\epsilon_\nu(k,\theta)}{\epsilon_J(k,\theta)} = \frac{2 \nu' k^2 E(\mathbf k)}{2 B_0'^2 \cos^2\theta E(\mathbf k)} =  \frac{2 \nu' k^2 }{2 B_0'^2 \cos^2\theta }.
\end{equation}
Thus, $\epsilon_\nu(k,\theta) = \epsilon_J(k,\theta)$ at
 \begin{equation}
 k^* = \frac{B_0' \cos\theta}{\sqrt{{\nu'}}}.
 \label{eq:kstar}
\end{equation}
Hence $\epsilon_J(k,\theta) > \epsilon_\nu(k,\theta)$ for $k <k^*$, consistent with the fact that $\epsilon_J$ dominates at low wavenumbers.   For a  sector of angle $\theta$, the ratio $\epsilon_\nu(k,\theta)/\epsilon_J(k,\theta) \propto k^2$.   In Fig.~\ref{fig:diss_spec} we plot the Joule dissipation spectrum~\cite{Reddy:thesis}
\begin{equation}
\epsilon_J(k)  = \int  2 B_0^{'2} \cos^2 \theta E(k,\theta)  d\theta
\end{equation}
and the viscous dissipation 
\begin{equation}
\epsilon_\nu(k)  =  2 \nu' k^2  E(k).
\end{equation}
Clearly the Joule dissipation dominates at small wavenumbers, while the viscous dissipation at large wavenumbers.   

\subsubsection{Anisotropy quantification using Legendre polynomials} \label{subsubsec:anis_group_theory}
Isotropic systems typically exhibit spherically symmetric correlations.  For example, we expect $E(k,\theta)$ of isotropic and homogeneous turbulence to be independent of $\theta$.  However, induction of external magnetic field or rotation breaks the spherical symmetry, and $E(k,\theta)$ becomes function of both $k$ and $\theta$.  Note however that the angular anisotropy could be scale-dependent, that is, the system may exhibit variations at different angular resolutions.  Such multi-resolution variations are not easily quantifiable using $E(k,\theta)$, but they are easier to quantify using polynomials.  For such analysis, it is customary to employ spherical harmonics, which are also eigenfunctions of the Laplacian operator ($\nabla^2$). This analysis is analogous to the Fourier transform in which the Fourier amplitudes for various $k$'s capture the scale-dependent features of the system.

QS MHD turbulence under the influence of a constant external magnetic field is azimuthally symmetric, hence the energy spectrum $E({\bf k})$ is independent of $\phi$.  Reddy and Verma~\cite{Reddy:PF2014} exploited this symmetry and employed Legendre polynomials to extract the angular dependence of the ring spectrum as
\begin{equation}
E(k, \theta) = \sum_l a_l P_{l}(\cos \zeta), \label{eq:legen}
\end{equation} 
where the angle $\zeta=\pi/2-\theta$ is chosen so as to keep the maximum of the function at  $\zeta=0$.   The coefficient $a_0$ represents the isotropic component of the flow or $E(k,\theta)$, while higher $a_l$'s provide information about the anisotropic components. Note that  odd-indexed $a_l$'s are negligible due to the $\theta \rightarrow \pi-\theta$ symmetry. 

%\begin{figure}[htbp]
\begin{figure}[htbp]
\begin{center}
\includegraphics[width=10cm]{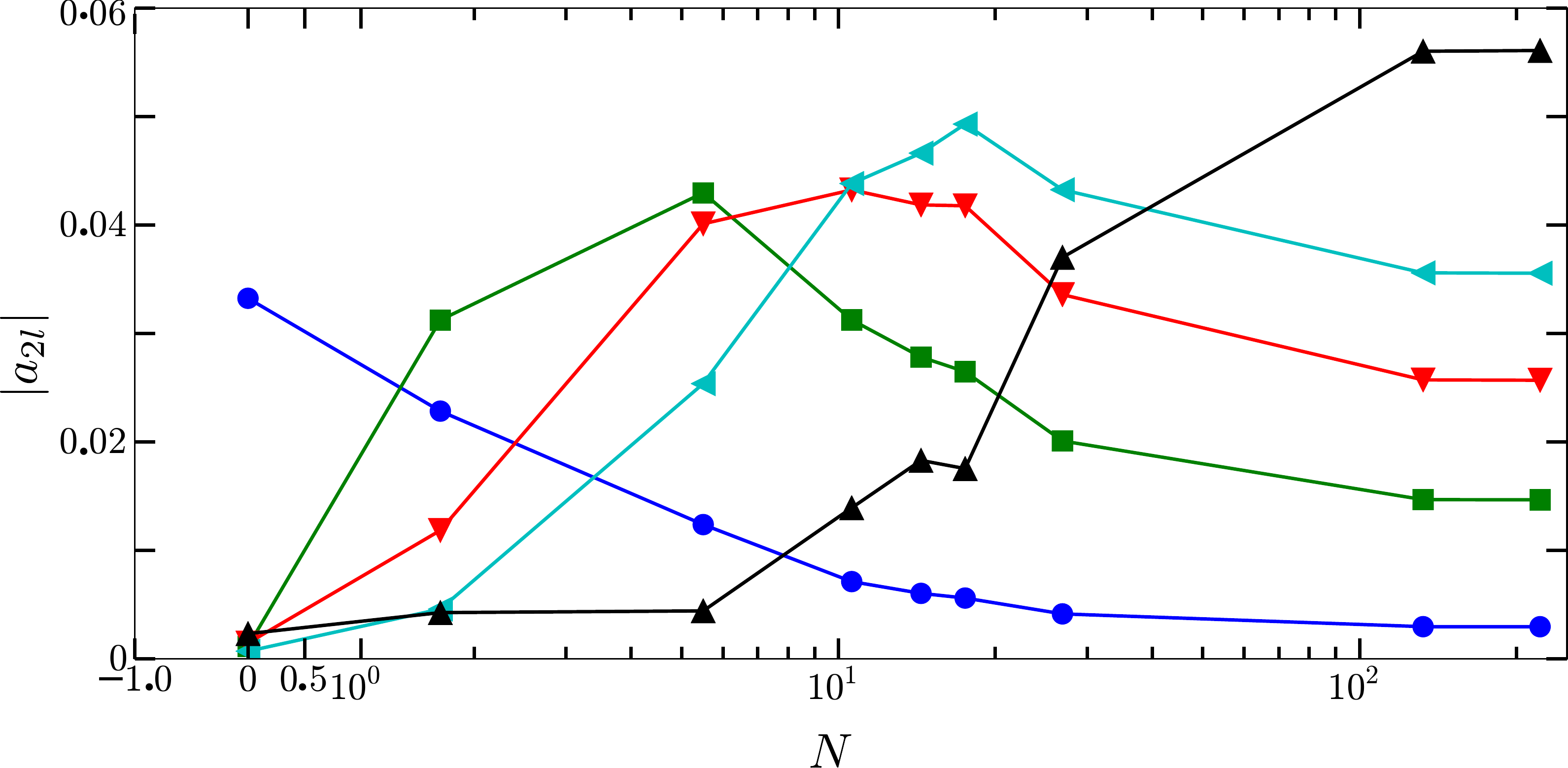}
\end{center}
\caption{The coefficients $a_l$ of the Legendre polynomials  of Eq.~(\ref{eq:legen}) for $N \in [0, 220]$, $k=20$. Here $a_0$: filled-circle (blue), $a_2$: $\blacksquare$ (green), $a_4$: $\blacktriangledown$ (red), $a_6$: $\blacktriangleleft$ (turquoise), and $a_{16}$: $\blacktriangle$ (black). From Reddy and Verma~\cite{Reddy:PF2014}. Reprinted with permission from AIP Publishing.}
\label{fig:Legen_coeff}
\end{figure}

Reddy and Verma~\cite{Reddy:PF2014} computed $a_{2l}$ using the numerical ring spectrum, $E(k,\theta)$; the results are exhibited in Fig.~\ref{fig:Legen_coeff}.  For $N=0$, $a_0$ dominates all other modes.  As exhibited in the figure, the amplitudes of $a_2$, $a_4$, $a_6$, and $a_{16}$ are most significant for $N = 5.5, 11, 18$, and 220 respectively.  Thus, larger $N$ have maximum amplitude at larger $l$.  This observation is consistent with the fact that the peak of the ring spectrum $E(k,\theta)$ shifts towards the equatorial region $(\theta \rightarrow \pi/2)$ as $N$ increases.  See Figs.~\ref{fig:ring_spectrum} and \ref{fig:EkDk_20}  for comparison. 

It is important to note that for spherically symmetric systems, the solution of the equations may not satisfy the spherical symmetry.  An well-known example is the  Hydrogen atom; here the potential $V({\bf r}) = -1/r$ is spherically symmetric, but all the wavefunctions of the electrons are not spherically symmetric.  In similar lines, it has been shown that the correlation function of the hydrodynamic turbulence contains anisotropic tensorial components in its expansion~\cite{Biferale:PR2005,Kurien:PRE2000}.  Note that in the QS MHD turbulence, the external mean magnetic field breaks the isotropy of the system as well as that of its equation.  Thus, the degree of anisotropy in QS MHD is  much larger than that in isotropic fluid turbulence, and the spherical harmonics are useful tools to specify the anisotropy.  We expect the anisotropy in QS MHD turbulence to be much stronger than that in the hydrodynamic turbulence.

%Group theory helps us characterise systems based on  symmetry properties. For example, the Coulomb potential in  Hydrogen atom is spherically symmetric and forms $SO$(3) group,  consequently, the wavefunction $\psi$, satisfying equation $\nabla^2 \psi = -E \psi$, is expanded using spherical harmonics.  Kurien~\cite{Kurien:PRE2000} and Biferale~\cite{Biferale:PR2005} exploited the rotation symmetry of the Navier Stokes equation and  expanded the velocity correlation function of isotropic fluid turbulence using spherical harmonics.  Thus various terms of anisotropic tensors appear in the expansion, with the spherically symmetric term appearing only to the lowest order. Experiments and numerical simulations~(see review~\cite{Biferale:PR2005} for references) reveal presence of higher-order terms of $SO$(3) decomposition in the velocity correlation function, thus they demonstrate deviations from the spherical symmetry or anisotropy in hydrodynamic turbulence.  

\subsubsection{Tensorial representation of anisotropy}  
\label{subsubsec:tensorial}
Researchers have also attempted to express the velocity correlation function using tensors.  One such attempt is by Verma~\cite{Verma:PR2004} who proposed a formula for the correlation function in the presence of an external field along $\hat{n}$ as
\be
\langle  \hat{u}_i({\mathbf k})  \hat{u}_j^*({\mathbf k})  \rangle =  \phi_{ij}({\mathbf k},t) = \left( \delta_{ij} - \frac{k_i k_j}{k^2} \right) C_1(k) + P'_{ij}({\bf k, n}) C_2(k),
\label{eq:verma_tensor}
\ee
where
\be
P'_{ij}({\bf k, n}) = \left( n_i - \frac{{\bf n \cdot k}}{k^2} k_i \right)\left( n_j - \frac{{\bf n \cdot k}}{k^2} k_j \right).
\ee
Here $C_1(k), C_2(k)$ are scalar functions of $k$.  From the above expressions, the modal energy is
\be
\frac{1}{2} \langle  |\hat{\bf u}({\mathbf k})|^2 \rangle = C_1(k) + \frac{1}{2} \sin^2\theta C_2(k),
\label{eq:Verma_PRanis_uu}
\ee
where $\hat{n} \cdot \hat{k} = \cos\theta$.  The above expression for the modal energy uses only $P_0(\cos\zeta)$ and $P_2(\cos\zeta)$, hence Eq.~(\ref{eq:verma_tensor}) cannot describe the velocity correlations  for moderate and large $N$ that involves higher $P_l$'s, as evident from Eq.~(\ref{eq:legen}) and Fig.~\ref{fig:Legen_coeff}.  

Ishida and Kaneda~\cite{Ishida:PF2007} employed perturbation method to derive a tensorial representation for the correlation function of QS MHD turbulence for small $N$; their arguments are based on symmetries.  Note that the formula of   Ishida and Kaneda~\cite{Ishida:PF2007} too is not applicable for general $N$.  Hence, general  tensorial expression for the energy spectrum needs to expanded using  Eq.~(\ref{eq:legen}). Note that Eq.~(\ref{eq:u_tensor}) of Sec.~\ref{sec:analytic}, which is a special case of Eq.~(\ref{eq:verma_tensor}) with $C_2=0$, is also inapplicable for strongly anisotropic QS MHD turbulence. 

In this section we showed that for large $N$, the kinetic energy is concentrated near the equator. We also observed that the  $E_\perp/E_\parallel$ peaks around $N \approx 18$ due to interesting exchange of energy between $U_\parallel$ and ${\bf U}_\perp$.  We will investigate these energy transfers in the next section, and explore further reasons for the quasi 2D behaviour of QS MHD turbulence.

\section{Energy Transfers in QS MHD turbulence} \label{sec:ET}

The nonlinear interactions among the Fourier modes of Eq.~(\ref{eq:k_NS})  yield energy transfers among the modes of QS MHD.   A major  effort in turbulent research is how to quantify  these transfers.  Kraichnan~\cite{Kraichnan:JFM1959} computed the energy transferred to one of the modes in a wavenumber triad  ${({\bf k}, {\bf p}, {\bf q)}}$ that satisfied ${{\bf k} = {\bf p}+ {\bf q}}$. Later,  Dar {\it et al}~\cite{Dar:PD2001} and Verma~\cite{Verma:PR2004}  developed a formalism in which  the energy transfer rate from  mode ${\bf p}$ to  mode ${\bf k}$ with mode ${\bf q}$ as a mediator is
\begin{equation}
S({\bf k|p|q}) = \mathrm{\Im} \{ [ {\bf k \cdot {\hat U}(q)] [\hat U^*(k) \cdot \hat U(p)}] \},
\label{eq:mode2mode}
\end{equation}
where $\Im$ is the imaginary part.   Reddy {\em et al}~\cite{Reddy:PP2014} used the above formula to compute the energy flux, shell-to-shell energy transfers, and ring-to-ring energy transfers  in QS MHD turbulence.  We will discuss these measures in the present section.

\subsection{Cumulative measures: Energy flux for QS MHD turbulence}  \label{subsec:flux}
\begin{figure}[htbp]
\begin{center}
\includegraphics[width=8cm]{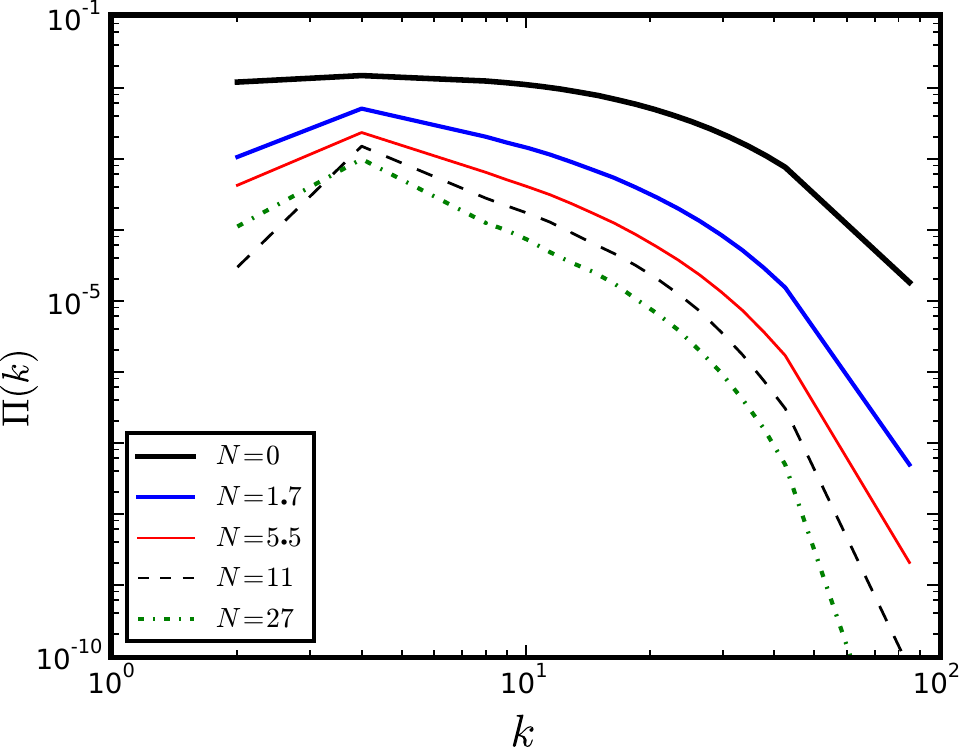}
\end{center}
\caption{Kinetic energy flux $\Pi(k)$ for $N=0, 1.7, 5.5, 11,$ and 27 for $k_f = (1,3)$.  Reprinted with permission from Reddy~\cite{Reddy:thesis}. }  
\label{fig:energy_flux}
\end{figure}

The energy flux $\Pi(k_0)$ is defined as the energy transferred from  the modes residing inside a sphere of wavenumber  radius $k_0$ to the modes outside the same sphere~\cite{Kraichnan:JFM1959,Lesieur:book:Turbulence,Verma:PR2004}, which is 
\begin{equation}
\Pi(k_0) = \sum_{|{\bf k}| > k_0}\sum_{|{\bf p}| \leq k_0}S({\bf k|p|q}).
\end{equation}
Using the statistical steady-state data of  forced QS MHD turbulence, Reddy {\em et al}~\cite{Reddy:PP2014} computed the energy flux  for $N = 1.7, 5.5, 11, 27, 130$ and 220, some of which are plotted in Fig.~\ref{fig:energy_flux}.   These plots are for the random forcing at $k_f =(1,3)$.  For the  hydrodynamic case ($N=0$), the energy flux is an approximate constant in the inertial range, which is consistent with the classical Kolmogorov flux~\cite{Kolmogorov:DANS1941c,Kolmogorov:DANS1941a}.   For $N > 0$, the Joule dissipation at different scales leads to a decrease of $\Pi(k)$ with $k$ as $d\Pi(k)/dk) = -(\epsilon_\nu(k) + \epsilon_J(k))$ [see Eq.~(\ref{eq:dPi_dk})].    The decay of the energy flux leads to steepening of the energy spectrum, as discussed in Sec.~\ref{subsec:Ek}, and as shown in Figs.~\ref{fig:vorobev}, \ref{fig:KE_spectrum_powerlaw}, and \ref{fig:KE_spectrum_semilog}.     In Sec.~\ref{subsec:VR_model} we will review the models that capture the variations of the kinetic energy flux in the presence of Joule dissipation.  According to these models,  $\Pi(k) \sim k^{-a}$ for small and moderate $N$'s, and $\Pi(k) \sim \exp(-ak)$ for large $N$'s ($N \ge 130$).  

Figure \ref{fig:energy_flux} does not capture an important aspect of QS MHD turbulence. The Lorentz force, which is proportional to $B^{'2}_0 \cos^2\theta$, vanishes at $\theta=\pi/2$.  Hence, in $k_z=0$ plane,  the nonlinear term ${\bf u} \cdot \nabla {\bf u}$ is the most effective term for large $\mathrm{Re}$, and the flow has behaviour similar to  2D hydrodynamic turbulence with an inverse cascade of energy at  wavenumbers lower than the forcing wavenumber band $k_f$.  For the runs described above, $k_f = (1,3)$, hence the energy flux does not exhibit inverse cascade. However, the small wavenumbers ($k \sim 1$) do receive energy due to nonlinearity, because of which $E_\perp/E_\parallel \gg 1$ for small $k$, as shown in Fig.~\ref{fig:Eperp_Epar}.  The inverse energy transfers lead to quasi 2D flow structures of Figs.~\ref{fig:isosurface_vorticity} and \ref{fig:N132_vectors}.

\begin{figure}
\begin{center}
\includegraphics[width=8.0cm,angle=0]{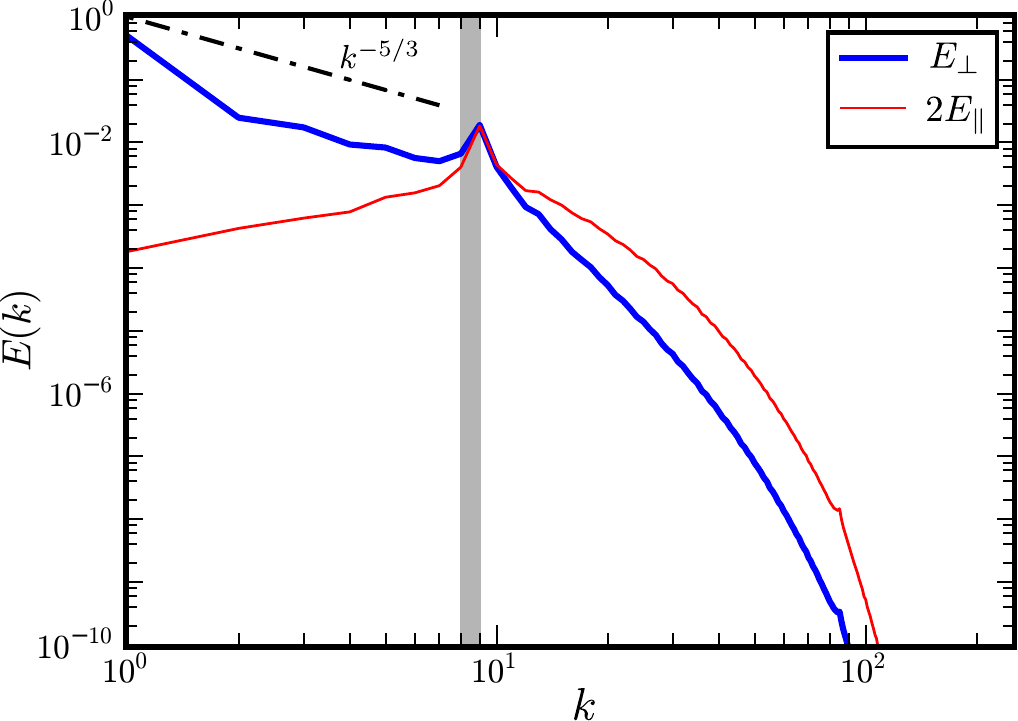}
\end{center}
\caption{ Plots of $E_{\perp}(k)$ and $2 E_{\parallel}(k)$ for $N=100$ when  $k_f \in [8,9]$ (the shaded region).  For $k< k_f$, $E_{\perp}(k) > E_{\parallel}(k)$ with $E_{\perp}(k)  \sim k^{-5/3}$, but  for $k> k_f$, $E_{\perp}(k) < E_{\parallel}(k)$.  From Reddy {\em et al}~\cite{Reddy:PF2014}. Reprinted with permission from AIP Publishing. }
\label{fig:spec}
\end{figure}

To explore the nature of the inverse energy cascade  in QS MHD turbulence,  Reddy {\em et al}~\cite{Reddy:PP2014} simulated forced QS MHD for $N=100$ with the forcing wavenumber band $k_f = (8,9)$.   Figure~\ref{fig:spec} illustrates the energy spectra of the parallel and perpendicular components of the velocity field for this simulation.  Here $E_\perp(k) \gg 2E_\parallel(k) $ for $k< k_f$, but $2E_\parallel(k)  \gg E_\perp(k)$ for  $k> k_f$.  Also, for $k<k_f$, $E_\perp(k) \sim k^{-5/3}$, thus exhibiting behaviour similar to 2D hydrodynamic turbulence.  

\begin{figure}
\begin{center}
\includegraphics[width=8.0cm,angle=0]{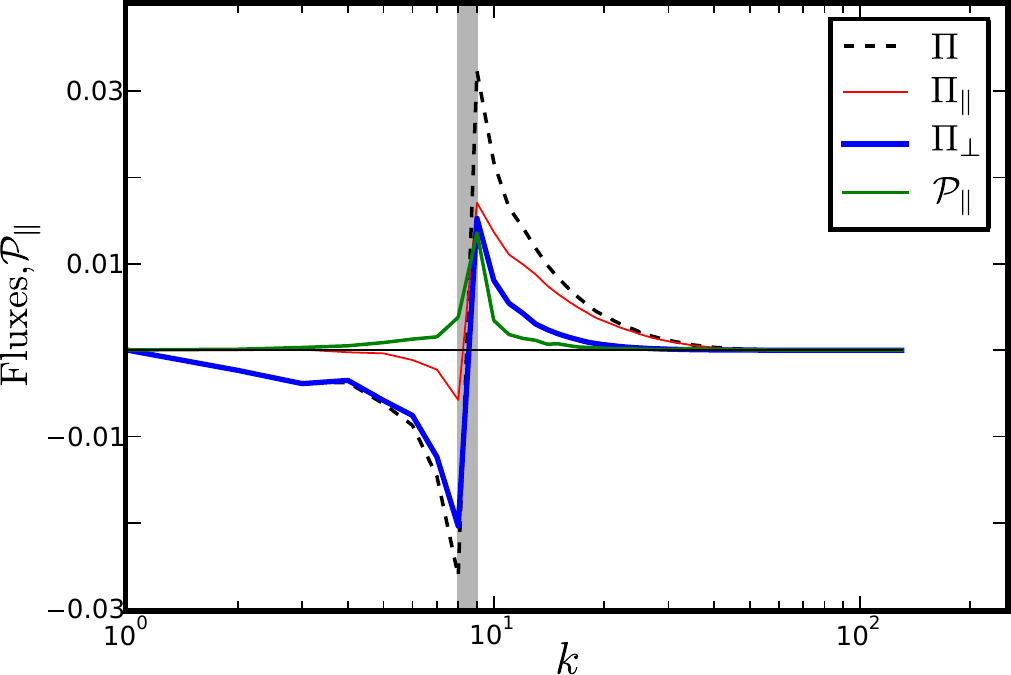}
\end{center}
\caption{Plots of the energy fluxes $\Pi(k)$,  $\Pi_{\parallel}(k)$,   $\Pi_{\perp}(k)$ and $P_\parallel(k)$ for $N=100$  when  $k_f \in [8,9]$ (the shaded region).   For $k<k_f$, $\Pi_{\perp}(k) < 0$ indicating an inverse cascade for ${\bf U}_\perp$, while for $k>k_f$, $\Pi_\parallel(k) > 0$ exhibiting a forward cascade for $U_\parallel$.  $\mathcal{P}_\parallel(k) > 0$ for $k> k_f$, indicating an energy transfer from ${\bf U}_\perp$ to $U_\parallel$ via pressure.  From Reddy {\em et al}~\cite{Reddy:PP2014}. Reprinted with permission from AIP Publishing.  }
\label{fig:flux_par_perp}
\end{figure}

Reddy {\em et al}~\cite{Reddy:PP2014} computed the energy flux $\Pi(k)$ (for $k_f = (8,9)$) which is exhibited in Fig.~\ref{fig:flux_par_perp} as a dashed curve. We observe a negative $\Pi(k)$ for $k<k_f$, thus demonstrating an inverse cascade of kinetic energy.  However  $\Pi(k) > 0$ for $k>k_f$ indicating a forward energy cascade in this range.   Note that 2D hydrodynamic turbulence predicts $\Pi(k) \approx 0$ for $k>k_f$~\cite{Boffetta:ARFM2012}.  The above observation that $\Pi(k) > 0$ for $k>k_f$, a major deviation from 2D hydrodynamics, is due to the forward cascade of $U_\parallel$.  The dominance of $E_\parallel$ over $E_\perp$ for $k>k_f$ (Fig.~\ref{fig:spec}) is due to this flux.   We will compute the fluxes of ${\bf U}_\perp$ and $U_\parallel$ in Sec.~\ref{sec:perp_parallel}.

The energy flux is a cumulative energy transfer from the modes inside a wavenumber sphere to the modes outside the sphere.  The shell-to-shell energy transfers, to be described in the next section, provides a more detailed picture of the energy transfer in turbulence.

\subsection{Shell-to-shell energy transfers of QS MHD turbulence}   \label{subsec:shell2shell}
The shell-to-shell energy transfer rate is another quantity used for quantifying the energy transfers.   The shell-to-shell energy transfer rate from all the modes of shell $m$  to the modes of shell $n$   is defined as 
\begin{equation}
T^m_n=\sum_{{\bf k} \in n}\sum_{{\bf p} \in m}S({\bf k|p|q}),
\end{equation}
where $S({\bf k|p|q})$ is given by Eq.~(\ref{eq:mode2mode}).  For hydrodynamics turbulence, $T^m_n$ has been shown to be local and forward in the inertial range, i.e., the maximal energy transfer takes place from shell $m$ to $m+1$~\cite{Davidson:book:Turbulence,Lesieur:book:Turbulence,Leslie:book,Verma:Pramana2005a}.   

\begin{figure}[ht]
\begin{center}
\includegraphics[width=10.0cm,angle=0]{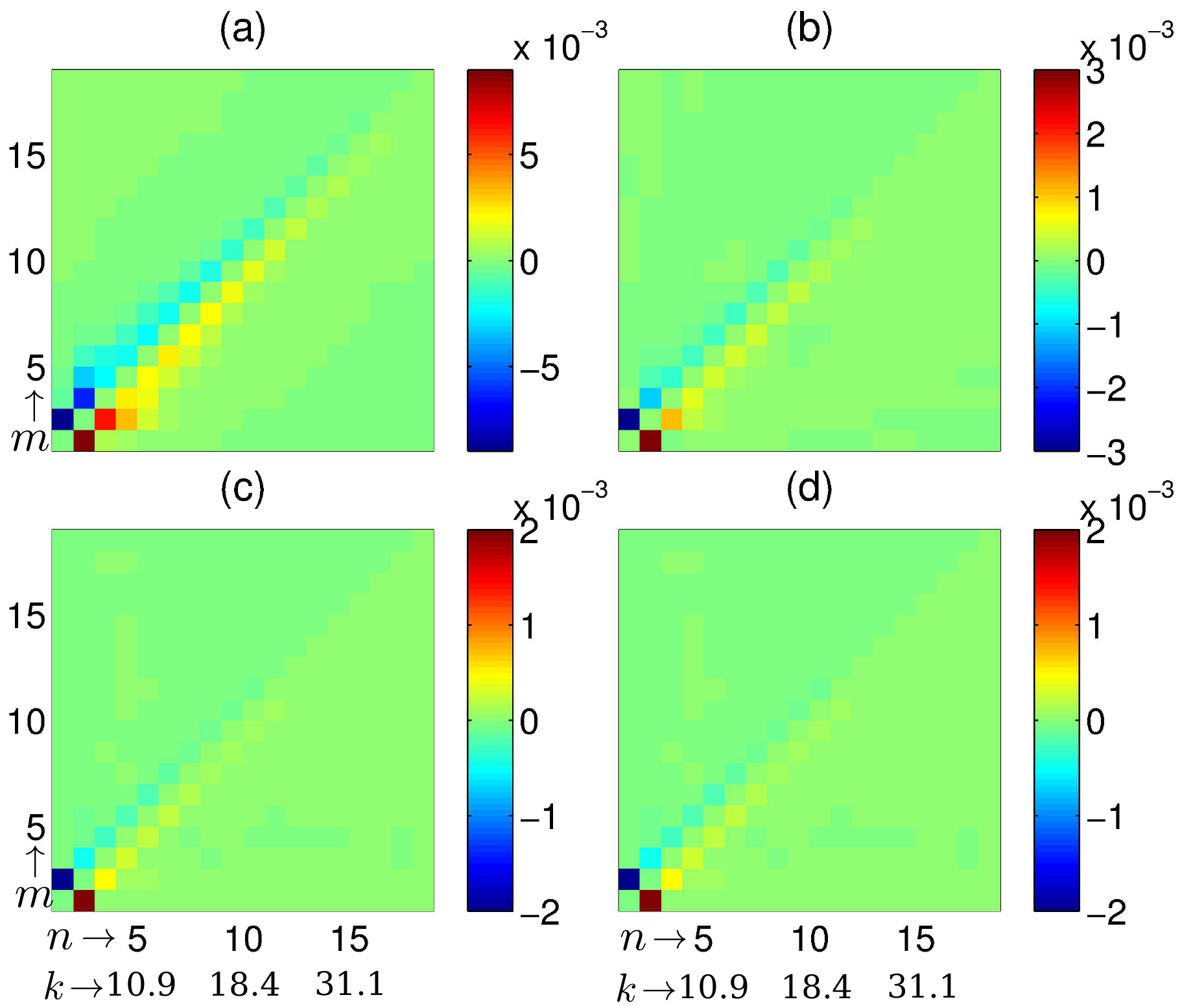}
\end{center}
\caption{Forward and local shell-to-shell energy transfer rates $T^m_n$ for (a) $N=1.7$, (b) $N=11$, (c) $N=18$, and (d) $N=130$. {\em Notation} $m$: giver shell,  $n$: receiver shells, $k$ the wavenumber of the outer radius of the corresponding shell.  From Reddy {\em et al}~\cite{Reddy:PP2014}. Reprinted with permission from AIP Publishing.  } 
\label{fig:shell}
\end{figure}

Reddy {\em et al}~\cite{Reddy:PP2014} computed the shell-to-shell energy transfers for QS MHD turbulence when $k_f = (1,3)$.  They  binned the Fourier space logarithmically with  the shell radii as 4.0, 8.0, 8.9, 9.9, 10.9, 12.2, 13.5, 14.9, 16.6, 18.4, 20.5, 22.7, 25.2, 28.0, 31.1, 34.5, 38.3, 42.5, and 85.0.   Figure~\ref{fig:shell} exhibits the shell-to-shell energy transfer rates for $N=1.7, 11, 18$, and $130$.  The figure shows that  shell $n$ gives energy to   shell $(n+l)$ with ($l>0$), and it receives energy from  shell $(n-l)$  indicating forward energy transfer.  Also, the maximum energy transfer is to the nearest neighbour, i.e., shell $n$ gives the maximum positive energy transfer to the shell $(n+1)$.  Therefore we conclude that the shell-to-shell energy transfer in QS MHD turbulence is  local and forward.  

The shell-to-shell  transfers are dominant for small $m$ and $n$, specially for large $N$.  This is because the energy dominantly resides in small wavenumber shells.  This feature differs from 3D hydrodynamic turbulence for which the shell-to-shell energy transfer is local for a larger range of wavenumbers.  Also, for $k < k_f$ we expect a backward energy transfer.  This feature of QS MHD needs to be investigated for large $k_f$.  

The energy flux and the shell-to-shell energy transfers provide an averaged measure over the polar angle $\theta$ of Fig.~\ref{fig:ring_decomposition}, hence they do not capture the anisotropic energy transfers which are $\theta$-dependent.  In the following, we present the ring-to-ring energy transfers using which we can quantify $\theta$-dependent anisotropic  energy transfers. 

\subsection{Anisotropic measures: Ring-to-ring energy transfers of QS MHD turbulence}
\label{subsec:ring2ring}

Reddy {\em et al}~\cite{Reddy:PP2014} divided the wavenumber shells into rings, as shown in Figure~\ref{fig:ring_decomposition}.    The ring-to-ring energy transfer rate from the ring $(m,\alpha)$ to the ring $(n,\beta)$ is~\cite{Teaca:PRE2009}
\begin{equation}
T_{(n,\beta)}^{(m,\alpha)}=\sum_{{\bf k} \in (n,\beta)} \sum_{{\bf p} \in (m,\alpha)}  S({\bf k}|{\bf p}|{\bf q}),
\end{equation}
where $S({\bf k|p|q})$ is given by Eq.~(\ref{eq:mode2mode}).   The ring-to-ring energy transfers are normalized using $A_{\alpha} = |\mathrm{cos}(\theta_\alpha)-\mathrm{cos}(\theta_{\alpha+1})|$ to compensate for the uneven distribution of modes among the rings~\cite{Teaca:PRE2009}; the rings closer to the equator have more Fourier modes than those near the poles. Therefore, the normalized ring  transfer  is
\begin{equation}
\overline{T}_{(n,\beta)}^{(m,\alpha)} = \frac{1}{A_{\alpha}A_{\beta}} T^{(m,\alpha)}_{(n,\beta)}.
\end{equation}
Teaca {\em et al}~\cite{Teaca:PRE2009} performed first such computations for MHD turbulence.

\begin{figure}[ht]
\begin{center}
\includegraphics[scale=1.2]{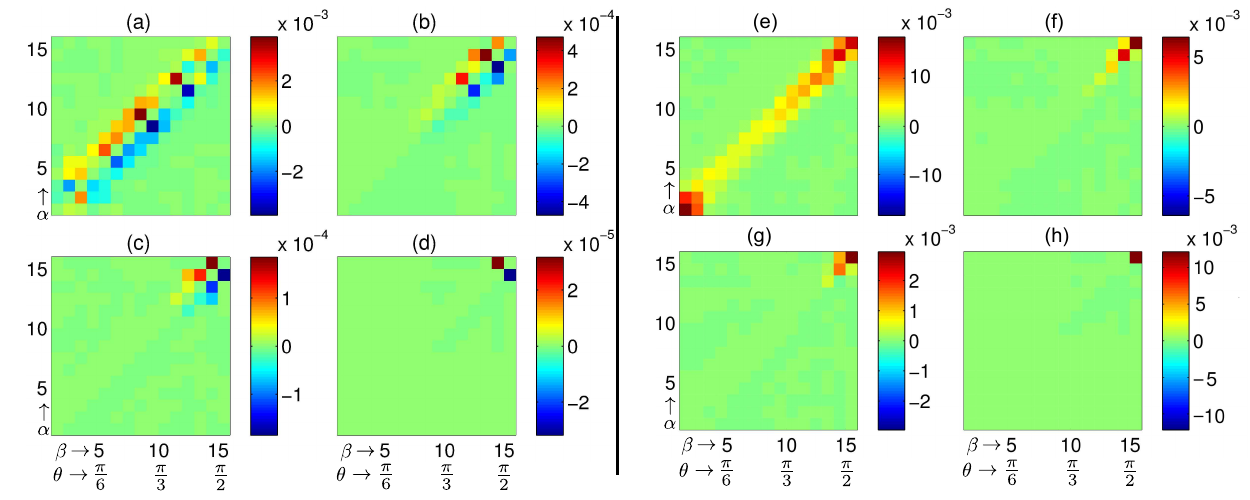}
\end{center}     
\caption{  Ring-to-ring energy transfers $\overline{T}_{(9,\beta)}^{(9,\alpha)}$ among the rings of the $9^{\rm th}$ shell for (a) $N=1.7$, (b) $N=11$, (c) $N=18$, and (d) $N=130$; and $\overline{T}_{(10,\beta)}^{(9,\alpha)}$ for (e) $N=1.7$, (f) $N=11$, (g) $N=18$, and (h) $N=130$. {\em Notation} $\alpha$: giver ring,  $\beta$: receiver  ring,  $\theta$:   the angle of the corresponding rings. $\overline{T}_{(9,\beta)}^{(9,\alpha)}$ is dominant for neighbouring rings  (local).  For large $N$, the ring transfers are dominant near the equator.  Figure (a) from Reddy {\em et al}~\cite{Reddy:PP2014}; reprinted with permission from AIP Publishing. Figure (b) reprinted with permission from Reddy~\cite{Reddy:thesis}. }
\label{fig:ring-2-ring}
\end{figure}

Reddy {\em et al}~\cite{Reddy:PP2014} computed the ring-to-ring energy transfers for $N=1.7, 11, 18$, and 130 using the steady-state numerical data.  They observed that the maximum energy transfer takes place among the neighbouring rings (nearest shells and sectors), thus the energy transfers among the rings is local.   In the following discussion, we focus on ring-to-ring transfers among the neighbouring shells, in particular, for the $9^\mathrm{th}$ and $10^\mathrm{th}$  shells.   In   Figs.~\ref{fig:ring-2-ring}  we present the results on the normalized ring-to-ring energy transfers $\overline{T}^{(m,\alpha)}_{(n,\beta)}$ from the rings of  $9^{\rm th}$ shell ($m=9$) to the rings of shells $n=9$ and 10.   In these figures, the vertical axis represents the sector indices of the giver rings ($\alpha$), while the horizontal axis represents the sector indices of the receiver rings ($\beta$). 

\vspace{1.0cm}
According to Fig.~\ref{fig:ring-2-ring}(a-d), $\overline{T}^{(9,\alpha)}_{(9,\beta)}$ (the energy transfers among the rings within the $9^{\rm th}$ shell) has maximum value for  $\overline{T}^{(9,\alpha)}_{(9,\alpha \pm 1)}$ with $\overline{T}^{(9,\alpha)}_{(9,\alpha-1)} > 0$ and $\overline{T}^{(9,\alpha)}_{(9,\alpha+1)} < 0$.  Thus the energy transfer is local among the rings, and they are from  larger $\theta$ to  smaller $\theta$.    Hence, the energy transfer is local in the angular direction as well (along with the local shell-to-shell transfers described in the previous subsection).   Another important feature of the ring-to-ring transfers is that for large $N$ ($N=11, 18, 130$), the dominant energy transfers takes place from the rings closer to the equator to their neighbours (lower $\theta$).  Thus, for large $N$, the dominant energy transfers take place near the equatorial region because the energy is concentrated near this region.

%\begin{figure}[htbp]
%\begin{center}
%\includegraphics[width=14.0cm,angle=0]{figures/R9_R10}
%\end{center}     
%\caption{Ring-to-ring energy transfers $\overline{T}_{(10,\beta)}^{(9,\alpha)}$ from the  rings of the $9^{\rm th}$ shell  to the rings of the $10^{\rm th}$ shell for (a) $N=1.7$, (b) $N=11$, (c) $N=18$, and (d) $N=130$.   Note that $\overline{T}_{(10,\beta)}^{(9,\alpha)} > 0$. The notation is same as that for Fig.~\ref{fig:ring-2-ring9_9}.   Taken from Reddy {\em et al}~\cite{Reddy:PP2014}.} 
%\label{fig:ring-2-ring9_10}
%\end{figure}

%\begin{figure}[ht]
%\begin{center}
%\includegraphics[width=14.0cm,angle=0]{C4/R9_R8}
%\end{center}     
%\caption{Local ring-to-ring energy transfers $\overline{T}_{(8,\beta)}^{(9,\alpha)}$ from the rings of the $9^{\rm th}$ shell to the rings of the $8^{\rm th}$ shell for  (a) $N=1.7$, (b) $N=11$, (c) $N=18$, and (d) $N=130$.   Note that $\overline{T}_{(8,\beta)}^{(9,\alpha)}<0$.}
%\label{fig:ring-2-ring9_8}
%\end{figure}

Figure~\ref{fig:ring-2-ring}(e-h) illustrates $\overline{T}^{(9,\alpha)}_{(10,\beta)}$, the energy transfers  from the rings in the $9^{\rm th}$ shell to those in the $10^{\rm th}$ shell (immediate neighbour of shell 9).  The figure shows that  $\overline{T}^{(9,\alpha)}_{(10,\beta)} > 0$ for all rings with the maximal transfers occurring for the equatorial rings ($\alpha, \beta \approx 15$).   Using these observations, we conclude that the energy is transferred dominantly along a sector near the equator, and that the energy transfers are from lower $k$ to larger $k$.     

\begin{figure}[htbp]
\begin{center}
\includegraphics[width=6.0cm,angle=0]{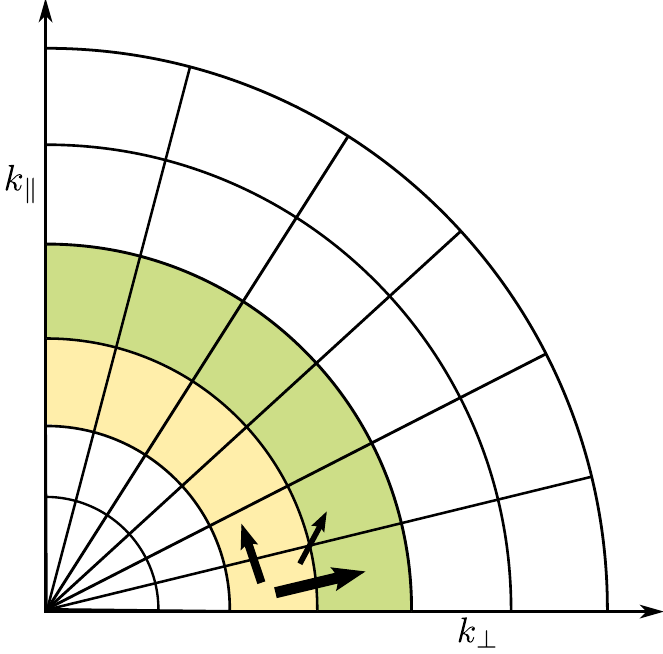}
\end{center}     
\caption{A schematic diagram exhibiting the dominant ring-to-ring energy transfers. The thickness of the lines is proportional to the intensity of the transfer.  Figure (b) reprinted with permission from Reddy~\cite{Reddy:thesis}.}
\label{fig:ring_schematic}
\end{figure}

We summarise the ring-to-ring energy transfers in QS MHD using a schematic diagram exhibited in Fig.~\ref{fig:ring_schematic}.  The ring-to-ring transfers are local and forward along $k$, but local and inverse along $\theta$ (transfers from larger $\theta$ to smaller $\theta$).  For large $N$, these transfers tend to be dominant near the equator because the energy is concentrated near the equator.  The energy transfer from larger $\theta$ to smaller $\theta$ is due to the stronger Joule dissipation at lower $\theta$ because of the $\cos^2\theta$ factor.

In an earlier discussions we had argued that $U_\perp$ feeds energy to $U_\parallel$ for large $k$'s.  In the following subsection we will compute these transfers.

\subsection{Energy transfers among the parallel and perpendicular components in QS MHD turbulence} \label{sec:perp_parallel}
 
The energy equations for the perpendicular and parallel components of the velocity field of QS MHD turbulence are~\cite{Reddy:PP2014}
\begin{eqnarray}
{{\partial E_{\perp}({\bf k})}\over {\partial t}} &=  & \sum_{\mathbf p} S_{\perp}({\bf k}|{\bf p}|{\bf q}) - 2 {{B{'}}_0^{2}}\cos^{2}(\theta)E_{\perp}({\bf k}) + \mathcal{P}_{\perp}({\bf k}) \nonumber \\
& &-\, 2 \nu' k^2 E_{\perp}({\bf k})+ \Re \{{ \bf \hat{F}_{\perp}(k) \cdot \hat{U}_{\perp}^*(k)} \}, \label{eq:Eperp_evol} \\ 
{{\partial E_{\parallel}({\bf k})}\over {\partial t}} &= & \sum_{\mathbf p} S_{\parallel}({\bf k}|{\bf p}|{\bf q}) - 2 {{B{'}}_0^{2}}\cos^{2}(\theta)E_{\parallel}({\bf k}) + \mathcal{P}_{\parallel}({\bf k})\nonumber \\
 && -\, 2 \nu' k^2 E_{\parallel}({\bf k})+ \Re \{{ \hat{F}_{\parallel}(k) \hat{U}_{\parallel}^*(k)}\}, \label{eq:Epar_evol}
\end{eqnarray}
where ${\bf q = k-p}$, $E_{\perp}({\bf k})$ $=\frac{1}{2}|\hat {\bf U }_{\perp}({\bf k})|^2$, $E_{\parallel}({\bf k})$ $=\frac{1}{2}|\hat U_{\parallel}({\bf k})|^2$, and
\begin{eqnarray}
\mathcal{P}_{\perp}({\bf k}) &=& \Im \{ [{\bf k \cdot \hat U^*_{\perp}(k)} ] \hat{P}({\bf k})\},\label{eq:defPperp} \\
\mathcal{P}_{\parallel}({\bf k}) &=& \Im \{ [  k_{\parallel}  \hat U^*_{\parallel}({\bf k}) ] \hat{P}({\bf k'})\}, \label{eq:defPpar}
\end{eqnarray}
where $ \hat{P}({\bf k})$ is Fourier transform of the pressure field $P$. 

 Reddy {\em et al}~\cite{Reddy:PP2014}  showed that ${\bf U}_\perp$ Fourier modes exchange energy among themselves (see~\ref{sec:ET_appendix}). The energy transfer rate from  ${\bf U}_\perp({\bf p})$ to ${\bf U}_\perp({\bf k})$ with  ${\bf U}({\bf q})$ acting as a mediator is given by
\begin{equation}
S_{\perp}({\bf k|p|q}) =  \mathrm{\Im} \{ [ {\bf k \cdot {\hat U}(q)] [\hat U^*_{\perp}(k) \cdot \hat U_{\perp}(p)}]\}, \label{eq:Sperp}
\end{equation}
while the energy transfer rate from $U_\parallel({\bf p})$ to   $U_\parallel({\bf k})$ with ${\bf U}({\bf q})$ acting as a mediator is given by
\begin{equation}
S_{\parallel}({\bf k|p|q}) =  \mathrm{\Im} \{ [ {\bf k \cdot {\hat U}({\bf q})}] [\hat U^*_{\parallel}({\bf k})  \hat U_{\parallel}({\bf p})] \}.  \label{eq:Spar}
\end{equation}
The energy fluxes, $\Pi_{\perp}(k_0)$ and $ \Pi_{\parallel}(k_0)$, of the perpendicular and parallel components of the velocity field can be defined using $S_{\perp}({\bf k|p|q})$ and $S_{\parallel}({\bf k|p|q})$.  They are  respectively the energy transfer rates of ${\bf U}_\perp$  and $U_\parallel$  out of the wavenumber sphere of radius $k_0$:
\begin{eqnarray}
\Pi_{\perp}(k_0) & = & \sum_{|{\bf k}| > k_0 } \sum_{|{\bf p}| \leq k_0} {  S_{\perp}({\bf k|p|q)}}, \label{eq:Pi_perp} \\
\Pi_{\parallel}(k_0) & = & \sum_{|{\bf k}| >  k_0 } \sum_{|{\bf p}| \leq k_0}  {  S_{\parallel}({\bf k|p|q)}}. \label{eq:Pi_pll}
\end{eqnarray} 
Note that the total energy flux $\Pi(k_0) = \Pi_{\perp}(k_0) + \Pi_{\parallel}(k_0)$. 

The energy equations (\ref{eq:Eperp_evol}, \ref{eq:Epar_evol}) reveal that ${\bf U}_\perp$ and $U_\parallel$ receive energy from the pressure as given by Eqs.~(\ref{eq:defPperp}) and (\ref{eq:defPpar}).  A closer inspection of Eqs.~(\ref{eq:defPperp}, \ref{eq:defPpar}) indicates that
\begin{equation}
 \mathcal{P}_{\perp}({\bf k}) +  \mathcal{P}_{\parallel}({\bf k}) = 0
\end{equation}
due to the incompressibility condition ${\bf k \cdot u(k)} = 0$.  Thus, the pressure acts as a mediator for the energy transfer between ${\bf U}_\perp$ and $U_\parallel$ (see~\ref{sec:ET_appendix}).  The parallel component $U_\parallel$  receives energy from ${\bf U}_\perp$ via pressure by an amount $\mathcal{P}_{\parallel}({\bf k}) $.   We also remark that there is no direct  energy exchange between ${\bf U}_\perp$ and $U_\parallel$ in the energy equation; and that the energy transfer via pressure is internal to a given mode.  Also note that the pressure does not transfer energy from a mode ${\bf p}$ to ${\bf k}$.    Another consequence of the above results is that  $\Pi_\parallel(k)$ and $\Pi_\perp(k)$ vary with $k$ due to the energy exchange among themselves via  pressure.

Reddy {\em et al}~\cite{Reddy:PP2014}  computed  $\Pi_\parallel(k)$, $\Pi_\perp(k)$, and $\mathcal{P}_{\parallel}({\bf k}) $ using the numerical data for the QS MHD run with $N=100$ and $k_f = (8,9)$.  The results illustrated in Fig.~\ref{fig:flux_par_perp} show that in the $k<k_f$ region, $\Pi_\perp$, which is negative, dominates other fluxes; this is consistent with the inverse cascade of $\mathbf U_\perp$.  In the $k>k_f$ region, $\Pi_\parallel$ dominates $\Pi_\perp$, and it is positive indicating a forward cascade for $ U_\parallel$~\cite{Favier:PF2010,Favier:JFM2011,Reddy:PP2014}.  The energy transfer from  $\mathbf U_\perp$ to $U_\parallel$ via pressure is the reason for quasi 2D nature of QS MHD turbulence described earlier.  Note that $\mathbf U_\perp$ is primarily dissipated by the Joule heating since it is active at small $k$, while $U_\parallel$ is dissipated by the viscous force due to its dominance at large $k$ (see Figs.~\ref{fig:diss_spec_theta} and \ref{fig:diss_spec}).

 The above phenomena along with its physical interpretation can be summarised as follows.  Equation~(\ref{eq:Epar_evol}) shows that $U_\parallel$ receives energy from ${\bf U}_\perp$ via pressure, and the rate of energy transfer is $\mathcal{P}_\parallel$ of Eq.~(\ref{eq:defPpar}). This is the reason why the flow of QS MHD turbulence is quasi two-dimensional (not two-dimensional) with significant $U_\parallel$. 

In addition, $U_\parallel$ is transported to larger wavenumbers by forward cascade.   The equation for $U_\parallel$ is 
\be
\frac{\partial U_\parallel}{\partial t} + {\bf U}  \cdot \nabla U_\parallel = -\frac{\partial P}{\partial z} - \frac{ \sigma } {\rho} \Delta^{-1} \left[ ({\mathbf B}_0 \cdot \nabla)^2  U_\parallel  \right] + \nu \nabla^2 U_\parallel,
\label{eq:Favier_Uparallel}
\ee
which is similar to the equation of passive scalar, except the  $-\partial P/\partial z$ and $(\sigma/\rho)\Delta^{-1} \left[ ({\mathbf B}_0 \cdot \nabla)^2  U_\parallel  \right] $ terms which do not appear for passive scalar. We  compute flux of $U_\parallel$, $\Pi_\parallel$, using Eq.~(\ref{eq:Pi_pll}) and find this to be significant for $k>k_f$, as shown in Fig.~\ref{fig:flux_par_perp}. 

Earlier, Favier {\em et al}~\cite{Favier:PF2010} had argued that ${\bf U}_\perp$ evolves as in 2D hydrodynamic turbulence since the Lorentz fore is absent in $k_z=0$ plane (note $\cos \theta =0$ at $k_z=0$ plane).  Therefore, ${\bf U}_\perp$ exhibits an inverse cascade. Favier {\em et al}~\cite{Favier:PF2010} proposed that $U_\parallel$ follows  the following equation:
\be
\frac{\partial U_\parallel}{\partial t} + {\bf U}_\perp \cdot \nabla_\perp U_\parallel = \nu \nabla^2 U_\parallel .
\label{eq:Favier_Uparallel}
\ee
As described above, $-\nabla P$ of the momentum equation plays a major role in the transfer of energy from ${\bf U}_\perp$ to $U_\parallel$.  It is easy to verify that without the $-\nabla P$ term, the total energy of the parallel component, $E_\parallel = (1/2)\int |U_\parallel|^2 d\tau$, of Eq.~(\ref{eq:Favier_Uparallel}) obeys
\be
\frac{d E_\parallel}{dt} = - \nu \int U_\parallel \nabla^2 U_\parallel d\tau.
\ee
Therefore, for $\nu \ne 0$, $E_\parallel$ vanishes and the flow will become two-dimensional, not quasi two-dimensional.  In addition, Eqs.~(12, 13) of Favier {\em et al}~\cite{Favier:PF2010} refer to the modes on the $k_z = 0$ plane only.  However, the other modes, specially those near $k_z = 0$ plane, are important in QS MHD turbulence.  The formalism presented in the present subsection   takes care of all the interactions of QS MHD in a consistent manner.

\begin{figure}
\begin{center}
\includegraphics[width=6.0cm,angle=0]{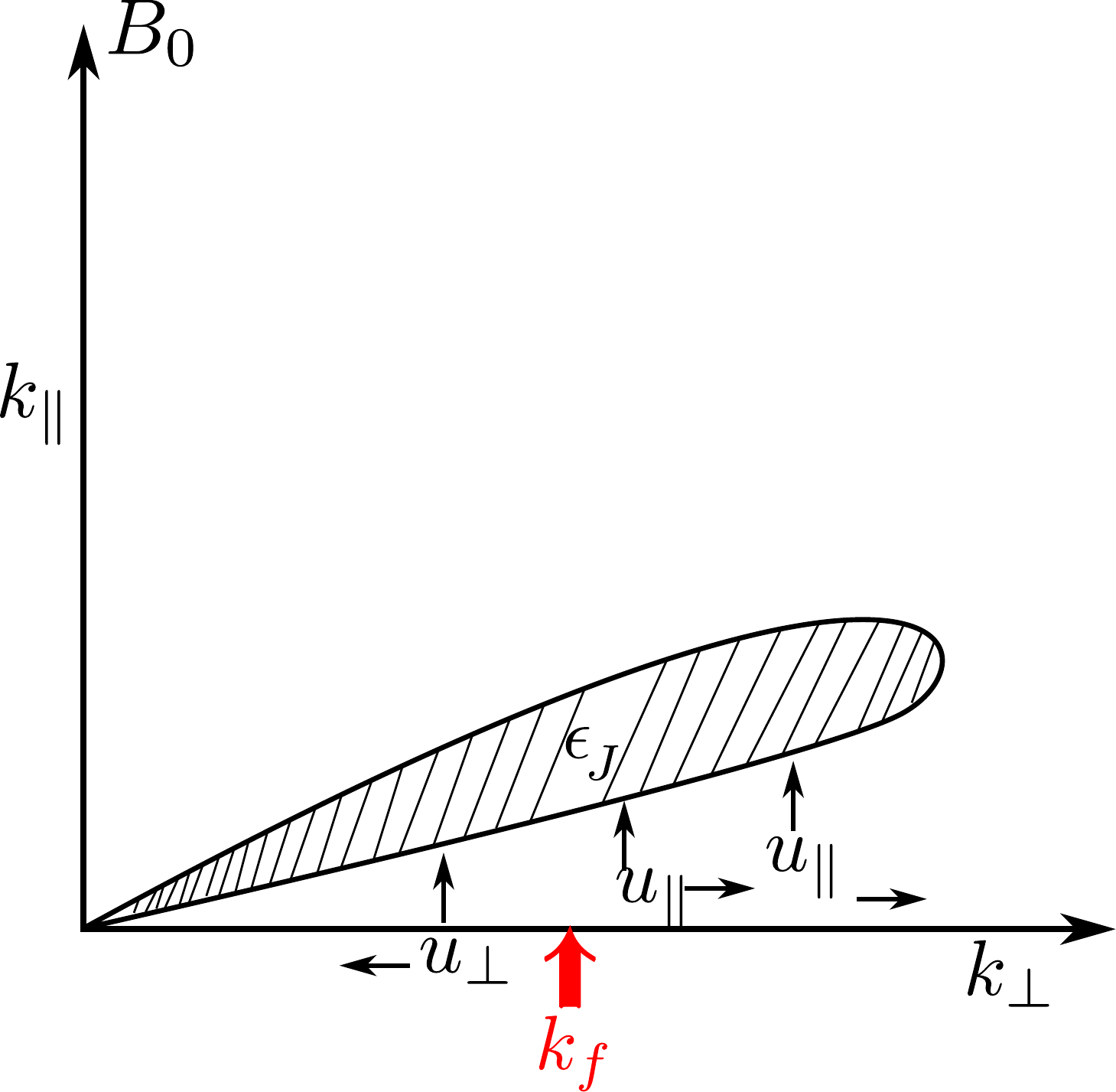}
\end{center}
\caption {For large $N$ in QS MHD turbulence, a schematic illustration of the energy transfers indicated by arrow.  The strong dissipation takes place in the shaded region.  ${\bf U}_\perp$ exhibits an inverse cascade for $k < k_f$, while $U_\parallel$ exhibits a forward cascade for $k>k_f$; here $k_f$ is the forcing wavenumber.  From Reddy {\em et al}~\cite{Reddy:PP2014}. Adopted with permission from AIP Publishing. }
\label{fig:mech}
\end{figure}

In Fig.~\ref{fig:mech} we summarise the energy transfers in QS MHD turbulence.  For $k<k_f$, the perpendicular component $U_\perp$ follows an inverse cascade due to the quasi two-dimensional nature of the flow, while for $k > k_f$, the parallel component $U_\parallel$ cascades forward, that is from small $k$ to large $k$.  Both $U_\perp$ and $U_\parallel$ transfer energy from sectors near the equator to the ones with lower $\theta$.  The energy is primarily dissipated by Joule heating near the equator.  There is an energy transfer from $U_\perp$ to $U_\parallel$ via pressure (not shown in the figure).  

In the next section, we describe several models of QS MHD turbulence that capture the steepening of $E(k)$ quite well.

\section{Modelling QS MHD turbulence}  \label{sec:model}

In Sec.~\ref{sec:analytic} we described some of the earlier models of QS MHD turbulence.   In this section we will review these models in the light of new findings.  Before such discussions, we describe in detail the models of Verma and Reddy~\cite{Verma:PF2015b} that successfully describes the energy spectrum and flux of QS MHD turbulence for small $N$ and very large $N$.

\subsection{Modelling QS MHD turbulence using variable energy flux}
\label{subsec:VR_model}

In this subsection, we describe the turbulence model of QS MHD turbulence constructed by Verma and Reddy~\cite{Verma:PF2015b}.  These models exploit the fact that the Joule dissipation depletes the energy flux and explain the energy flux and spectrum observed in numerical simulations for small $N$ and very large $N$.

As discussed earlier, the  QS MHD turbulence is anisotropic, hence the energy spectrum   $E({\bf k})$ is function of $k$ and $\theta$.  However we make a simplification that
 \begin{equation}
E(k,\theta) = E(k) \frac{g(\theta)}{\pi},
\label{eq:Ek5}
\end{equation}
where $E(k)$ is the one-dimensional energy spectrum, and  $g(\theta)$ is the angular dependence of the energy spectrum.  Integration of Equation~(\ref{eq:Ek5}) over $\theta$ yields
\begin{equation}
\int_0^{\pi}  d\theta E(k,\theta) = E(k) \int_0^{\pi}  \frac{g(\theta)}{\pi} d\theta = E(k).
\end{equation}
Therefore, 
\begin{equation}
  \int_0^{\pi}  \frac{g(\theta)}{\pi} d\theta = 1.
\end{equation}
 For the isotropic case, $g(\theta) = \mathrm{const}=1$.

The inertial-range energy flux $\Pi(k)$ decreases  with the increase of $k$ due to the Joule and viscous dissipation.  Quantitatively, the difference between the energy fluxes $\Pi(k+dk)$ and $\Pi(k)$ is due to the energy dissipation in the shell $(k,k+dk)$, i.e., 
 \bea
 \Pi(k+dk) - \Pi(k) & = &  - \epsilon(k) dk  \nonumber \\
 & = &  -\left\{ \int_0^\pi d\theta \left[ 2 \nu  k^2 + 2 \frac{\sigma B_0^2}{\rho} \cos^2 \theta \right] E(k,\theta) \right\} dk,
\eea
or
 \begin{equation}
 \frac{d\Pi(k)}{dk} = -\left[ 2c_1  \nu k^2 +2c_2 \frac{\sigma B_0^2}{\rho}  \right] E(k),
 \label{eq:dPidk}
\end{equation}
with 
 \begin{eqnarray}
 c_1 &  = & \frac{1}{\pi} \int_0^\pi g(\theta) d \theta = 1 \\ \label{eq:c1}
 c_2 &  = &  \frac{1}{\pi} \int_0^\pi g(\theta) \cos^2 \theta d \theta.  \label{eq:c2}
 \end{eqnarray}
Based on Eq.~(\ref{eq:dPidk}) Verma and Reddy~\cite{Verma:PF2015b} constructed two  models for QS MHD turbulence: model $A$ for small $N$'s for which the energy spectrum is still a power law but steeper than Kolomogorov's $k^{-5/3}$ spectrum; and model $B$ for large $N$'s for which the energy spectrum is exponential.   

\subsubsection{Model $A$ for small interaction parameters}
\label{sec:modelA}
As discussed in Sec.~\ref{subsec:Ek}, for small and moderate interaction parameters, the energy spectrum is a power law but with spectral indices lower than $-5/3$ (see Table~\ref{tab:energy}).   Verma and Reddy~\cite{Verma:PF2015b} employed a modified form of Pope's shell spectrum~\cite{Pope:book} to model the energy spectrum for small and moderate $N$.  In particular,
\begin{equation}
E(k,\theta) = E(k)  \frac{g(\theta)}{\pi} = K_ {Ko} [\Pi(k)]^{2/3} k^{-5/3}  f_L(k L) f_\eta(k \eta) \frac{g(\theta)}{\pi},
\label{eq:Ek_smallN}
\end{equation}
where $K_ {Ko}$ is the Kolmogorov constant with an approximate value of 1.5.  Since $N$ is small, the flow is nearly isotropic and $g(\theta) \approx 1$, hence $c_2 \approx 1/2$.  The functions $f_L(kL)$ and $f_\eta(k \eta)$ specify the  large-scale and dissipative-scale components, respectively, of the energy spectrum:
\begin{eqnarray}
f_L(kL)  & = & \left( \frac{kL}{[(kL)^2 + c_L]^{1/2}} \right)^{(5/3)+p_0}, \\
\label{eq:fL}
 f_\eta(k\eta)  & = & \exp \left[ -\beta \left\{ [ (k\eta)^4 + c_\eta^4 ]^{1/4}   - c_\eta \right\} \right].
\label{eq:feta}
\end{eqnarray}
Here  $c_L, c_\eta, p_0$ and $\beta$ are constants used by Pope~\cite{Pope:book}: $C_L \approx 6.78$, $c_\eta \approx 0.40$, $\beta \approx 5.2$, and $p_0 =2$.   Since the focus of the review is on  the inertial and dissipative range, we choose  $f_L(kL) = 1$.  

It is important to contrast Eq.~(\ref{eq:Ek_smallN}) with Pope's original formula.  In Eq.~(\ref{eq:Ek_smallN}), $\Pi(k)$ is $k$-dependent in contrast to a constant $\epsilon$ in Pope's formula.  By making the flux variable, we can model the behaviour of QS MHD quite accurately.   We substitute the energy spectrum of the form Eq.~(\ref{eq:Ek_smallN}) into  Eq.~(\ref{eq:dPidk}), which yields
 \begin{equation}
 \frac{d\Pi(k)}{dk} = -\left[ 2c_1  \nu k^2 +2c_2 \frac{\sigma B_0^2}{\rho}  \right] K_ {Ko} (\Pi(k))^{2/3} k^{-5/3} f_\eta(k \eta).
 \label{eq:dPidk_smallN}
\end{equation}
We integrate the above equation from $k=k_1$,  the starting wavenumber of the inertial range, and assume that $\Pi(k_1) = \Pi_0$.  With this, the solution of Eq.~(\ref{eq:dPidk_smallN}) is
\begin{eqnarray}
\left[ \frac{\Pi(k)}{\Pi_0} \right]^{1/3} & = & 1- \frac{2 K_ {Ko} c_1}{3} 
			\left( \frac{\nu^3}{\Pi_0 \eta^4} \right)^{1/3} I_1(k \eta)
			-\frac{2 c_2 K_ {Ko} \sigma B_0^2}{3\rho} \frac{\eta^{2/3}}{\Pi_0^{1/3}} I_2(k\eta) 
			\nonumber \\
		& = & 1-\frac{2c_1 c_3 K_ {Ko} }{3}   I_1(k \eta) - \frac{2}{3}\frac{c_2 K_ {Ko} N}{\sqrt{c_3 Re}}   I_2(k\eta),
		\label{eq:flux_exact}
\end{eqnarray}
 where $\eta$ is the Kolmogorov length,  the dimensionless constant $c_3 = (\nu^3/\Pi_0 \eta^4)^{1/3}$, and the integrals $I_1$ and $I_2$ are, respectively,
\begin{eqnarray}
I_1(k \eta)  & =  & \int_{k_1 \eta}^{k \eta} dk' k'^{1/3} f_\eta(k' ) \\
I_2(k \eta)  & = & \int_{k_1 \eta}^{k \eta}  dk' k'^{-5/3} f_\eta(k').
\end{eqnarray}
Verma and Reddy~\cite{Verma:PF2015b} choose $c_3 = 3.1$ in order to achieve $\Pi(k) \rightarrow 0$ for $k \eta \gg 1$ when $N=0$ (the isotropic case).  Given $\Pi(k)$ of Eq.~(\ref{eq:flux_exact}), the computation of the energy spectrum is straight forward:
\begin{equation}
E(k) =  \left\{ \begin{array}{lll}
K_ {Ko} \Pi_0^{2/3} k^{-5/3} f_\eta(k \eta) \left[ \frac{\Pi(k)}{\Pi_0} \right]^{2/3}, & \mathrm{if}~~k>k_1,\\
K_ {Ko} \Pi_0^{2/3} k^{-5/3}  f_L(k L)  & \mathrm{otherwise.} 
\end{array}  \right.
\label{eq:spectrum_exact}
\end{equation}

\begin{table}[htbp]
\begin{center}
\caption{\label{tab:table}Table depicting various parameters used: the grid size, non-dimensional magnetic field $B'_0$,  the interaction parameter $N$ calculated at the steady state, the  energy spectrum, and $c_2$ of Eq.~(\ref{eq:c2}).} 
\vspace{0.5cm}
\begin{tabular}{|c|c|c|c|c|c|c|}
\hline
Grid 	& $B'_0$		& $N$	 & scaling law	 &  $c_2$\\ 
\hline
\hline
$512^3$	& 0		& 0		& $k^{-5/3}$   & 0.35\\
$512^3$	& 0.739		& 0.10	& $k^{-1.8}$	  & 0.34\\
$512^3$	& 1.65  		& 0.64	& $k^{-2.0}$	  & 0.34\\
$512^3$	& 2.34    	& 1.6	& $k^{-2.8}$	   		& 0.23\\ 
$256^3$	& 25.1		& 130	& $\mathrm{exp}(-0.18k)$	   & $1.4\times 10^{-4}$\\
$256^3$	& 32.6		& 220	& $\mathrm{exp}(-0.18k)$	  & $1.3\times 10^{-4}$ \\
\hline
\end{tabular}
\label{eq:table_model}
\end{center}
\end{table}

\begin{figure}[!h]
\begin{center}
\includegraphics[width=10cm]{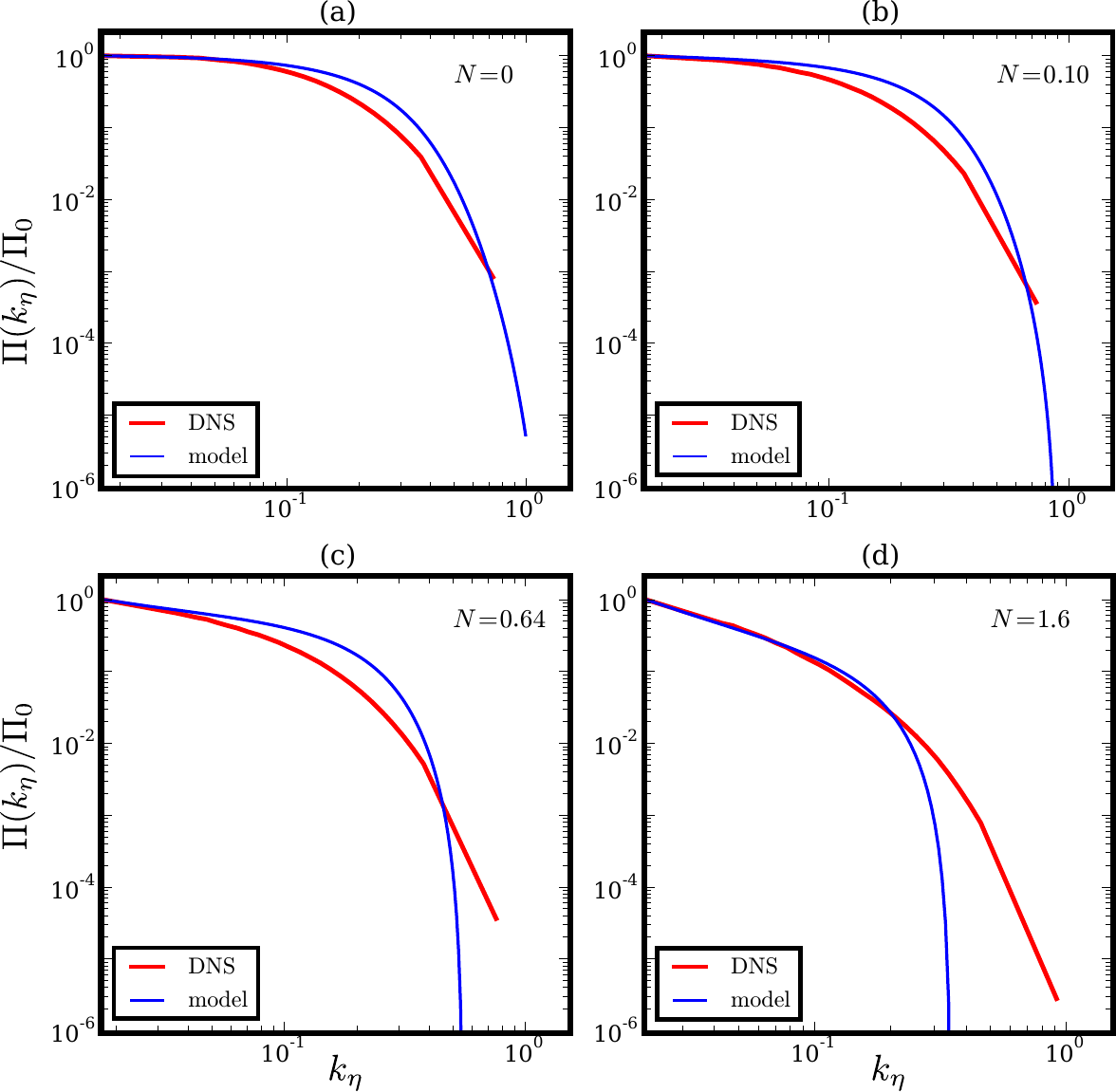}
\caption{ Plots of the normalized energy flux $\Pi(k\eta)/\Pi_0$ for (a) $N=0$, (b) $N=0.10$, (c) $N=0.64$ and (d) $N=1.6$. The energy flux decreases with $k$ due to Joule dissipation. From Verma and Reddy~\cite{Verma:PF2015b}. Reprinted with permission from AIP Publishing.  }
\label{fig:flux}
\end{center}
\end{figure}

\begin{figure}[!h]
\begin{center}
\includegraphics[width=10cm]{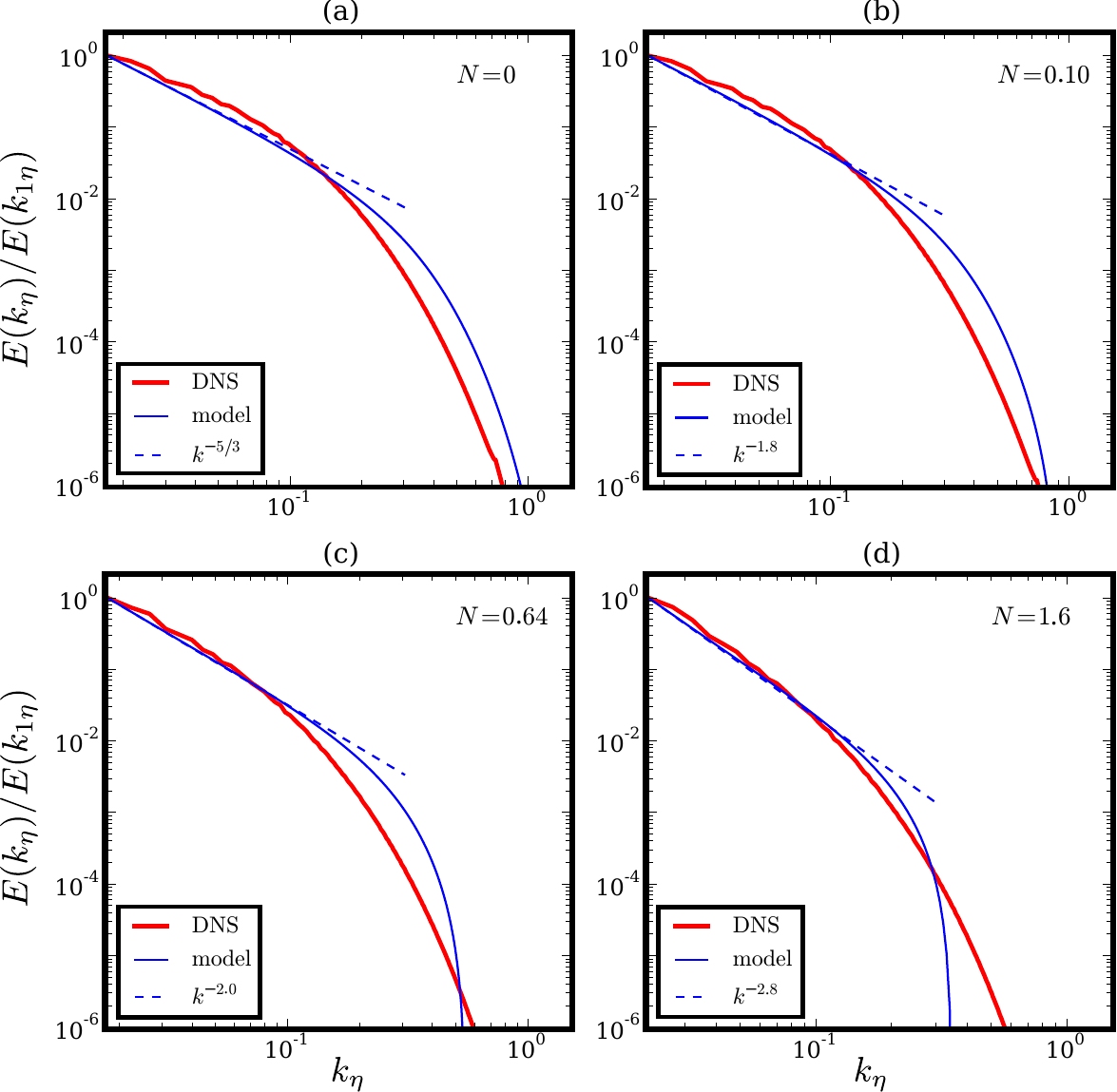}
\caption{ Plots of the normalized energy spectra $E(k\eta)$ for (a) $N=0$, (b) $N=0.10$, (c) $N=0.64$ and (d) $N=1.6$.  The dashed lines are the best fit curves.  Adopted with permission from Reddy~\cite{Reddy:thesis}.}
\label{fig:spectrum}
\end{center}
\end{figure}

Verma and Reddy~\cite{Verma:PF2015b}  then compared the aforementioned model predictions with the numerical  results discussed in Sec.~\ref{subsec:Ek}.  The summary of their parameter values are listed in Table~\ref{eq:table_model}.   In Figs.~\ref{fig:flux} and \ref{fig:spectrum}, we plot the numerically computed energy fluxes and the spectra   for $N=0, 0.10,0.64$ and 1.6.  In the same figure we also plot the model predictions. The figures show that the model describes the numerical data quite well, and that both the energy flux and energy spectrum steepen with the increase of $N$.  The spectral indices for various $N$'s are listed in Table~\ref{eq:table_model}.  In the Table we also list $c_2$ for various $N$'s and find them to be close to $1/2$ for small $N$ (0---1.6). Thus $g(\theta) \approx 1$, or that the flow is nearly isotropic is  a good approximation till $N \approx 1.6$.

 We remark that model $A$ works well only for small  $N$'s ($N \lessapprox 1$) for which Kolmogorov's spectrum is a good starting point.    This assumption breaks down for large $N$ since the flow becomes quasi 2D.  We also remark that Pao's model for fluid turbulence~\cite{Pao:PF1965} could also be used in Eq.~(\ref{eq:Ek_smallN}) as an alternative to the Pope's model.   In the following subsection we construct another simple model that can explain the energy spectrum for very large $N$.
 
\subsubsection{Model $B$ for very large interaction parameters}\label{subsec:modelB}
\label{subsec:modelB}

\begin{figure}[ht]
\begin{center}
\includegraphics[width=10.0cm]{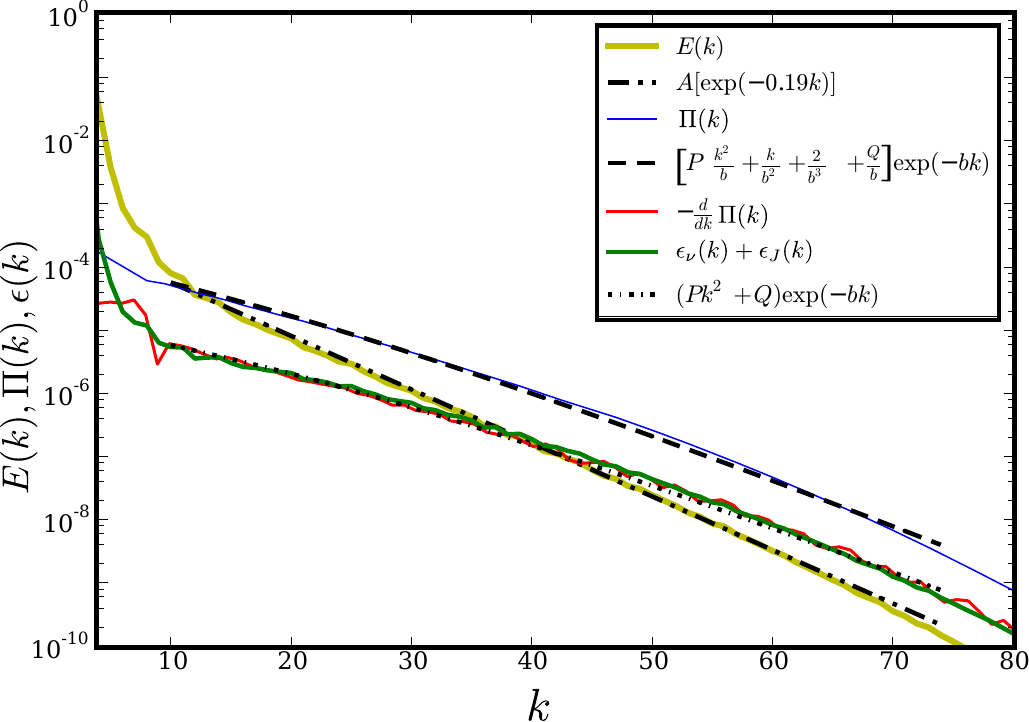}
\caption{For $N= 220$, plot of the kinetic energy spectrum $E(k)$, flux $\Pi(k)$, total dissipation $\epsilon(k) = \epsilon_J(k) + \epsilon_\nu(k)$, and $-\frac{d}{dk}\Pi(k)$. Note that $-\frac{d}{dk}\Pi(k) \approx \epsilon(k) $. The black double dot-single dash,  dashed and dash-dot lines are the best fit curves for $E(k)$, $\Pi(k)$ and $\epsilon(k)$, respectively.  Adopted with permission from Reddy~\cite{Reddy:thesis}.}
\label{fig:N220}
\end{center}
\end{figure}

The Joule dissipation is  strong for very large $N$, and it causes a rapid decrease of  the energy flux with $k$, resulting in an exponential behaviour for the energy spectrum and energy flux (see Sec.~\ref{subsec:Ek}).  This behaviour is similar to the dissipative fluid flows. Therefore, for $N \gg 1$, Verma and Reddy~\cite{Verma:PF2015b} postulated that the energy spectrum follows~\cite{Verma:PF2015b}
\begin{eqnarray}
E(k) & = &  A \exp(-b k),  \label{eq:Ek_largeN} 
\end{eqnarray}
where $A$ and $b$ are constants. Using 
\be 
\epsilon(k)=   - \frac{d\Pi(k)}{dk} =  (P k^2 + Q) \exp(-b k), \label{eq:epsilonk_largeN}
\ee
 where $P$ and $Q$ are constants, we compute $\Pi(k)$ by integration:
\begin{equation}
\Pi(k)  =  \left\{ P \left(\frac{k^2}{b}+\frac{2k}{b^2}+\frac{2}{b^3}\right) + \frac{Q}{b} \right\} \exp(-b k).
\label{eq:Pik_largeN}
\end{equation}
A comparison of Eq.~(\ref{eq:epsilonk_largeN}) with Equation~(\ref{eq:dPidk}) yields
\begin{eqnarray}
P & = & 2Ac_1 \nu,  \label{eq:largeN_P}\\
Q  & =  &2Ac_2 \frac{\sigma B_0^2}{\rho}. 
\label{eq:largeN_Q}
\end{eqnarray}
Thus,  the exponential energy spectrum and flux are consistent solutions of the variable flux equation [Eq.~(\ref{eq:dPidk})], as shown in Fig.~\ref{fig:N220}.

It is important to note that the two models discussed above predict the energy spectra and fluxes for $k>k_f$ where the energy cascades in the forward direction. The regime $k <k_f$ would be affected by inverse cascade of energy, and it needs to be worked out.  Unfortunately the aforementioned models $A$ and $B$ cannot be used for intermediate $N$ say $N \sim 5$.  A  general and comprehensive model needs to be constructed for moderate $N$'s.  

\subsection{Review of the existing models of QS MHD turbulence}
\label{subset:critique}
Researchers have constructed models to understand the dynamics of QS MHD turbulence. One of the critical puzzle in the field has been the steepening of the energy spectrum. It has been postulated that for large $N$, the QS MHD turbulence  has behaviour similar to 2D hydrodynamic turbulence, hence its spectrum is expected to be close to $k^{-3}$ rather than $k^{-5/3}$~\cite{Hossain:PFB1991,Kit:MG1971,Kolesnikov:FD1974}.  Experiments~\cite{Eckert:IJHFF2001} and numerical simulations~\cite{Burattini:PD2008,Reddy:PF2014,Zikanov:JFM1998} however reveal that the spectral index changes monotonically with $N$.  Hence the hypothesis that $E(k) \sim k^{-3}$ is ruled out.  Reddy and Verma~\cite{Reddy:PF2014}, and Verma and Reddy~\cite{Verma:PF2015b} showed that the steepening of the energy spectrum is due to the decrease in energy flux with $k$, which occurs because of the Joule dissipation.  Note that the Joule dissipation is active at all scales, unlike viscous dissipation that acts primarily at small scales.  We also remark that QS MHD turbulence with  very large $N$ has exponential spectrum ($E(k) \sim \exp(-ak)$) due to extreme Joule dissipation.

Some of the early models~\cite{Moffatt:JFM1967,Schumann:JFM1976} of QS MHD turbulence focus on the decay laws of QS MHD.  Moffatt~\cite{Moffatt:JFM1967}  started with a linear equation for the decay of kinetic energy, and then derived a decay law for energy as $E(t) \sim t^{-1/2}$ [see Eq.~(\ref{eq:Ek_decaying}, \ref{eq:E_decaying}) of Sec.~\ref{sec:analytic}].  Present set of calculations show that the nonlinear term, possibly very weak, is active at all time.  Note however that the nonlinear term dominates other terms in the $k_z=0$ plane of the Fourier space.  In Eq.~(\ref{eq:dEk_dt}), the term $T(k)$ yields a nonzero forward flux in the dissipation range.  Thus the variable energy flux is present at all $k$, see for example Model $B$ discussed in Sec.~\ref{subsec:modelB}.     These features invalidate  Eq.~(\ref{eq:Ek_decaying}) as a starting point for modelling the decay law for QS MHD turbulence.  

In Sec.~\ref{sec:analytic} we discussed how Moffatt~\cite{Moffatt:JFM1967} derived $E_\parallel = E_\perp= E/2$ for the asymptotic state ($t \rightarrow \infty$). This prediction is contrary to the the steady-state flow profile shown in Figs.~\ref{fig:isosurface_vorticity} and \ref{fig:N132_vectors} for which $E_\perp \gg E_\parallel$. The discrepancy  is due to the assumptions of initial isotropic energy spectrum [Eq.~(\ref{eq:u_tensor})] and the linear decaying equation for the energy spectrum [Eq.~(\ref{eq:Ek_decaying})].  We believe that careful simulations and modelling are required for deriving a definitive decay law for QS MHD turbulence.  
 
Favier {\em et al}~\cite{Favier:PF2010,Favier:JFM2011} and  Reddy and Verma~\cite{Reddy:PF2014} studied the anisotropy in QS MHD turbulence by dividing the spectral space into rings.  Favier {\em et al}~\cite{Favier:PF2010,Favier:JFM2011} presented the poloidal and toriodal components of the spectrum at different angles, while Reddy and Verma presented ring spectrum.  Both the groups showed that QS MHD turbulence is quasi two-dimensional, however their equations for $U_\perp$ and $U_\parallel$  differ in a critical manner.  The equation for $U_\parallel$ by Favier {\em et al}~\cite{Favier:PF2010}, Eq.~(\ref{eq:Favier_Uparallel}), does not contain $-\nabla P$ term.  Reddy {\em et al}~\cite{Reddy:PP2014}  showed that $-\nabla P$ facilitate the transfer of energy from ${\bf U}_\perp$ to $U_\parallel$; without $-\nabla P$ term, the total energy of the parallel component, $E_\parallel = (1/2)\int |U_\parallel|^2 d\tau$, will vanish when $\nu \ne 0$ and the flow will become two-dimensional, not quasi two-dimensional.   This is a crucial factor that is expected to play a major role in other anisotropic flows, such as rotating, convective, and MHD turbulence.  Refer to Sec.~\ref{sec:perp_parallel} for details on the energy transfers from ${\bf U}_\perp$ to $U_\parallel$ via pressure.

In summary, there is a convergence in the community that the QS MHD turbulence is quasi two-dimensional, and the energy spectrum is steeper than the hydrodynamic turbulence due to Joule dissipation.  Recent turbulence models are able to explain these phenomena.

\section{Bounded QS MHD flows}   \label{sec:wall}

in this review we focus on the bulk flow of QS MHD turbulence.  In the present section we provide a brief overview of QS MHD flows in bounded geometries---channel and closed box.  Such flows are common in industrial applications, as well as in planetary interiors.  For detailed discussion, the reader is referred to Davison~\cite{Davidson:book:MHD}, Moreau~\cite{Moreau:book:MHD},  M\"{u}ller and B\"{u}her~\cite{Muller:book},  and Zikanov {\em et al}~\cite{Zikanov:AMR2014}.

Bounded flows have boundary layers near the walls.  The boundary layer of QS MHD is called {\em Hartmann layer}. The bounded flow of QS MHD differs from hydrodynamic flows in several aspects.  The Amp\'{e}re force (${\bf J \times B}$)  in the bulk produces additional drag force in QS MHD.  Also, the conducting walls support electric current, which is not the case for hydrodynamic flows; such currents tend to accelerate the flow near the wall leading to jets.  In the following discussion, we only touch upon some of these features.

\subsection{QS MHD over a plate and in a channel: linear limit}  \label{subset:channel}

\begin{figure}[H]
\begin{center}
\includegraphics[scale=0.8]{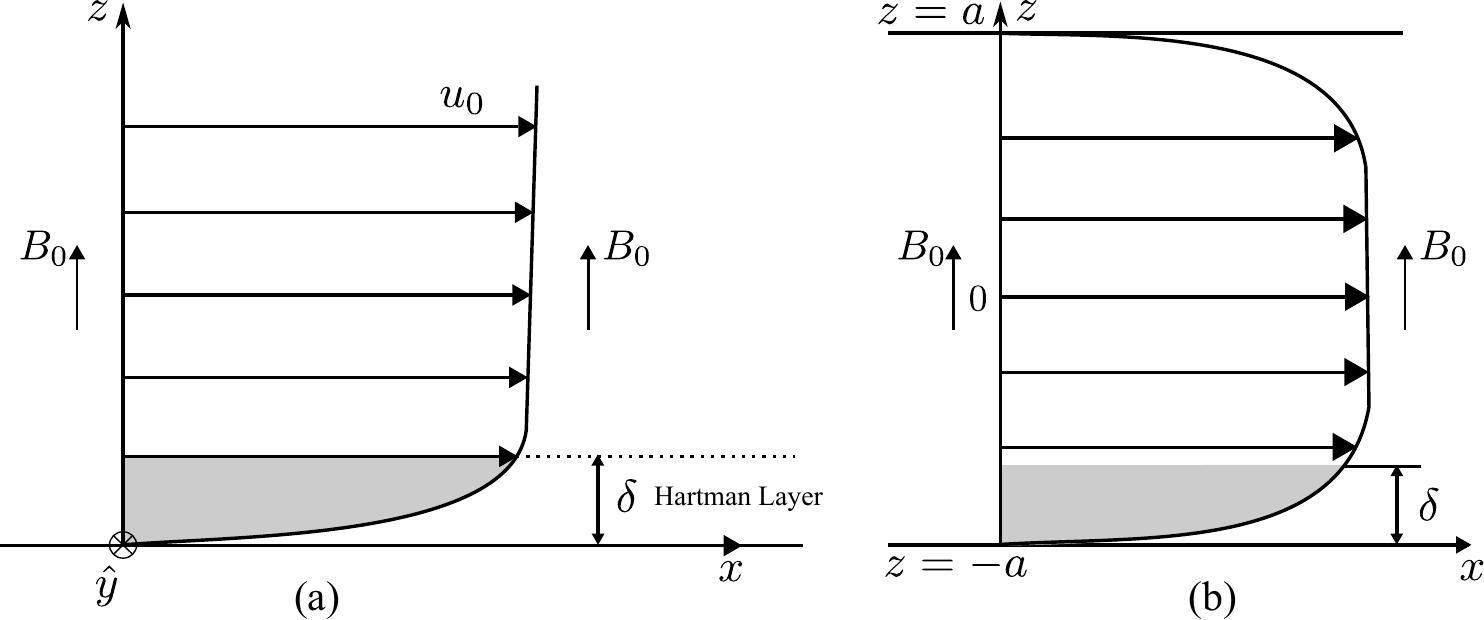}
\end{center}
\caption{(a) QS MHD flow over a flat plate.  (b) QS MHD flow in a channel. }  
\label{fig:plate_channel}
\end{figure}

First, we study the flow over a flat plat as shown in Fig.~\ref{fig:plate_channel}(a).  We assume that ${\bf u} = u_0 \hat{x}$ as $z \rightarrow \infty$, and ${\bf u} = 0$ (no-slip boundary condition) at $z=0$.  We also assume  presence of an external magnetic field ${\bf B} = B_0 \hat{z}$, absence of an external electric field, and that the fluid is forced by a constant pressure gradient $-\partial p/\partial x$.  Hence the electric current density is
\be
{\bf J} = \sigma {\bf u \times B}_0 = -\sigma u B_0 \hat{y},
\ee
where $\sigma$ is the electric conductivity of the fluid.   Therefore  the force density is
\be
{\bf f}_L = {\bf J \times B}_0 = -\sigma u B_0^2 \hat{x}.
\ee
For simplicity, we assume that the nonlinear term is negligible. Under steady state, the force balance yields
\be
\rho \nu \frac{\partial^2 u}{\partial z^2}  - \sigma B_0^2 u = \frac{\partial p}{\partial x},
\label{eq:QS_wall}
\ee
where $\rho,\nu$ are respectively the  density and kinematic viscosity of the fluid.   For a constant $\partial p/\partial x$, the above equation has the following particular solution
\be 
u_\mathrm{particular} = u_0 = \frac{1}{\sigma B_0^2} \left(-\frac{\partial p}{\partial x} \right),
\label{eq:particular}
\ee
and the following homogeneous solution
\be 
u_\mathrm{homog} =  A \exp(-z/\delta) + C \exp(z/\delta),
\ee
where $A,B$ are constants, and $\delta = [\rho \nu/(\sigma B_0^2)]^{1/2}$.  The solution $u = u_\mathrm{homog} + u_\mathrm{particular}$ with boundary condition $u=0$ at $z=0$, and   finite $u$ as $z \rightarrow \infty$ yields $C=0$. Hence the solution is
\be
u = u_0[1-\exp(-z/\delta)].
\ee
It is easy to note that the Lorentz force induces additional suppression in the flow.

In Fig.~\ref{fig:plate_channel}, the region from $z=0$ to $z=\delta$ where the velocity increase exponentially from 0 to  $\approx u_0$ is called the {\em Hartmann layer}, and the quantity $\delta$ is the {\em width of the Hartmann layer}.  The Hartmann number is defined as the square root of the ratio of the Lorentz force and the viscous force:
\be
\mathrm{Ha} = \sqrt{\frac{\sigma B_0^2 u/\rho}{\nu u/a^2}} =B_0 a  \sqrt{ \frac{\sigma}{\rho \nu }}.
\ee
Clearly
\be 
\frac{a}{\delta} = \mathrm{Ha}.
\label{eq:delta_Ha}
\ee
Note that
\be
N \times \mathrm{Re} = \mathrm{Ha}^2.
\ee
Using Eq.~(\ref{eq:delta_Ha}) we deduce that 
\be
\delta = \frac{a}{\sqrt{\mathrm{Re}}} \frac{1}{\sqrt{N}}  \approx \frac{\delta_\nu}{\sqrt{N}},
\label{eq:delta_Ha2}
\ee
where $\delta_\nu$ is the thickness of the viscous boundary layer.  Thus, the Hartmann layer is thinner than the viscous boundary layer by a factor of $\sqrt{N}$.

A related problem is the flow in a channel shown in Fig.~\ref{fig:plate_channel}(b).  The walls are located at $z=\pm a$ at which we employ no-slip boundary condition (${\bf u} = 0$).  The equation of the flow under the linear approximation is same as Eq.~(\ref{eq:QS_wall}), whose homogeneous solution is
\be 
u_\mathrm{homog} =  A \cosh(z/\delta),
\ee
and the particular solution is same as that of Eq.~(\ref{eq:particular}).  The odd solution $\sinh(z/\delta)$ is absent due to the even symmetry about $z=0$.
Therefore, for the no-slip condition at $z=\pm a$, the solution for the velocity is ${\bf u} = u \hat{x}$ with
\be
u = u_0\left[1-\frac{\cosh(z/\delta)}{\cosh(a/\delta)} \right].
\ee
The above solution yields $u = u_0[1-1/\cosh(a/\delta)] \approx u_0$ at $z=0$.  

The above equations are linear hence they could be solved analytically.  The full QS MHD equations with nonlinearity are more difficult, and they are typically solved using numerical simulations.  We describe thees issues briefly in the next subsection.

\subsection{QS MHD turbulence in a channel}  \label{subset:channel}
For large $\mathrm{Re}$ and $N$, Eq.~(\ref{eq:delta_Ha2}) shows that the Hartmann layer is very thin, whose resolution in numerical simulations  is one of the most difficult issues of QS MHD turbulence research.    Boeck {\em et al}~\cite{Boeck:PRL2008} and Krasnov {\em et al}~\cite{Krasnov:PF2008} performed spectral simulations using Fourier basis for the periodic direction, and Chebyshev polynomials for the wall direction(s).  Boeck {\em et al}~\cite{Boeck:PRL2008}  simulated low-Rm MHD flows in a channel with no-slip boundary conditions for $\mathrm{Re}=8000$ and $\mathrm{Ha}=80$ and observed recurring transitions between two-dimensional and three-dimensional  states in the flow.  Dymkou and Poth\'{e}rat~\cite{Dymkou:TCFD2009}  and Kornet and Poth\'{e}rat~\cite{Kornet:JCP2015} invented a new spectral scheme based on basis functions suitable for QS channel flow; their method is quite efficient since  their basis functions contain the Hartmann layer.   Finite difference, finite volume, and finite element methods are also used to solve flows with walls~\cite{Vantieghem:TCFD2009}. Some of these methods are discussed in Sec.~\ref{sec:numerical}.

A natural question is  the role of nonlinearity in QS MHD turbulence.  The QS MHD flows above a flat plate and in a channel discussed in Sec.~\ref{subset:channel} are quite stable, primarily due to the fact that the Couette and channel flows are quite stable~\cite{Drazin:book:new}.  In Sec.~\ref{sec:experiment} we described how perturbed (e.g., grid-generated turbulence in experiments) QS MHD turbulence exhibit powerlaw and exponential energy spectra.  However, it will be interesting to compare  the properties of turbulence in the bulk part of a channel (e.g,  as in Boeck {\em et al}~\cite{Boeck:PRL2008}) with the spectral results discussed in the present paper.

A Generic feature of the aforementioned flows is that the Lorentz force provides additional drag (in addition to the viscous drag), and hence the flow is slower than its hydrodynamic counterpart. This is due to the induced electric currents in the flow.  Additional side walls create further complexity due to the wall currents;  for such flows, the induced electric currents accelerate the flow near the wall  leading to strong near-wall jets.  For details, refer to Moreau~\cite{Moreau:book:MHD} and M\"{u}ller and B\"{u}hler~\cite{Muller:book}.

\subsection{QS MHD in a box}  \label{subset:box}
Sommeria~\cite{Sommeria:JFM1986}, Herault {\em et al}~\cite{Herault:EPL2015}, Klein and Proth\'{e}rat~\cite{Klein:PRL2010}, and Proth\'{e}rat and Klein~\cite{Potherat:JFM2014} performed experiments on liquid metals in a closed box (see  Fig.~\ref{fig:schmatic_real_space}), and analysed the flow in great detail.  They reported quasi two-dimensionalization of the flow for strong $B_0$.  Klein and Poth\'{e}rat~\cite{Klein:PRL2010} and  Poth\'{e}rat and Klein~\cite{Potherat:JFM2014} also discussed how the flow becomes three-dimensional  due to inertia and induced currents.  In this review we do not cover these topics in detail, and only discuss some of the patterns observed in two-dimensional QS MHD flows.

Sommeria~\cite{Sommeria:JFM1986} and Sommeria and Moreau~\cite{Sommeria:JFM1982} studied a forced liquid-metal flow in which magnets with alternating polarities located at the bottom of the box drive the flow.  They employed the following equation of motion for the flow:
\be 
\frac{\partial{\bf u}_\perp}{\partial t} + ({\bf u}_\perp\cdot\nabla){\bf u}_\perp = -\nabla_\perp{(p/\rho)} -\frac{ {\bf u}_\perp}{\tau_H}  + \nu\nabla^2 {\bf u}_\perp , \label{eq:NS0}
\label{eq:Sommeria}
\ee
where Hartmann damping time $\tau_H = \delta/u_\mathrm{small} =(a/B_0) (\rho a/\sigma \nu)^{1/2} $ with $u_\mathrm{small}  = \nu/a$.  The Lorentz force in Eq.~(\ref{eq:Sommeria}) is suppressed by a factor $\delta/a = \mathrm{Ha}^{-1}$ compared to that in a periodic box [see Eq.~(\ref{eq:k_NS})].  A nondimensional version of the above equation is
\be 
\frac{\partial{\bf U}_\perp}{\partial t} + ({\bf U}_\perp\cdot\nabla){\bf U}_\perp = -\nabla_\perp{P} -\frac{ {\bf U}_\perp}{\mathrm{Rh}}  + \frac{1}{\mathrm{Re}} \nabla^2 {\bf U}_\perp , \label{eq:NS0}
\label{eq:Sommeria2}
\ee
where $\mathrm{Re} = U_0L/\nu$, $\mathrm{Rh} = \tau_H L/U_0$,  and ${\bf U}_\perp,P$ are nondimensional variables.  

Sommeria~\cite{Sommeria:JFM1986} and Heralt {\em et al}~\cite{Herault:EPL2015} performed experiments on mercury contained in box of size $12 \mathrm{cm} \times 12 \mathrm{cm} \times 2 \mathrm{cm}$.  The fluid was forced using 36 electromagnets of alternating magnetic polarities.  For a set of parameters, Sommeria observed $k^{-5/3}$ energy spectrum corresponding to two-dimensional hydrodynamic turbulence (the inverse cascade regime as predicted by Kraichnan~\cite{Kraichnan:PF1967b}); he also reported large-scale structures for another set of parameters.

Mishra {\em et al}~\cite{Mishra:PRE2015}   simulated the  setup  of Sommeria~\cite{Sommeria:JFM1986} and Heralt {\em et al}~\cite{Herault:EPL2015} for a set of $\mathrm{Re}$  and $\mathrm{Rh}$.  For a given $\mathrm{Re}$, with the increase of $\mathrm{Rh}$, Mishra {\em et al} observed the following set of bifurcations: stable $6\times 6$ vortex structures, temporal and spatial chaos, flow reversals, and large-scale circulation at the box size (the condensate state).  Note that large $\mathrm{Rh}$ corresponds lower $B_0$ or lower Lorentz force.  Thus, Mishra {\em et al}~\cite{Mishra:PRE2015}   show that  large $B_0$ yields same patterns as the forcing configuration ($6\times 6$ vortex structures), but lower $B_0$ leads to inverse cascade of energy and consequent coalescence of flow structures, similar to those observed in 2D hydrodynamic turbulence.

 Here we close our discussion on QS MHD in a bounded box.  In the next section we summarise the present status of the field.

\section{Conclusions}\label{chap6}

In this review we describe the main results of  QS MHD turbulence obtained using experimental, numerical, and modelling.  A summary of the results presented in the review is as follows:
\begin{enumerate}
\item The imposed external magnetic field creates flow anisotropy  that  increases with the increase of the external magnetic field $B_0$ or the interaction parameter ($N$).   For moderate and large $N$, the QS MHD flow is quasi two-dimensional with strong $U_\perp$ and weak $U_\parallel$, where $U_\perp$ and $U_\parallel$ are the perpendicular and parallel components of the velocity field  relative to  the mean magnetic field ${\bf B}_0$. The energy spectrum $E(k)$ is steeper than Kolmogorov's spectrum ($k^{-5/3}$) with the spectral index decreasing with the increase of $N$.  For very large $N$, $E(k) \sim \exp(-bk)$, where $b$ is a constant.  The results from numerical simulations, experiments, and model of Sec.~\ref{sec:model} are in good agreement with each other.

\item In QS MHD turbulence, the energy flux $\Pi(k)$ decreases with $k$ due to the Joule dissipation.  The steepening of the energy spectrum in comparison to Kolmogorov's spectrum is due to this variable energy flux.

\item  For large $N$, the energy is concentrated near the equatorial region (near $k_z =0$ plane).  In this regime, $U_\perp$ dominates at small $k$, while $U_\parallel$ dominates at large $k$.  The pressure facilitates energy transfer from $U_\perp$ to $U_\parallel$.

\item The anisotropy in QS MHD turbulence is quantified using the ring spectrum and ring-to-ring energy transfers.  Studies reveal that energy flows from  $\theta \approx \pi/2$  (near the equatorial plane) to lower $\theta$.  Also, the energy flux of the perpendicular component, $\Pi_\perp$, is negative at small $k$ ($k < k_f$ where $k_f$ is the forcing wavenumber) indicating an inverse cascade of $U_\perp$.  However the energy flux  of parallel component, $\Pi_\parallel$, is positive for  $k>k_f$, thus $U_\parallel$ exhibits forward cascade.
\end{enumerate}

In this review we focus on  the turbulence phenomenology of the bulk flow in QS MHD turbulence.  There are however many experiments~\cite{Lielpeteris:book} and numerical simulations ~\cite{Boeck:PRL2008} that focus on the flow in a channel that includes a Hartman layer; researchers have not studied the spectrum and flux of such flows in detail. It will be interesting to compare $E(k)$ in a channel  with the theoretical results presented in this review. For example, it is reasonable to conjecture that in a channel flow involving QS MHD, $E(k)$  could be exponential as reported in Sections \ref{sec:numerical} and \ref{sec:model}.  Such studies may prove very useful for modelling channel flows, specially for the liquid metal blanket of the ITER (International Thermonuclear Experimental Reactor) project.

  The model of Sec.~\ref{sec:model} needs further refinements.  For example, the assumption that $E(k,\theta) = E(k) g(\theta)$ needs to be validated against numerical results. In addition, the model needs to be extended to intermediate $N$ (for example, $1 < N < 100$) where we observe steep power-law  and quasi 2D behaviour.  For such flows,  we need to start with $E(k) \sim k^{-3}$ that corresponds to constant enstrophy flux.  Also,  the model of  Sec.~\ref{sec:model} would be  useful for large-eddy simulations of QS MHD turbulence, as well as for constructing turbulence models for realistic QS MHD flows.

 Most flows in nature are anisotropic due to (a) external applied field, e.g. buoyancy, external magnetic field, or (b) inhomogeneous boundary conditions, e.g. in a channel with no-slip boundary conditions at the top and bottom plates. Researchers have developed tools to study anisotropy, for example, see Sagaut and Cambon~\cite{Sagaut:book}, Davidson~\cite{Davidson:book:TurbulenceRotating}, Shabalin~\cite{Shebalin:JPP1983} and references therein.  They have proposed poloidal and toroidal decomposition, ring decomposition similar to ours, mean angle $\theta_Q$ of Eq.~(\ref{eq:thetaQ}), etc.
The tools described in this review complement the tools proposed by these authors, and they would be useful for studying anisotropy in generic turbulent systems.  We also remark that such  quantification of turbulence will help us model the diffusion of particles in turbulent MHD turbulence as well~\cite{Lalescu:AUC2014,Negrea:AUC2004}.

 It is interesting to compare the anisotropy in QS MHD with other related systems.  MHD turbulence with strong external magnetic field, and strongly rotating turbulence tend to exhibit  quasi 2D behaviour~\cite{Davidson:book:MHD,Davidson:book:TurbulenceRotating,Oughton:JFM1994,Teaca:PRE2009,Sagaut:book} with $U_\perp \gg U_\parallel$.  This feature is similar to that of QS  MHD turbulence.   In Rayleigh-B\'{e}nard flow however $U_\perp < U_\parallel$, but $U_\perp $ and $ U_\parallel$ are comparable, hence  the flow is nearly isotropic~\cite{Nath:PRF2016}.  The above conclusions have been drawn using the tools discussed in this review.
 
{\color{blue} The external fields like magnetic field (in MHD and QS MHD) or buoyancy, as well as rotation, makes the flow anisotropic.  It has been observed that the external magnetic field and rotation makes the flow quasi 2D~\cite{Davidson:book:MHD,Davidson:book:TurbulenceRotating,Oughton:JFM1994,Sagaut:book,Sundar:PP2017,Teaca:PRE2009} with  $U_\perp \gg U_\parallel$, that is, the flow perpendicular to the anisotropic axis is stronger than its parallel counterpart. However in Rayleigh-B\'{e}nard convection, the behaviour is quite different.  The thermal plumes accelerate the flow along the buoyancy direction, which yields  $U_\parallel > U_\perp$. However, recently Nath {\em et al}~\cite{Nath:PRF2016} and Verma {\em et al}~\cite{Verma:NJP2017} showed that $U_\parallel$ and $U_\perp$ are comparable, hence the flow is much less anisotropic than QS MHD turbulence for large interaction parameters.  Thus the nature of anisotropy is different in these systems. Yet, the tools discussed in this paper have been applied to study  diverse anisotropic systems~\cite{Davidson:book:MHD,Davidson:book:TurbulenceRotating,Nath:PRF2016,Oughton:JFM1994,Sagaut:book,Sundar:PP2017,Teaca:PRE2009,Verma:NJP2017}.}

Turbulence remains an unsolved phenomena.  We hope that the tools described in this review will provide further insights into this phenomena, specially those related to anisotropy.

\section*{Acknowledgement}
Some of the material of the review is based on the thesis work of Sandeep Reddy, and collaborative work with Raghwendra Kumar.  I am grateful to both of them for the fruitful collaboration and exciting discussions. I thank the colleagues of our laboratory at IIT Kanpur specially Anando Chatterjee and Abhishek Kumar for ideas and help. I  am grateful to colleagues of ULB Brussels---Daniele Carati, Bernard Knaepen,  Xavier Albets, Paolo Burattini, and Maxime Kinet---for valuable ideas and suggestions when I was new to this field.  I also benefitted from the conversations with Thomas Boeck, J\"{o}rg Schumacher, Andre Thess,  Stephan Fauve, and V. Eswaran, and from the useful comments of the anonymous referees.    I thank Roshan Bhaskaran for help in making the figures.

Some results of the review are based on simulations  performed on the HPC system and Chaos cluster of IIT Kanpur, India. This work was supported by a research grants 2009/36/81-BRNS from Board of Research in Nuclear Science, Department of Atomic Energy, Government of India, and Indo-Russian project (DST-RSF) project INT/RUS/RSF/P-03 from Department of Science and Technology, India.

\appendix

\newpage
\appendix
\section[Appendix]{Energy transfers in anisotropic turbulence}  \label{sec:ET_appendix}
 
The turbulence in QS MHD turbulence is anisotropic due to the external magnetic field.  In this appendix we quantify the energy transfers between the perpendicular and parallel components of the velocity field.  To  derive the formulae for the energy transfers, we focus on a wavenumber triad $({\bf k',p,q})$  that satisfies ${\bf k'+p+q}= 0$.  For convenience we denote ${\bf k'} = -{\bf k}$.  

Following Dar {\it et al}~\cite{Dar:PD2001} and Verma~\cite{Verma:PR2004}, we derive the following equations  from Eqs.~(\ref{eq:k_NS}) and (\ref{eq:k_NS2}):
\begin{eqnarray}
{{\partial E_{\perp}({\bf k'})}\over {\partial t}} &=&   S_{\perp}({\bf k'}|{\bf p}|{\bf q}) + S_{\perp}({\bf k'}|{\bf q}|{\bf p}) +  \mathcal{P}_{\perp}({\bf k'}) -D_\perp({\bf k'}), \label{eq:dt_Eperp} \\
{{\partial E_{\parallel}({\bf k'})}\over {\partial t}} &=&   S_{\parallel}({\bf k'}|{\bf p}|{\bf q}) + S_{\parallel}({\bf k'}|{\bf q}|{\bf p}) +  \mathcal{P}_{\parallel}({\bf k'})  -D_\parallel({\bf k'}), \label{eq:dt_Epll} \
\end{eqnarray}
\noindent 
where $E_{\perp}({\bf k}) =E_{\perp}({\bf k'})  =\frac{1}{2}|\hat {\bf U}_{\perp}({\bf k})|^2$ and $E_{\parallel}({\bf k}) = E_{\parallel}({\bf k'})  = \frac{1}{2}|\hat { U}_{\parallel}({\bf k})|^2$ are respectively the energies of the perpendicular and parallel components of the velocity field, and
\begin{eqnarray}
S_{\perp}({\bf k'|p|q})&=& -\mathrm{\Im} \{ [ {\bf k' \cdot {\hat U}(q)] [\hat U_{\perp}(k') \cdot \hat U_{\perp}(p)}]\},\label{eq:Sperp}\\
S_{\parallel}({\bf k'|p|q})&=& -\mathrm{\Im} \{ [ {\bf k' \cdot {\hat U}({\bf q})}] [\hat U_{\parallel}({\bf k'})  \hat U_{\parallel}({\bf p})] \},\label{eq:Spar}\\
\mathcal{P}_{\perp}({\bf k'}) &=& \Im \{ [{\bf k \cdot \hat U^*_{\perp}(k)} ] \hat{P}({\bf k})\},\label{eq:defPperp_appendix} \\
\mathcal{P}_{\parallel}({\bf k'}) &=& \Im \{ [  k_{\parallel}  \hat U^*_{\parallel}({\bf k}) ] \hat{P}({\bf k})\}, \label{eq:defPpar_appendix}
\end{eqnarray}
where $\Re,\Im$, * represent the real part, imaginary part, and complex conjugate of a complex number respectively.    The terms $D_\perp({\bf k'})$ and $D_\parallel({\bf k'})$ are the total dissipation rates (viscous + Joule) of $E_{\perp}({\bf k'})$ and $E_{\parallel}({\bf k'})$ respectively.  Equations~(\ref{eq:dt_Eperp}) and (\ref{eq:dt_Epll}) indicate that the mode ${\bf k^\prime}$ receives energy from the modes {\bf p} and {\bf q}. Similarly, we can also derive 
\begin{eqnarray}
{{\partial E_{\perp}({\bf p})}\over {\partial t}} &=&   S_{\perp}({\bf p}|{\bf q}|{\bf k'}) + S_{\perp}({\bf p}|{\bf k'}|{\bf q}) +  \mathcal{P}_{\perp}({\bf p}) -D_\perp({\bf p}) ,\\
{{\partial E_{\parallel}({\bf p})}\over {\partial t}} &=&   S_{\parallel}({\bf p}|{\bf q}|{\bf k'}) + S_{\parallel}({\bf p}|{\bf k'}|{\bf q}) +  \mathcal{P}_{\parallel}({\bf p})  -D_\parallel({\bf p}) ,\\
{{\partial E_{\perp}({\bf q})}\over {\partial t}} &=&   S_{\perp}({\bf q}|{\bf k'}|{\bf p}) + S_{\perp}({\bf q}|{\bf p}|{\bf k'}) +  \mathcal{P}_{\perp}({\bf q}) -D_\perp({\bf q}) ,\\
{{\partial E_{\parallel}({\bf q})}\over {\partial t}} &=&   S_{\parallel}({\bf q}|{\bf k'}|{\bf p}) + S_{\parallel}({\bf q}|{\bf p}|{\bf k'}) +  \mathcal{P}_{\parallel}({\bf q})  -D_\parallel({\bf q}).
\end{eqnarray}
Using ${\bf k \cdot {\hat U}(k)}=0$, we can show that 
\begin{eqnarray}
\mathcal{P}_{\perp}({\bf k'})+\mathcal{P}_{\parallel}({\bf k'}) = 0,\\
S_{\perp}({\bf k'}|{\bf p}|{\bf q}) = -S_{\perp}({\bf p}|{\bf k'}|{\bf q}),\\
S_{\parallel}({\bf k'}|{\bf p}|{\bf q}) = -S_{\parallel}({\bf p}|{\bf k'}|{\bf q}).
\end{eqnarray}
and
\begin{eqnarray}
S_{\perp}({\bf k'}|{\bf p}|{\bf q}) + S_{\perp}({\bf k'}|{\bf q}|{\bf p})  + S_{\perp}({\bf p}|{\bf k'}|{\bf q})  & & \nonumber \\
+\, S_{\perp}({\bf p}|{\bf q}|{\bf k'}) + S_{\perp}({\bf q}|{\bf k'}|{\bf p})  + S_{\perp}({\bf q}|{\bf p}|{\bf k'})  &=& 0, \\
S_{\parallel}({\bf k'}|{\bf p}|{\bf q}) + S_{\parallel}({\bf k'}|{\bf q}|{\bf p})  + S_{\parallel}({\bf p}|{\bf k'}|{\bf q}) & & \nonumber \\
+\,S_{\parallel}({\bf p}|{\bf q}|{\bf k'}) + S_{\parallel}({\bf q}|{\bf k'}|{\bf p})  + S_{\parallel}({\bf q}|{\bf p}|{\bf k'})  &=& 0.
\end{eqnarray}

Using the above equations and ignoring the dissipation terms, we  conclude that 
\begin{eqnarray}
{{\partial }\over {\partial t}}  \left[ E_{\perp}({\bf k'}) + E_{\perp}({\bf p}) + E_{\perp}({\bf q}) \right]  &=&  \mathcal{P}_{\perp}({\bf k'}) + \mathcal{P}_{\perp}({\bf p})+\mathcal{P}_{\perp}({\bf q}), \label{eq:Eperp} \\
{{\partial }\over {\partial t}}  \left[ E_{\parallel}({\bf k'}) + E_{\parallel}({\bf p}) + E_{\parallel}({\bf q}) \right] &=&  -\left[ \mathcal{P}_{\perp}({\bf k'}) + \mathcal{P}_{\perp}({\bf p})+\mathcal{P}_{\perp}({\bf q}) \right]. \label{eq:Epll}
\end{eqnarray}

Therefore, we can make the following conclusions regarding the energy transfers for the parallel and perpendicular components of the velocity field:
\begin{enumerate}
\item The sum of Equations~(\ref{eq:Eperp}) and (\ref{eq:Epll}) shows that the total energy (sum of the perpendicular and parallel components) for a triad is conserved. However, there is an energy transfer between the perpendicular and parallel components via pressure.

\item The perpendicular component ${\bf \hat U_\perp(k')}$ receives  energy by an amount $S_{\perp}({\bf k'|p|q})$ from ${\bf \hat U_\perp(p)}$ with ${\bf \hat U(q)}$ as a mediator. Symmetrically, it also receives energy by an amount $S_{\perp}({\bf k'|q|p})$ from ${\bf \hat U_\perp(q)}$ via ${\bf \hat U(p)}$.  The parallel component ${ \hat U_\parallel({\bf k')}}$ receives energy by amounts $S_{\parallel}({\bf k'|p|q})$ and $S_{\parallel}({\bf k'|q|p})$ respectively from the modes ${ \hat U_\parallel({\bf p})}$ and ${ \hat U_\parallel({\bf q})}$ with ${\bf \hat U(q)}$ and ${\bf \hat U(p)}$ acting as the respective mediators.

\item  Equation~(\ref{eq:dt_Eperp}) implies that the perpendicular component ${\bf \hat U_\perp(k')}$ gains energy from the $\mathcal{P}_{\perp}({\bf k'})$ term, which arises due to the pressure.  Since  $\mathcal{P}_{\perp}({\bf k'}) = -\mathcal{P}_{\parallel}({\bf k'}) $, the energy gained by  ${\bf \hat U_\perp(k')}$ via pressure is the same as the energy lost by ${ \hat U_\parallel({\bf k})}$ [see Equation~(\ref{eq:dt_Epll})].  Hence, the energy transfer between the parallel and perpendicular components occur via pressure.

\item Since ${\bf k'} = -{\bf k}$, $E({\bf k'}) = E({\bf k})$.  It is customary to express the energy transfers in terms of $({\bf k,p,q})$.  For the same we replace ${\bf {\hat U}(k') = {\hat U}^*(k)}$.
\end{enumerate}

We use the aforementioned formulas to compute the energy fluxes of the perpendicular and parallel components of the velocity field.    The energy flux $\Pi_\perp(k_0)$ for the perpendicular component of the velocity field for a wavenumber sphere of radius $k_0$  is defined as the net energy transferred  from the modes $\mathbf{U}_\perp(\mathbf p)$  residing inside the sphere to the modes $\mathbf{U}_\perp(\mathbf k)$ outside the sphere,  i.e.,
\begin{equation}
\Pi_{\perp}(k_0) =  \sum_{|{\bf k}| \geq k_0 } \sum_{|{\bf p}| < k_0}  {  S_{\perp}({\bf k|p|q)}}.
\end{equation}
A similar formula for $\Pi_{||}(k_0)$, the flux of the parallel velocity component, is
\begin{equation}
\Pi_{\parallel}(k_0) =  \sum_{|{\bf k}| \geq k_0 } \sum_{|{\bf p}| < k_0}  {  S_{\parallel}({\bf k|p|q)}}.
\end{equation} 
The total flux is the sum of the above two fluxes.

\printindex
\cleardoublepage
%\phantomsection

%\addcontentsline{toc}{chapter}{Bibliography}%
%\input{bib.tex}
%\input{Untitled.bib}
%\input{Untitled2.bib}

{\bf References} \\

%\input{main.bbl}
%\bibliographystyle{abbrv_abhishek}
% USE THIS TO SUPPRESS MONTH
%\AtEveryBibitem{\clearfield{month}}
%\bibliography{/Users/mkv/Dropbox/docs-pub/bib/journal,/Users/mkv/Dropbox/docs-pub/bib/book,/Users/mkv/Dropbox/docs-pub/bib/conf,/Users/mkv/Dropbox/docs-pub/bib/preprint,/Users/mkv/Dropbox/docs-pub/bib/thesis,/Users/mkv/Dropbox/docs-pub/bib/reports}

%\bibliography{Untitled2}

\end{document}